\documentclass[a4paper,11pt]{article}
\usepackage{jcappub}
\usepackage{lineno}
\usepackage{adjustbox}
\usepackage{tabularx}
\usepackage{lscape}

\newcommand\J {\ensuremath{J}}
\def\GS {\ensuremath{\mathcal{AGS}}}
\def\LS {\ensuremath{\mathcal{PS}}}
\newcommand{\sv}{$\langle \sigma \mathit{v} \rangle$}
\newcommand\CellTopTwo{\rule{0pt}{2.8ex}}

\usepackage[capitalize]{cleveref}
\crefformat{section}{Sec.~#2#1#3}
\Crefformat{section}{Section~#2#1#3}
\crefformat{appendix}{App.~#2#1#3}
\Crefformat{appendix}{Appendix~#2#1#3}
\crefformat{table}{Tab.~#2#1#3}
\Crefformat{table}{Table~#2#1#3}
\crefformat{pluralequation}{#2\black{Eqs.~(}#1\black{)}#3}
\Crefformat{pluralequation}{#2\black{Equations~(}#1\black{)}#3}
\crefformat{pluralfigure}{#2\black{Figs.~}#1#3}
\Crefformat{pluralfigure}{#2\black{Figures~}#1#3}

\title{Improved Heavy Dark Matter Annihilation Search from Dwarf Galaxies with HAWC.}
\collaboration{The High Altitude Water Cherenkov (HAWC) Collaboration}
\collaborationImg{
  \includegraphics[height=4cm]{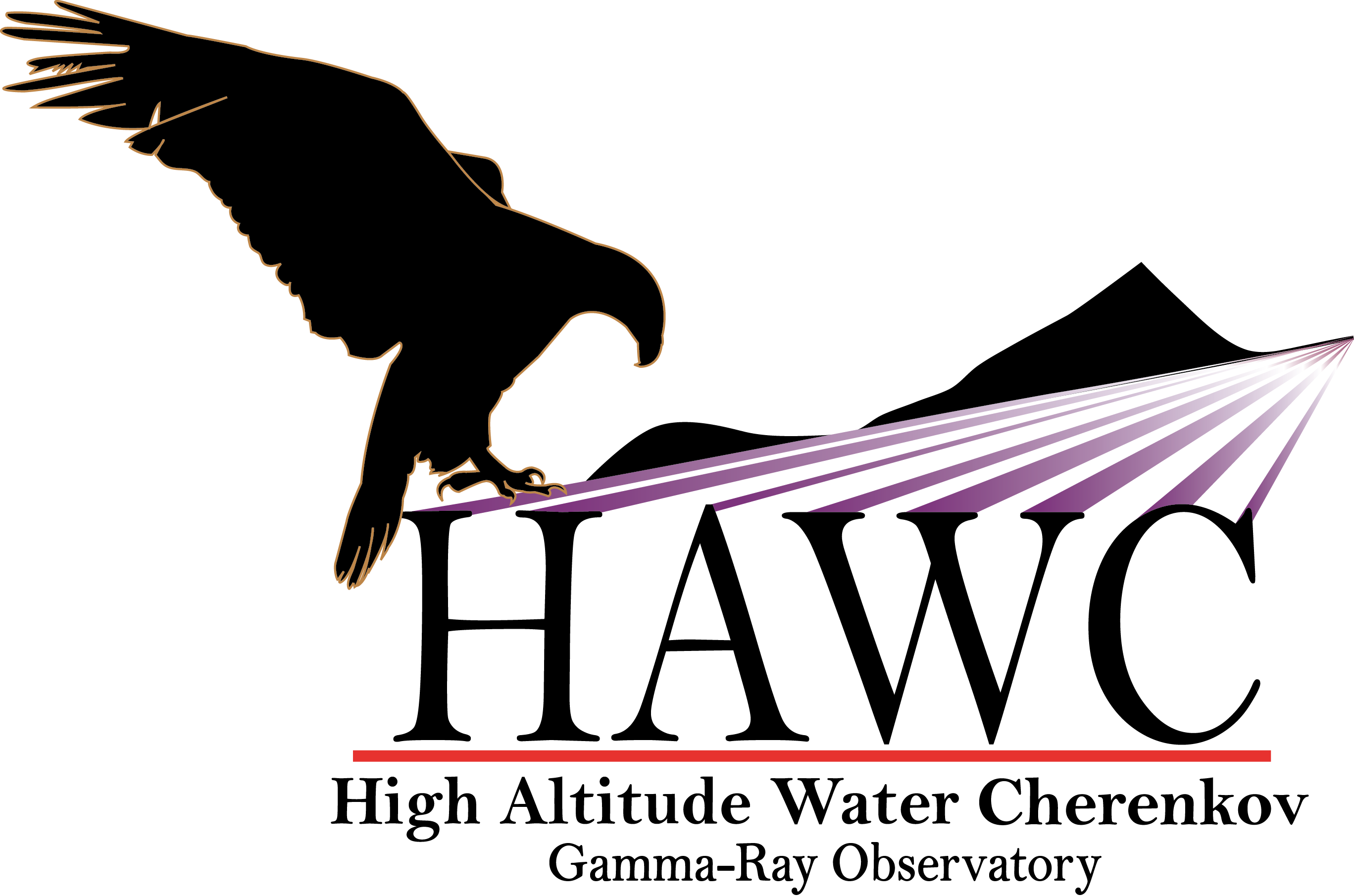}
}

\abstract{Understanding dark matter’s elusive nature is crucial for the framework of particle physics and expanding the Standard Model. This analysis utilizes the High Altitude Water Cherenkov (HAWC) gamma ray Observatory to indirectly search for dark matter (DM) by studying gamma ray emission from dwarf spheroidal galaxies (dSphs). Selected for their high ratio of dark matter to baryonic matter, dSphs are useful for this type of search owing to the low background emission. In comparison to previous HAWC studies, we significantly improve our sensitivity to DM from dSphs due to improvements to our event reconstruction and reduced hadronic contamination. We expanded the number of dSphs studied, DM annihilation channels into the Standard Model (SM), and the amount of data collected on each previously studied dSph. We searched for DM signals in each dSph using the latest version of the algorithms used to reconstruct data from the primary detector of the HAWC instrument. We report that we do not detect evidence of DM from dSphs, so we place upper limits for the velocity-weighted DM annihilation cross-section ($\langle\sigma v \rangle$) on the order of $10^{-23}~\text{cm}^3\text{s}^{-1}$ for a DM mass range of $1-10^4$ TeV.}

\author[1]{A.~Albert}

\author[2]{R.~Alfaro}

\author[3]{C.~Alvarez}

\author[4]{A.~Andrés}

\author[4]{E.~Anita-Rangel}

\author[5]{M.~Araya}

\author[6]{J.C.~Arteaga-Velázquez}

\author[4]{D.~Avila Rojas}

\author[7]{H.A.~Ayala Solares}

\author[8]{R.~Babu}

% ...

\affiliation[1]{Los Alamos National Laboratory, Los Alamos, NM, USA }

\affiliation[2]{Instituto de F\'{i}sica, Universidad Nacional Autónoma de México, Ciudad de Mexico, Mexico }

\affiliation[3]{Universidad Autónoma de Chiapas, Tuxtla Gutiérrez, Chiapas, México}

\affiliation[4]{Instituto de Astronom\'{i}a, Universidad Nacional Autónoma de México, Ciudad de Mexico, Mexico }

\affiliation[5]{Universidad de Costa Rica, San José 2060, Costa Rica}

\affiliation[6]{Universidad Michoacana de San Nicolás de Hidalgo, Morelia, Mexico }

\affiliation[7]{7 University, Department of Physics, 1925 N. 12th Street, Philadelphia, PA 19122, USA}

\affiliation[8]{Department of Physics and Astronomy, Michigan State University, East Lansing, MI, USA }

% ...

\emailAdd{hawc-publications@umdgrb.umd.edu}
\emailAdd{salaza82@msu.edu}
\emailAdd{claire.lundy@tufts.edu}
\emailAdd{tollefs2@msu.edu}

\begin{document}

\maketitle
\flushbottom

\section{Introduction}\label{sec:intro} %%%%%%%%%%%%%%%%%%%%%%%%%%%%%%%%%%%%%%%%%
While evidence for the existence of dark matter (DM) has been observed clearly within the universe, its true nature is yet to be uncovered.
The large rotational velocities of galaxies compared to their visible mass content \cite{HistoryOfDM} is just one piece of evidence for DM’s existence.
There are numerous particle candidates that could potentially make up all or a fraction of DM \cite{HistoryOfDM}.
To narrow our search and discuss more specific candidates, we assume a Cold Dark Matter (CDM) structure formation history.
This defines dark matter as non-relativistic, or “cold”.
Within these constraints, there is motivation for the so-called Weakly Interacting Massive Particles (WIMPs) as a potential dark matter candidate.
WIMPs interact only through the weak and gravitational forces and exist at mass scales from GeV to TeV \cite{STRIGARI20131}.
In this study, we presume heavy WIMP DM can annihilate into Standard Model (SM) particles which decay to final state photons that HAWC can detect.

Our search focuses solely on dwarf spheroidal (dSphs) galaxies.
These make ideal targets for a DM search because of their high dark matter to baryonic matter ratio, low overall baryonic matter content, and position as satellites of the Milky Way \cite{arg_for_dsph}.
We selected 17 dSphs for this analysis due to their positions in HAWC's field of view (FOV) from about -26$^\circ$ to +64$^\circ$ \cite{Abeysekara_2023} and well-studied dark matter parameters.
Previous HAWC studies were limited to 15 dSphs \cite{Albert_2018} because observations of Willman 1 and Draco II had significant uncertainties \cite{Geringer-Sameth_2015}.
Recent observations \cite{DM_Strigari20,Ando_2020} constrain the DM densities around these two dSphs, so we include them for this study.
The dSphs considered in this analysis are Boötes I, Canes Venatici I, Canes Venatici II, Coma Berenices, Draco, Draco II, Hercules, Leo I, Leo II, Leo IV, Leo V, Pisces II, Segue 1, Sextans, Ursa Major I, Ursa Major II, and Willman 1.
The properties of these sources are listed in \cref{tab:mtd_J_factor}.

This work is built off previous HAWC analyses \cite{GloryDuck, Albert_2018,HAWC_dm_gammalines}, and this study improves on the previous in the following ways.
The particle physics model used for gamma ray spectra was updated to incorporate recent measurements and constraints in particle physics.
The details of this are discussed in \cref{sec:mtd_particlephysics}.
For this study, HAWC samples DM masses up to $10$ PeV, where previously we stopped at 1 PeV \cite{HAWC_dm_gammalines}.
To model the DM spatial profiles, we use two different DM density publications: from A. Pace and L. Strigari \cite{DM_Strigari20} and S. Ando and A. Geringer-Sameth \cite{Ando_2020}.

We also implement significant updates to HAWC's data and reconstruction that improved on our energy estimation and gamma/hadron separation.
Unlike previous HAWC dSphs analyses that relied on an older ``Pass 4'' reconstruction \cite{Albert_2018, HAWC_dm_gammalines, Abeysekara_2017} and used the fraction of the PMT hits on the main array ($f_{hit}$)  \cite{GloryDuck}, we utilize the newer ``Pass 5'' reconstructed data \cite{HAWC_Pevatron} that significantly improves on HAWC's angular resolution and sensitivity.
We also use a Neural Network (NN) based energy estimator, improving on the energy resolution and the maximum energy sensitivity in this work.
Additionally, HAWC introduced a Machine Learning (ML) optimization suite for selecting cuts on event data in the detector that better discriminates between gamma rays versus hadrons \cite{P5MLcuts}.
These newer datasets improve our sensitivity by more precisely estimating the energies of photon events and reducing the cosmic ray contamination in the gamma ray data.

In this paper, we construct spatial models for the dark matter content of each dSph and annihilation spectra for 11 SM channels to calculate the expected gamma ray flux due to dark matter annihilation. \Cref{sec:data_bkgd} describes the HAWC observatory and our method of background estimation. \Cref{sec:mtd_analysis} details our method of determining DM annihilation using the DM density calculated from two different dSph catalogs and gamma ray spectra. \Cref{sec:mtd_results} contains the results of our analysis across both catalogs of 17 dwarfs for each SM channel. While no gamma ray excess was found in the dSphs, we present the upper limits on annihilation cross-section for each dSph and SM channel and compare to previous searches and sensitivities. Finally, \cref{sec:mtd_conclusion} contains the conclusion and discussion of our results.

\section{The HAWC Observatory}\label{sec:hawc}

Located near the Sierra Negra volcano in Puebla, Mexico, the High-Altitude Water Cherenkov (HAWC) Observatory detects high-energy gamma rays and cosmic rays. Its main array has been operating since March 2015 with an uptime of over 95\%. HAWC is positioned at $4,100$ meters above sea level and covers over $100,000$ m$^2$ \cite{Abeysekara_2023}.
Its FOV captures a declination range from about -26$^\circ$ to +64$^\circ$ \cite{Abeysekara_2023}. HAWC is composed of a number of water Cherenkov detectors (WCDs) that are separated into a main array and outriggers. The main array covers an area of $22,000$ m$^2$ and is composed of $300$ WCDs, each equipped with $200$ kL of purified water and $4$ photomultiplier tubes (PMTs) to detect charged particle showers from cosmic- or gamma rays of energies from $0.1$ TeV to over $100$ TeV \cite{outrigger_paper}. Outside the main array are the outriggers. These cover an area of $100,000$ m$^2$ and are $350$ smaller WCDs containing only $1$ PMT each and spaced much farther apart. These provide additional resolution to air showers whose cores fall outside the main array by improving our overall angular reconstruction and energy estimation \cite{outrigger_paper}.

When a very high-energy gamma ray enters Earth’s atmosphere, it interacts with the atmospheric particles and causes an extensive air shower. The showers are composed of secondary particles that "rain" down on Earth. When the charged particles enter into the WCDs they produce Cherenkov light detected by the PMTs.
HAWC records the time and charge measured by the PMTs \cite{Abeysekara_2023}.
We estimate the energy, reconstruct the direction of the primary particle, and discriminate between cosmic and gamma rays from the ensemble of PMT hit data \cite{Abeysekara_2017}.
These events are sorted into analysis bins developed using a NN energy reconstruction algorithm as described in \cite{Abeysekara_2019,P5MLcuts}.

The gamma ray dataset used in this study includes 3070 days of HAWC operation, compared to 507 days from \cite{Albert_2018}, and 1038 days from \cite{GloryDuck} and \cite{HAWC_dm_gammalines}.
This study also utilizes events which have reconstructed shower cores off the main and implements highly sensitive cuts discerned with ML algorithms \cite{P5MLcuts}, neither of which has been done in previous DM dSph studies.
In our analysis framework, we used HAWC public likelihood fitting software \texttt{HAL} and \texttt{threeML} \cite{3MLandHAL}.
dSphs are treated as extended sources in HAWC data and are analyzed as described in \cite{Albert_2018}.

\section{Background} \label{sec:data_bkgd} %%%%%%%%%%%%%%%%%%%%%%%%%%%%%%%%%

We consider the following for estimating the background around dSphs.
The dSphs' angular extent are similar to HAWC's spatial resolution, so the analysis is not contaminated by large-scale anisotropies in the cosmic ray background.
Most sources are fully contained within 1-5 \texttt{HEALpix} pixels with only a couple (Sextans and Ursa Major II) dSphs requiring more, with a pixel size of 0.025 steradians.
The dSphs used in this analysis are off the galactic plane and therefore not contaminated by diffuse emission from the Milky Way.
Moreover, they are baryonically faint relative to their expected dark matter content and do not host bright high-energy gamma ray sources.
Therefore, we make no additional assumptions on the background from our sources and use HAWC's standard direct integration method for low energy, gamma ray background estimation \cite{Abeysekara_2017,CRAB5}, and background randomization \cite{Abeysekara_2019} for the highest energy gamma rays.

The largest background under this consideration is from an isotropic flux of cosmic rays.
The contamination of this hadronic flux is worse at lower energies where HAWC's gamma ray and hadronic events appear identical.

\section{Analysis}\label{sec:mtd_analysis} %%%%%%%%%%%%%%%%%%%%%%%%%%%%%%%%%%%%%%%

The expected differential photon flux from DM-DM annihilation to standard model
particles is described as

\begin{equation}\label{eq:id_dm_flux}
	\frac{d\Phi_\gamma}{dE_{\gamma}} = \frac{\langle \sigma \mathit{v} \rangle}{8 \pi m^2_{\chi}} \frac{dN_\gamma}{dE_\gamma}\times J,
\end{equation}
where \sv~is the velocity weighted annihilation cross-section, $\frac{dN}{dE}$ is the expected energy-weighted differential number of photons produced at annihilation, $m_\chi$ is the rest mass of the supposed DM particle, and $J$ is the astrophysical \J-factor, integrated over a solid angle, $\Omega$. \J~is defined as

\begin{equation}\label{eq:jfactor}
	J = \int d\Omega \int_{l.o.s} dl \rho^2_{\chi}(r, \theta'),
\end{equation}
where $\rho_{\chi}$ is the DM density, $l$ is the line of sight distance to the source from Earth, $r$ is the radial distance from the center of the source, and $\theta'$ is the half angle defining a cone containing the DM source.
How each component is synthesized and considered for HAWC's analysis is presented in the following sections.
\Cref{sec:mtd_particlephysics} presents the particle physics model for DM annihilation.
\Cref{sec:mtd_spatialmodel} presents the spatial distributions built for each dSph.

%%%%%%%%%%%%%%%%%%%%%%%%%%%%%%%%%%%%%%%%%%%%%%%%
\subsection{DM to gamma ray Spectra}\label{sec:mtd_particlephysics}
%%%%%%%%%%%%%%%%%%%%%%%%%%%%%%%%%%%%%%%%%%%%%%%%
We use \texttt{HDMSpectra} (HDMS) \cite{Rodd:HDM_spec} for the estimation of the gamma ray spectra for each DM annihilation channel to the SM.
This repository is comparable to the \textit{Poor Particle Physicist Cookbook} (PPPC) \textit{for Dark Matter Indirect Detection} \cite{PPPC} DM spectral models but includes electro-weak (EW) corrections and higher-order loop corrections from $W$ and $Z$ bosons \cite{Rodd:HDM_spec}. These corrections allow for production of photon spectra from the EW to Planck scale, making it ideal for our mass range of heavy DM.

\begin{figure}[htb!]
	\centering{
        \includegraphics[width=0.6\textwidth]{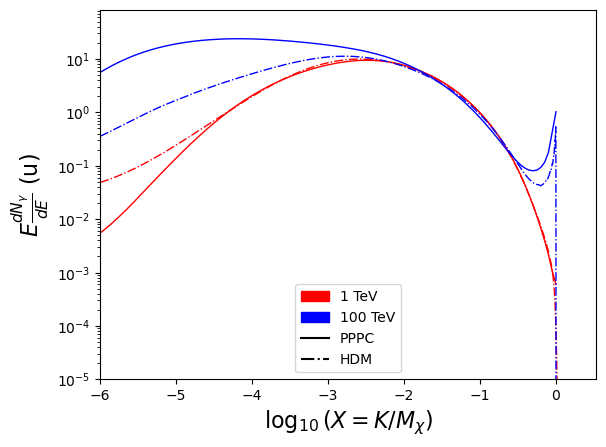}
	}
    \caption{Spectral hypotheses from PPPC and HDMS for DM annihilation: $\chi\chi \rightarrow W^-W^+$. The x-axis is the $\text{Log}_{10}$ of X, where X is the ratio of gamma ray energy to DM mass. The y-axis is the unitless, energy-weighted gamma ray counts. Solid lines are spectral models with EW corrections from the PPPC. Dash-dot lines are spectral models from HDMS. Red lines are models for $M_\chi = 1$ TeV, the lowest in our studied range. Blue lines represent models for $M_\chi = 100$ TeV, which are well above \texttt{PYTHIA}'s limits, where corrections are necessary \cite{Rodd:HDM_spec}.}
    \label{fig:pppc_vs_hdm}
\end{figure}

\Cref{fig:pppc_vs_hdm} demonstrates the impact of changes implemented with HDMS on DM annihilation to $W$ bosons in contrast to the PPPC. At masses around $1$ TeV, the spectra fall in general agreement while at masses of $100$ TeV, the HDMS spectrum initially falls an order of magnitude below the PPPC line and maintains a more conservative model. The SM DM annihilation channels studied here are $\chi\chi \rightarrow$: $e^+e^-$, $\mu^+\mu^-$, $\tau^+\tau^-$,$b\bar{b}$, $t\bar{t}$, $W^+W^-$, $Z^0Z^0$,  $u\bar{u}$, $d\bar{d}$, $\nu_e \overline{\nu}_e$, and $\gamma\gamma$.

\begin{figure}[htbp!]
    \centering{
    \includegraphics[scale=0.3]{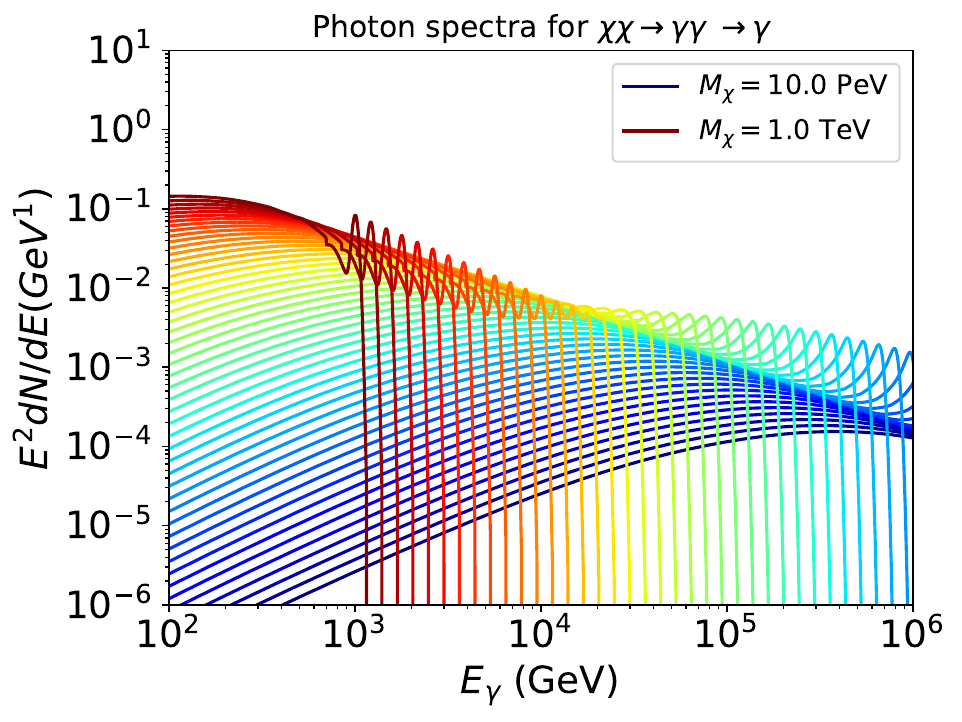}
    \includegraphics[scale=0.3]{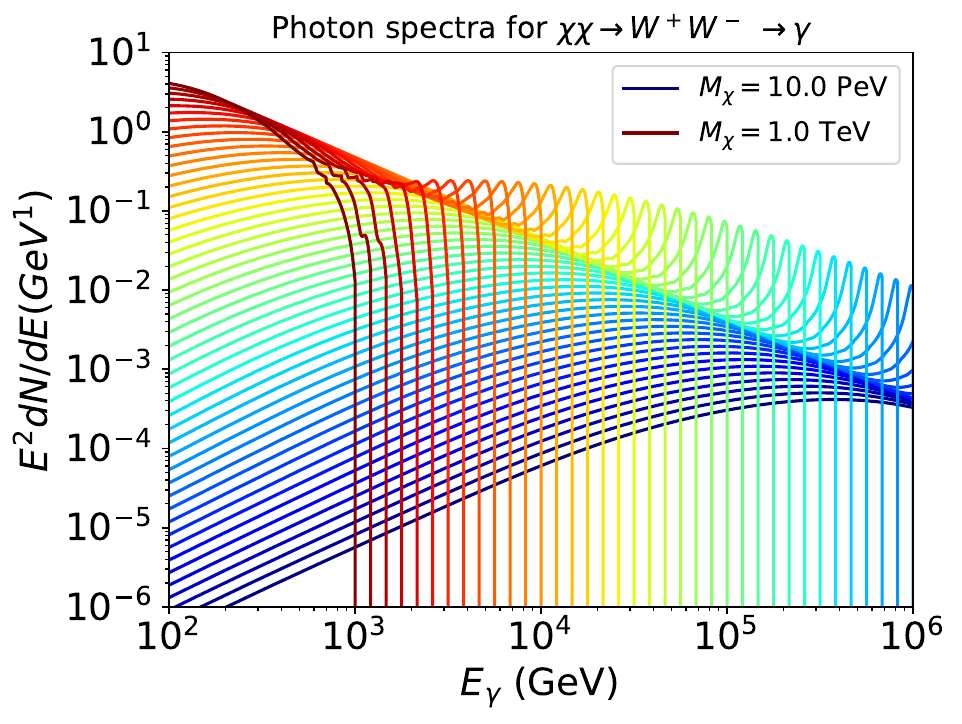}
    \includegraphics[scale=0.3]{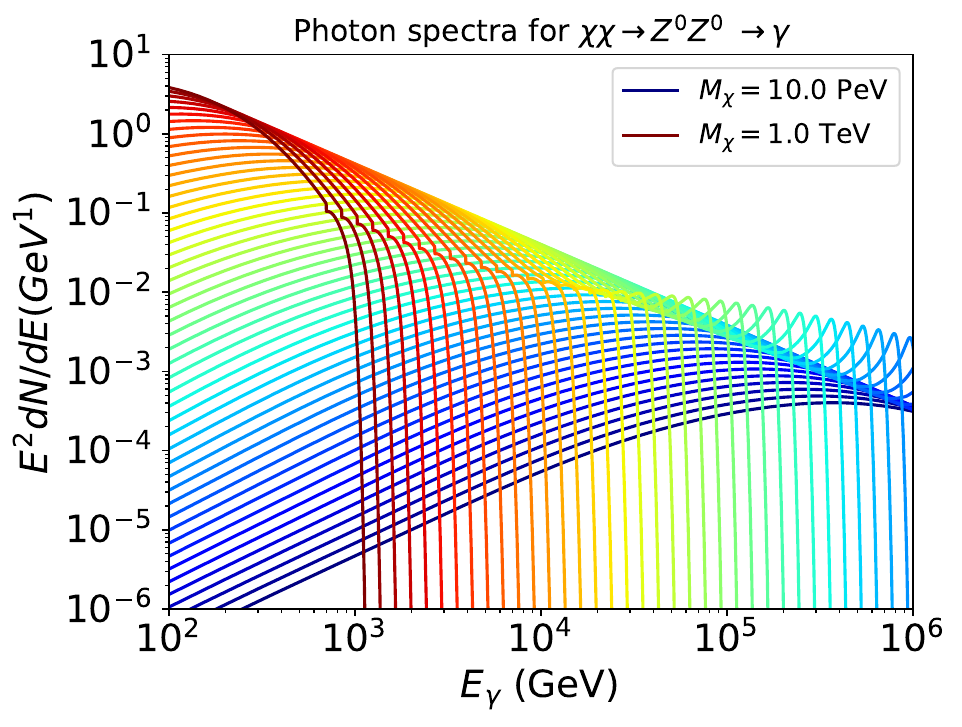}
    }
    \caption{Photon spectra for $\chi\chi \rightarrow \gamma\gamma$ (left), $\chi\chi \rightarrow WW$ (middle), and $\chi\chi \rightarrow Z^0Z^0$ (right) after Gaussian convolution of line features. Both the 1 TeV and 10 PeV spectra have $\delta$-line features at $E_\gamma = m_\chi$. Redder lines are annihilation spectra with lower DM mass. Bluer lines are spectra from larger DM mass. All spectral models are sourced from the HDMS models \cite{Rodd:HDM_spec}. Axes are drawn according to the energy sensitivity of HAWC.}
    \label{fig:hdm_gamma_lines1}
\end{figure}

For the $\gamma\gamma$, $W^+W^-$, and $Z^0Z^0$ annihilation channels, a substantial fraction of the signal photons are expected to have $E_\gamma \approx m_\chi$ \cite{Rodd:HDM_spec}.
This introduces a spectral line that is much narrower than the energy resolution of the HAWC detector.
To ensure that we do not underestimate the gamma ray contribution to the flux, the line feature is convolved with a Gaussian kernel with a $1\sigma$ one-sided width of $5\%~m_\chi$ and total convolution window of $\pm4\sigma$.
The kernel width was chosen based on the energy resolution of the utilized energy estimator (similar to \cite{HAWC_dm_gammalines}).
The NN energy estimator's (see \cite{thesis_SamM}) improved energy resolution compared to $f_\mathrm{hit}$, an alternative method of binning data by fraction of PMTs hit \cite{Abeysekara_2017}, at low gamma ray energy enables narrower kernels.
This kernel is not well represented at lower photon energies and underestimates the photon spectra, but agrees at high energies.
Ultimately, because our analysis is focused on constraining the upper end of the mass range, this approximation is adequate for our purposes.
The annihilation spectra after Gaussian smoothing for $\chi\chi \rightarrow \gamma\gamma$, $W^+W^-$, and $Z^0Z^0$ spectral hypotheses are shown in \cref{fig:hdm_gamma_lines1}.

All spectra used in this analysis are prompt only, and we do not correct for secondary emission from inverse Compton scattering (ICS) or electromagnetic (EM) cascades.
The impact of ICS and EM cascades requires additional layered assumptions of the radiation environment and electron transport.
Secondary gamma-ray emission from the aforementioned processes is expected to enhance the gamma ray flux and change the apparent source morphology for higher dark matter mass models.
The limits presented in this work are therefore conservative and we invite future studies to investigate the impact of secondary emission on our DM limits.

%%%%%%%%%%%%%%%%%%%%%%%%%%%%%%%%%%%%%%%%%%%%%%%%%
\subsection{Dark Matter Density Distributions and Source Selection}\label{sec:mtd_spatialmodel}
%%%%%%%%%%%%%%%%%%%%%%%%%%%%%%%%%%%%%%%%%%%%%%%%%

J-factors are the largest source of model uncertainty in our analysis.
The J-factor spatial templates for each source are created using the scale radius and scale density parameters of each dSph (see \cref{eq:jfactor,eq:nfw_density}) with a region of interest of 4$^\circ$ centered on each dSph.
Because the angular size of the dSphs are close to HAWC's angular resolution, we are not as sensitive to the underlying DM density distribution.
The cumulative \J-factor, however, is sensitive to the stellar selection and statistical priors that inform the fit (compare \cite{DM_Strigari20} to \cite{Ando_2020}).
Therefore, we selected dwarf studies with diverse methods of deriving \J-factors (see \cref{sec:LS,sec:GS}).
We implement dSph DM density parameters from both Pace-Strigari et al. (referred to with \LS{}) \cite{DM_Strigari20} and Ando et al. (referred to with \GS{}) \cite{Ando_2020}.
The parameters used for this study are summarized in \cref{tab:mtd_J_factor}.

\begin{table}[htb!]
\centering
\begin{adjustbox}{width=.95\textwidth}% angle=90}
    \small{\begin{tabular}{ccccccc}
    \hline
    \hline
    \CellTopTwo{}
    Name & \LS{} Distance & \GS{} Distance & $l, b$ & $\alpha_c$ & $\log_{10}J$~(\LS{} set) & $\log_{10}J$~(\GS{} set) \\
    & \scriptsize{(kpc)} & \scriptsize{(kpc)} & \scriptsize{($^\circ$)} & \scriptsize{($^\circ$)} & \scriptsize{$\log_{10}(\mathrm{GeV}^2 \mathrm{cm}^{-5}\mathrm{sr})$} & \scriptsize{$\log_{10}(\mathrm{GeV}^2 \mathrm{cm}^{-5}\mathrm{sr})$} \\
    \hline
    \CellTopTwo{}
    Boötes I & $66.0 \pm 3.0$ & $66.0$ & $358.08,\: 69.62$ & 0.325 & $17.83^{+0.44}_{-0.45}$ & $17.79^{+0.23}_{-0.25}$ \\
    \CellTopTwo{}
    Canes Venatici I & $210.0 \pm 6.0$ & $218.0$ & $91.39,\: 73.24$ & 0.231 & $17.24^{+0.36}_{-0.34}$ & $17.38^{+0.12}_{-0.11}$ \\
    \CellTopTwo{}
    Canes Venatici II & $160.0 \pm 7.0$ & $160.0$ & $113.58,\: 82.70$ & 0.049 & $17.43^{+0.61}_{-0.62}$ & $17.20^{+0.40}_{-0.44}$ \\
    \CellTopTwo{}
    Coma Berenices & $42.0 \pm 1.5$ & $44.0$ & $241.89,\: 83.61$ & 0.156 & $18.67^{+0.36}_{-0.35}$ & $18.39^{+0.30}_{-0.33}$ \\
    \CellTopTwo{}
    Draco & $76.0 \pm 6.0$ & $76.0$ & $86.37,\: -34.72$ & 0.274 & $18.80^{+0.26}_{-0.25}$ & $ 18.75^{+0.20}_{-0.21}$ \\
    \CellTopTwo{}
    Draco II & $20.0 \pm 3.0$ & $20.0$ & $98.29,\: 42.88$ & 0.069 & $19.43^{+0.98}_{-1.35}$ & $ 18.31^{+0.60}_{-0.72}$ \\
    \CellTopTwo{}
    Hercules & $132.0 \pm 6.0$ & $132.0$ & $28.73,\: 36.87$ & 0.085 & $17.12^{+0.71}_{-0.71}$ & $16.94^{+0.33}_{-0.38}$ \\
    \CellTopTwo{}
    Leo I & $258.2 \pm 9.5$ & $254.0$ & $226.00,\: 49.10$ & 0.130 & $17.51^{+0.31}_{-0.31}$ & $17.74^{+0.09}_{-0.09}$ \\
    \CellTopTwo{}
    Leo II & $233.0 \pm 15.0$ & $233.0$ & $220.17,\: 67.23$ & 0.078 & $17.53^{+0.30}_{-0.30}$ & $17.59^{+0.11}_{-0.11}$ \\
    \CellTopTwo{}
    Leo IV & $154.0 \pm 5.0$ & $154.0$ & $265.44,\: 56.51$ & 0.083 & $16.69^{+0.95}_{-1.41}$ & $16.59^{+0.52}_{-0.64}$ \\
    \CellTopTwo{}
    Leo V & $173.0 \pm 5.0$ & $178.0$ & $261.86,\: 58.54$ & 0.020 & $17.79^{+0.91}_{-0.76}$ & $16.58^{+0.62}_{-0.70}$ \\
    \CellTopTwo{}
    Pisces II & $183.0 \pm 15.0$ & $182.0$ & $79.21,\: -47.11$ & 0.030 & $17.50^{+1.06}_{-1.06}$ & $16.57^{+0.65}_{-0.77}$ \\
    \CellTopTwo{}
    Segue 1 & $23.0 \pm 2.0$ & $23.0$ & $220.50,\: 50.40$ & 0.105 & $18.90^{+0.58}_{-0.62}$ & $18.91^{+0.39}_{-0.45}$ \\
    \CellTopTwo{}
    Sextans & $92.5 \pm 2.2$ & $86.0$ & $243.50,\: 42.27$ & 0.648 & $17.74^{+0.33}_{-0.33}$ & $18.32^{+0.23}_{-0.23}$ \\
    \CellTopTwo{}
    Ursa Major I & $97.3 \pm 5.85$ & $97.0$ & $159.40,\: 54.40$ & 0.234 & $18.16^{+0.47}_{-0.45}$ & $18.18^{+0.22}_{-0.25}$ \\
    \CellTopTwo{}
    Ursa Major II & $34.7 \pm 2.1$ & $32.0$ & $152.50,\: 37.40$ & 0.327 & $19.34^{+0.59}_{-0.61}$ & $18.78^{+0.36}_{-0.42}$ \\
    \CellTopTwo{}
    Willman 1 & $38.0 \pm 7.0$ & $38.0$ & $158.58,\: 56.78$ & 0.054 & $19.18^{+0.41}_{-0.35}$ & $18.02^{+0.59}_{-0.69}$ \\
    
    \hline
    \hline
    \CellTopTwo{}
\end{tabular}}
\end{adjustbox}
    \caption{Summary of the relevant properties of the dSphs used in the present work. Columns 2 and 3 present their heliocentric distances as reported by \LS{} and \GS{}, respectively, where uncertainties are included when reported. Column 4 is the galactic coordinates of each dwarf. Column 5 is $\alpha_c=2r_{1/2}/d$, the profile cut off angle where \LS{} J-factor uncertainties are minimized \cite{DM_Strigari20}. Columns 6 and 7 give the \J-factors of each source from the \LS{} and \GS{} studies, respectively, and estimated $\pm 1\sigma$ uncertainties. The values $\log_{10}J$~(\LS{} set) \cite{DM_Strigari20} correspond to the mean \J-factor values for a source extension truncated at $\alpha_c$ while the values $\log_{10}J$~(\GS{} set) \cite{Ando_2020} correspond to the median J-factor values for a source extension truncated at $0.5^\circ$. }
    \label{tab:mtd_J_factor}
\end{table}

Both catalogs fit a Navarro–Frenk–White (NFW) \cite{NFWProfile} spatial DM distribution to the dSphs which has a DM density of

\begin{equation}\label{eq:nfw_density}
        \rho (r) = \frac{\rho_0}{\frac{r}{R_s}\left( 1 + \frac{r}{R_s} \right)^2 }.
\end{equation}
The scale density, $\rho_0$, and the scale radius, $R_s$ are free parameters fit for each dSph in the \LS{} and \GS{} catalogs.
$r$ is the distance from the center of the dSph.
Using these reported parameters, we extract median and $\pm 1\sigma$ \J-factor distributions for each dwarf and build spatial templates of each dSph.
How each selection for each catalog is made is described in the following sections \cref{sec:LS,sec:GS}.
\Cref{sec:djdomdists} includes figures of both the median and $\pm1\sigma$ for the $\frac{d\J}{d\Omega}$ values as functions of $\theta$.
Our 2D templates for this study are fixed, and we do not marginalize over the \J-factor uncertainties in our likelihood analysis.

%%%%%%%%%%%%%%%%%%%%%%%%%%%%%%%%%%%%%%%%%%%%%%%%%
\subsubsection{\LS{} Catalog} \label{sec:LS}
%%%%%%%%%%%%%%%%%%%%%%%%%%%%%%%%%%%%%%%%%%%%%%%%%

The \LS{} catalog \cite{DM_Strigari20} determines the parameters for 43 dwarfs through log-uniform priors.
This publication postulates that because the \J-factor can be thought of as a ``flux", it will scale as the inverse of distance squared.
They further develop a scaling relation of the form

\begin{equation}
	J_\text{model}(\theta_\text{max})=J_0\left(\frac{\sigma_\text{los}}{5\text{ km s}^{-1}}\right)^{\gamma_{\sigma_\text{los}}}\left(\frac{d}{100\text{ kpc}}\right)^{\gamma_d}\left(\frac{r_{1/2}}{100\text{ pc}}\right)^{\gamma_{r_{1/2}}}
\end{equation}
to model the \J-factor as a function of the kinematic parameters of the dwarf where $\sigma_\text{los}$ is the line-of-sight velocity dispersion, $d$ is the heliocentric distance, $r_{1/2}$ is the half-light radius, and $\gamma_{\sigma_\text{los}}=4$, $\gamma_d=-2$, and $\gamma_{r_{1/2}}=-1$ are fixed slope parameters.
The uncertainty of this \J-factor is minimized at the angle $\alpha_c= 2r_{1/2}/d$ \cite{DM_Strigari20}.

The 2D maps used as input for HAWC analysis were constructed using the \texttt{Gammapy} Python package \cite{gammapy2023}.
The dwarf parameters can be used to determine the scale radius $r_s = 4r_{1/2}$ and the scale density

\begin{equation}
	\rho_s r_s=\frac{5\sigma^2_{\text{los}}r_{1/2}}{2G}\frac{1}{4\pi r^2_s\mathbf{g(r_{1/2}/r_s)}},
\end{equation}

where $\mathbf{g(x)}=log(1+x)-x/(1+x)$.
Spatial templates in $\frac{d\J}{d\Omega} (\theta)$ up to an angular separation of $\alpha_c$ were created to minimize \J-factor uncertainties \cite{DM_Strigari20}.

%%%%%%%%%%%%%%%%%%%%%%%%%%%%%%%%%%%%%%%%%%%%%%%%%
\subsubsection{\GS{} Catalog} \label{sec:GS}
%%%%%%%%%%%%%%%%%%%%%%%%%%%%%%%%%%%%%%%%%%%%%%%%%

The \GS{} catalog is an important addition to this analysis because it provides kinematic priors for dSphs rather than just cosmological log-uniform priors as in \cite{DM_Strigari20}. \GS{} adopts a WIMP dark matter model rather than maintaining model independence. This is not a problem for our Heavy WIMP search but extremely high mass becomes harder to motivate with this catalog. For classical dwarfs, this change is mostly inconsequential, as their well-defined velocity dispersion profiles allow for an accurate posterior even in the case of log-uniform priors. Yet for ultra-faints, the low statistics benefits from kinematic priors. These new priors ultimately lower the \J-factors across the catalog and weaken the dark matter annihilation cross-section limits \cite{Ando_2020}.

This catalog is also constructed using \texttt{Gammapy}. The $\rho_s$ and $r_s$ priors are not directly reported, however, and must be generated through Monte Carlo (MC) simulation \cite{Ando_2020}. \GS{} reports a third parameter $r_t$, the tidal truncation radius, which we use to truncate the \J-factor distributions. From there, the \J-factor can be approximated as described in \cite{Evans_2016} for each MC sample. We then use the median \J-factor to produce the spatial template up to an angular separation of $0.5^\circ$ as reported in \cite{Ando_2020}.

\subsection{Systematics}\label{sec:mtd_systemaics} %%%%%%%%%%%%%%%%%%%%%%%%%%%%%%%%%%%
The largest uncertainties in this analysis are the choices of \J-factor model, which has been discussed at length in \cref{sec:mtd_spatialmodel}.
The second-largest contributor is the systematics from the detector response (DR).
We study the impact of detector uncertainties from PMT efficiencies, late light, and calibration epochs.
When considering calibration epochs, we model our response for HAWC as early as first light in 2015.
Since then, there have been a varying number of PMTs and WCDs online.
We studied the effect of these different HAWC snapshots and varied signal acceptance and charge deposition rates via detector simulation \cite{Abeysekara_2019}.
We found the effects of these systematics to fall within our 68\% containment band and in most cases only deviating by an average factor of $1.02$.
In the worst case, a simulation of early HAWC runs differed by a factor of $1.81$ in the $\tau^-\tau^+$ channel at a mass of 681 TeV of the \LS{} catalog, which falls within the 68\% containment for this model.

\section{Results}\label{sec:mtd_results} %%%%%%%%%%%%%%%%%%%%%%%%%%%%%%%%%%%
\begin{figure}[htbp]
\centering{
	\begin{tabular}{ccc}
	\includegraphics[width=0.3\textwidth]{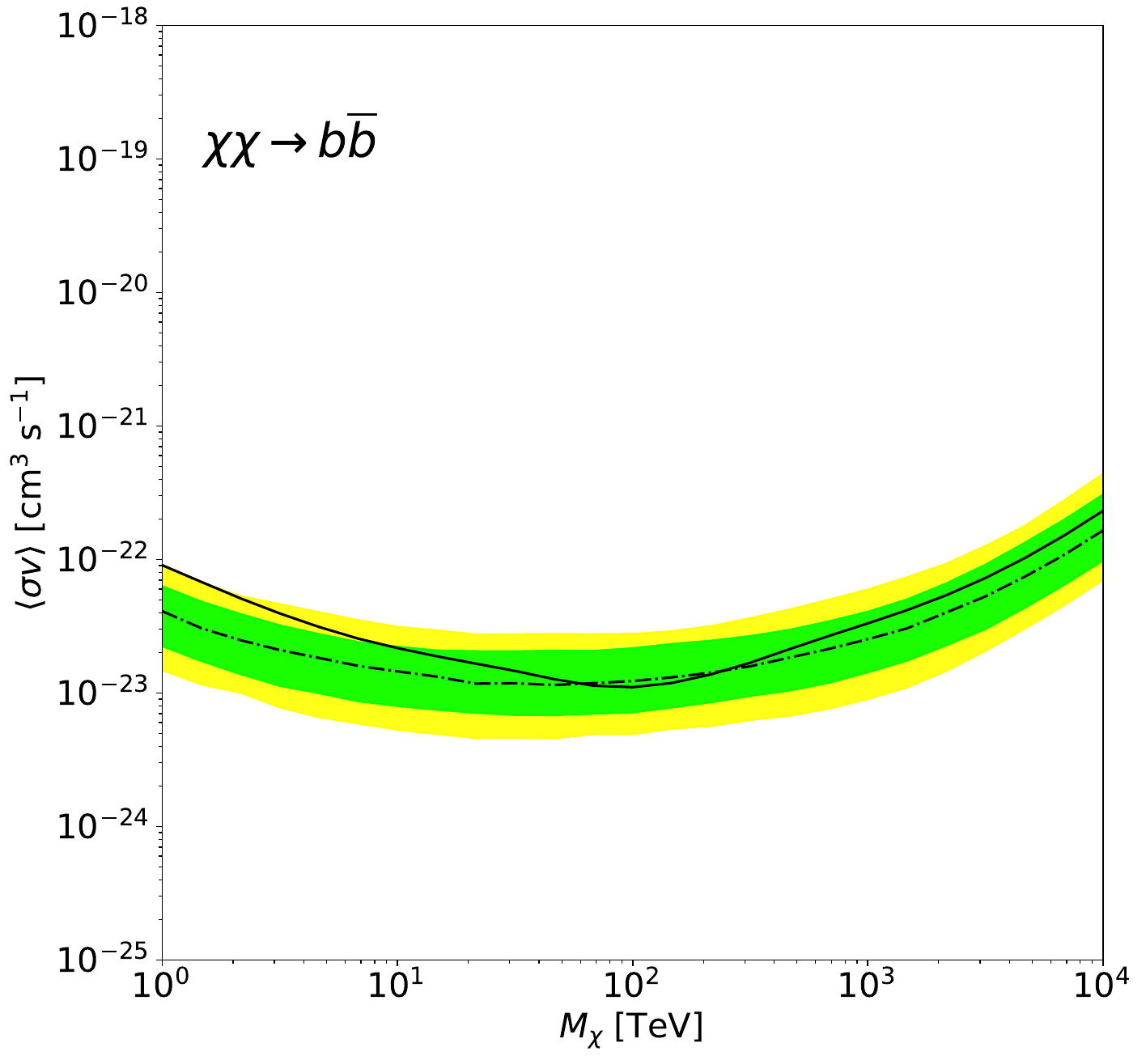} &
   	\includegraphics[width=0.3\textwidth]{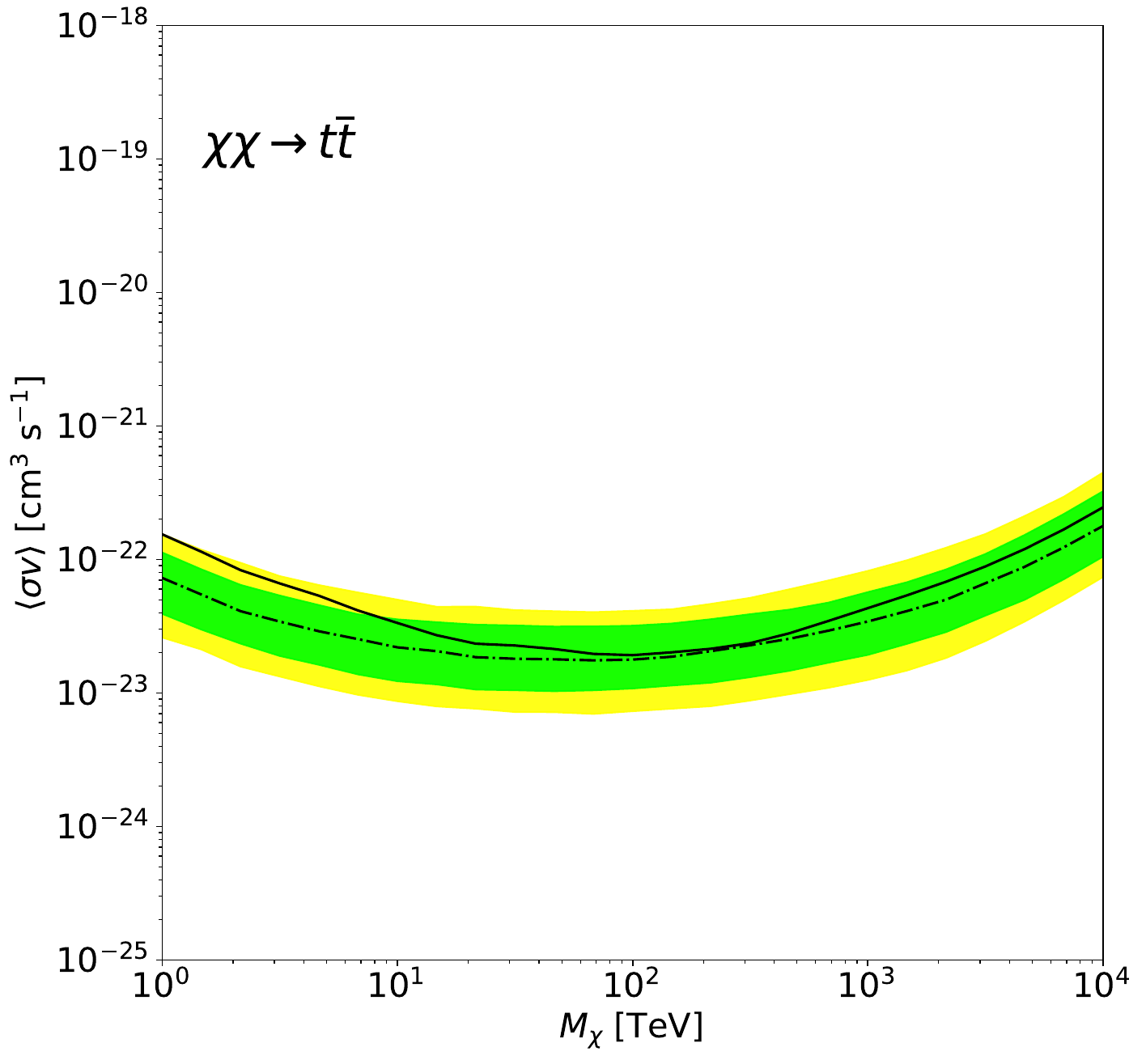} &
   	 \includegraphics[width=0.3\textwidth]{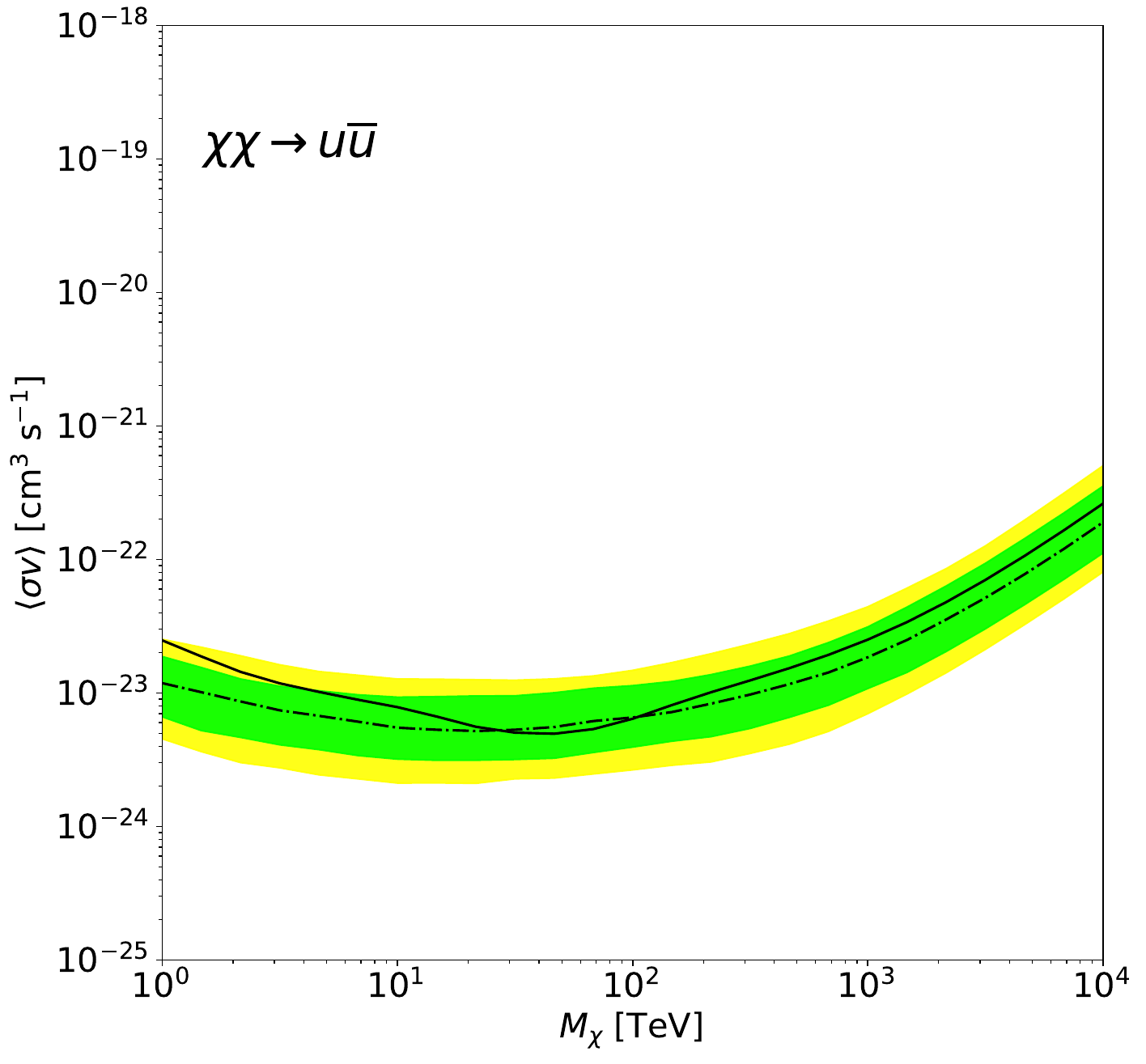} \\
   	 \includegraphics[width=0.3\textwidth]{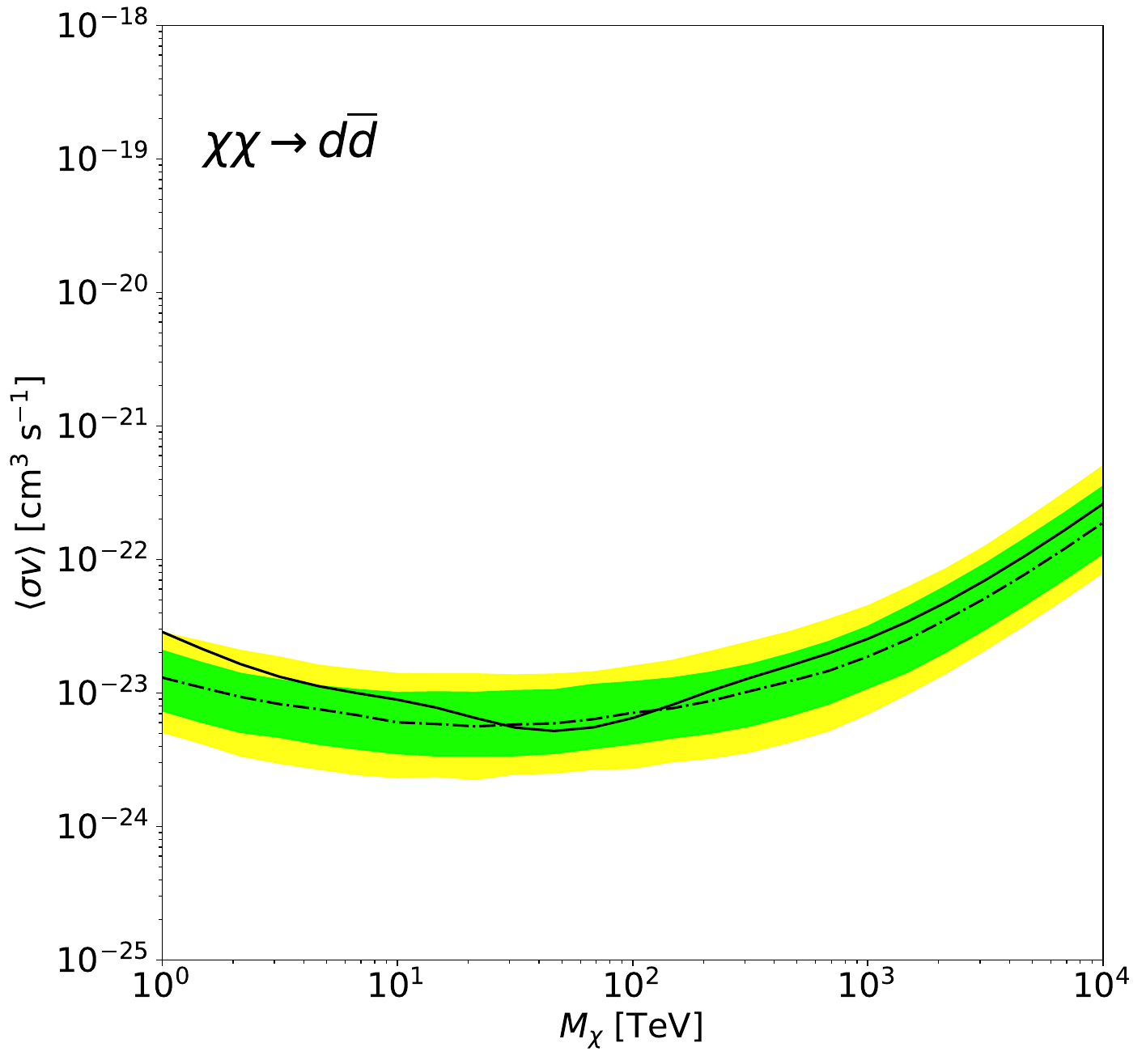} &
	\includegraphics[width=0.3\textwidth]{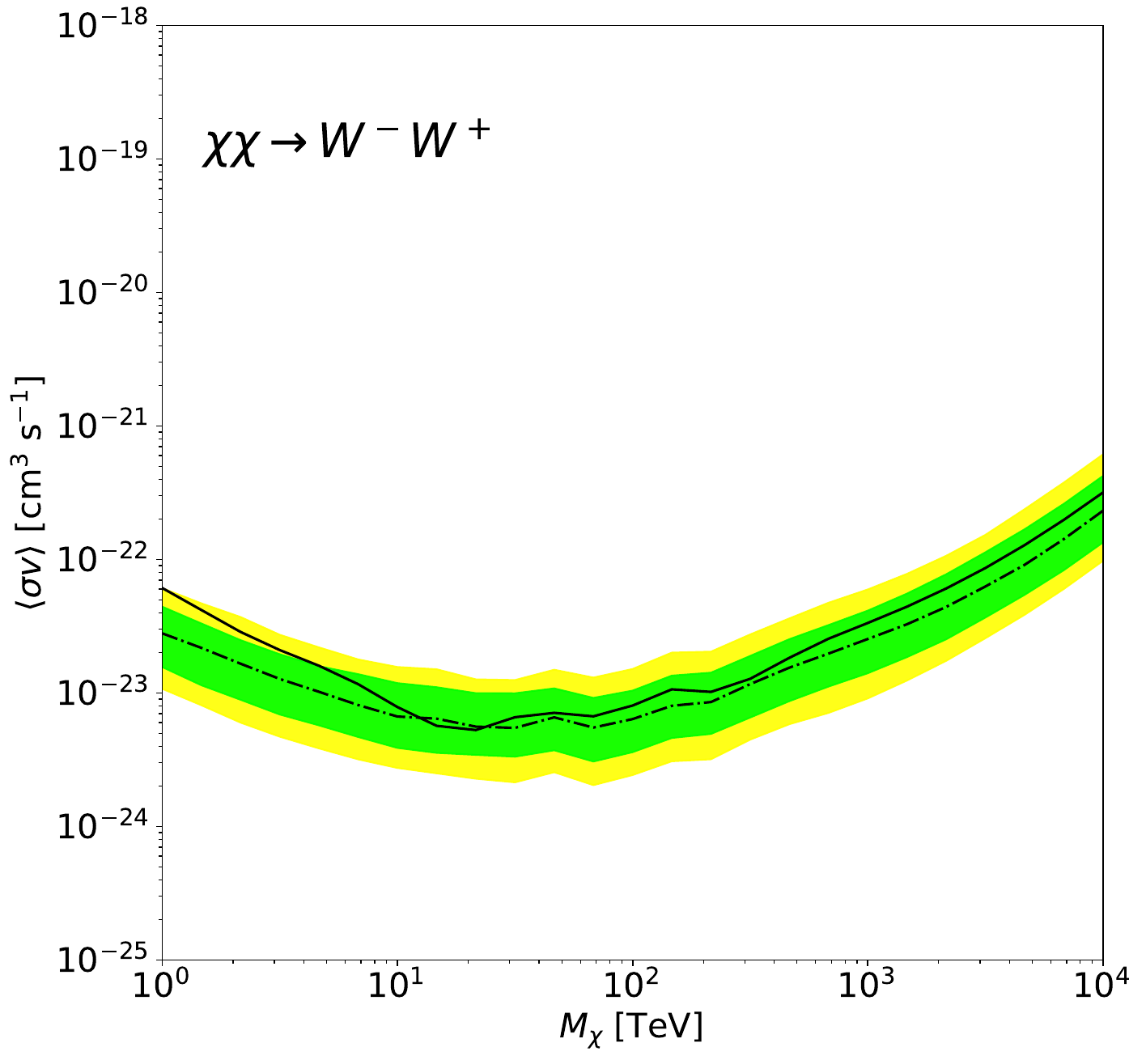} &
   	\includegraphics[width=0.3\textwidth]{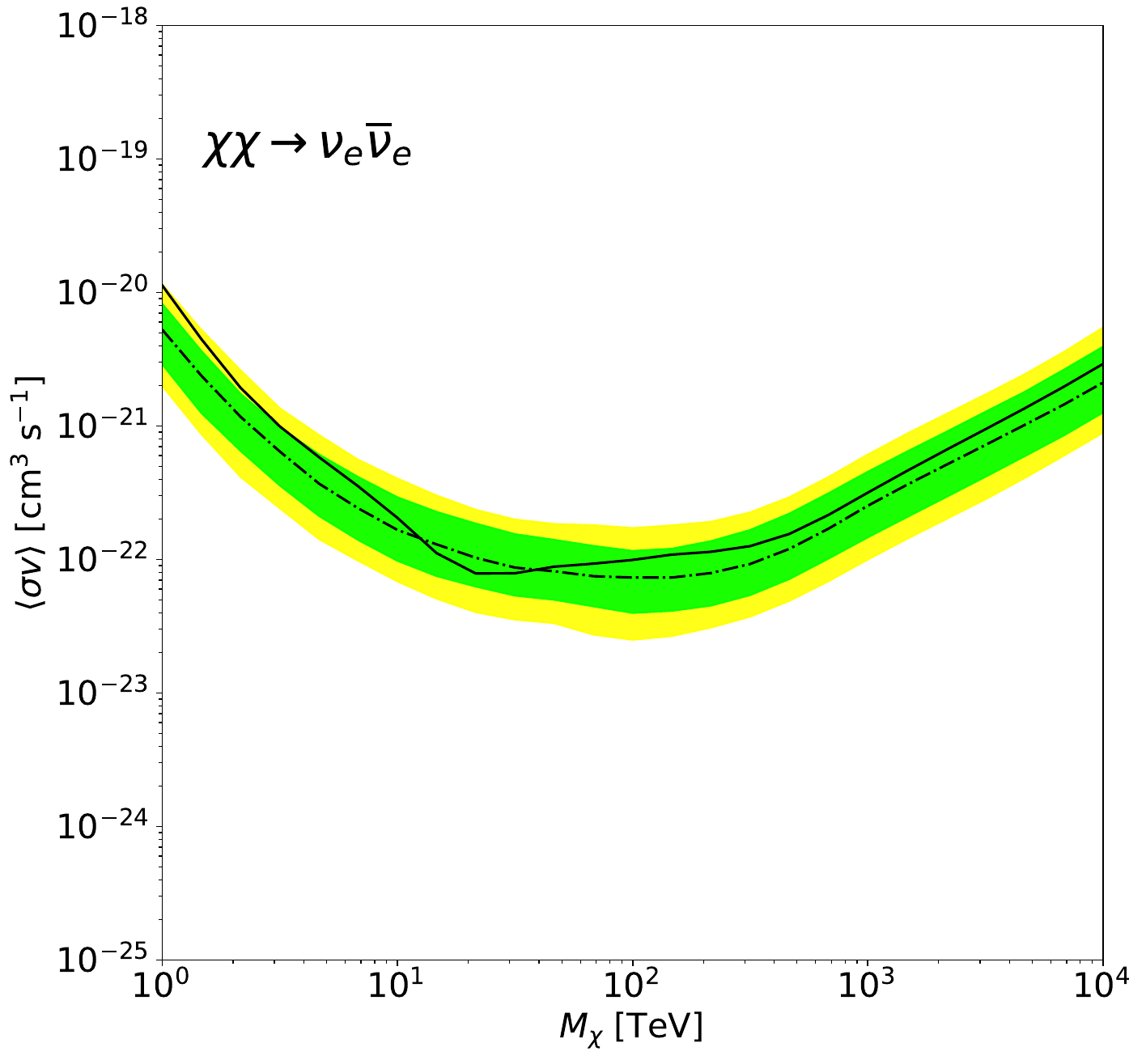} \\
	\includegraphics[width=0.3\textwidth]{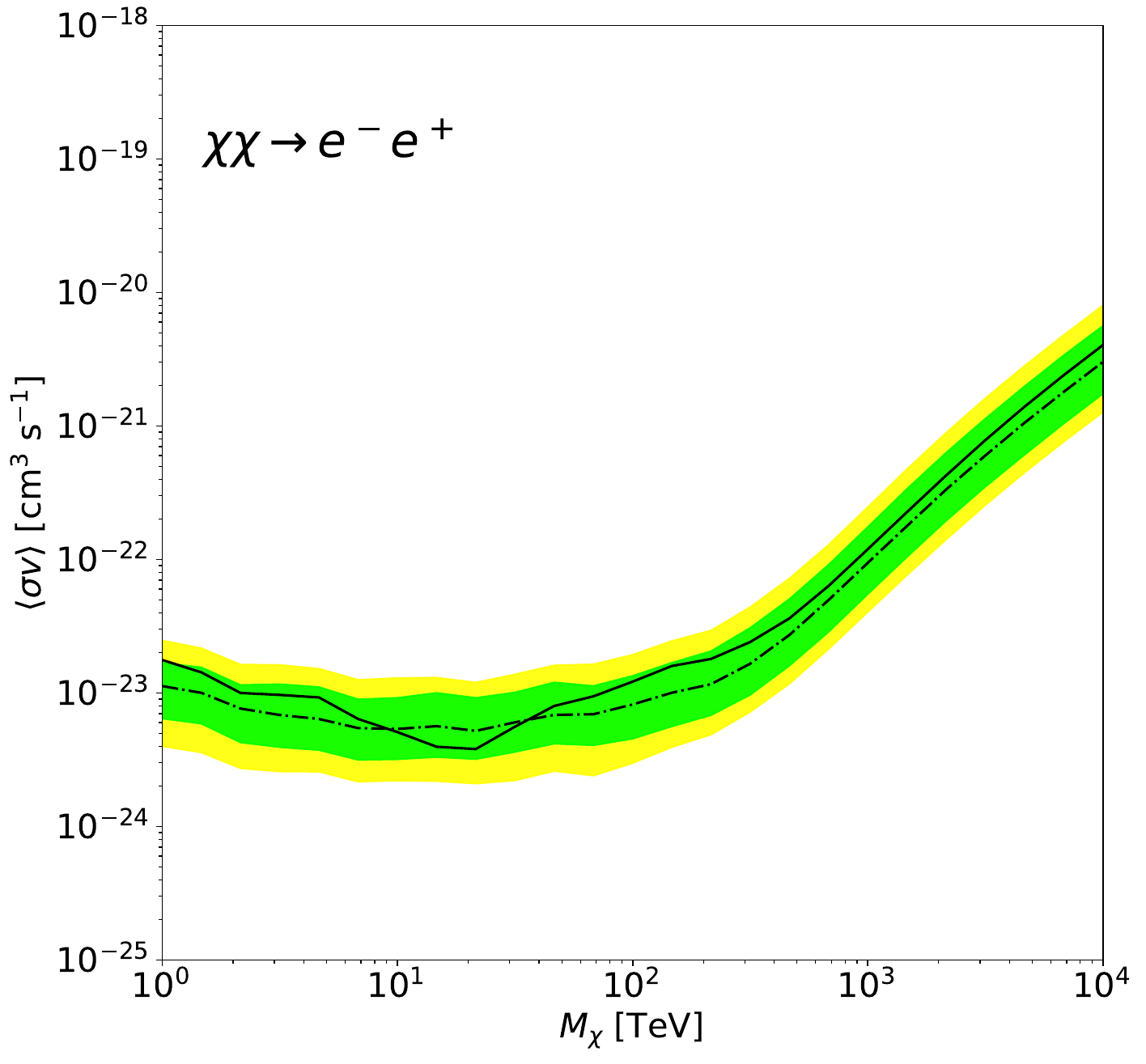} &
   	\includegraphics[width=0.3\textwidth]{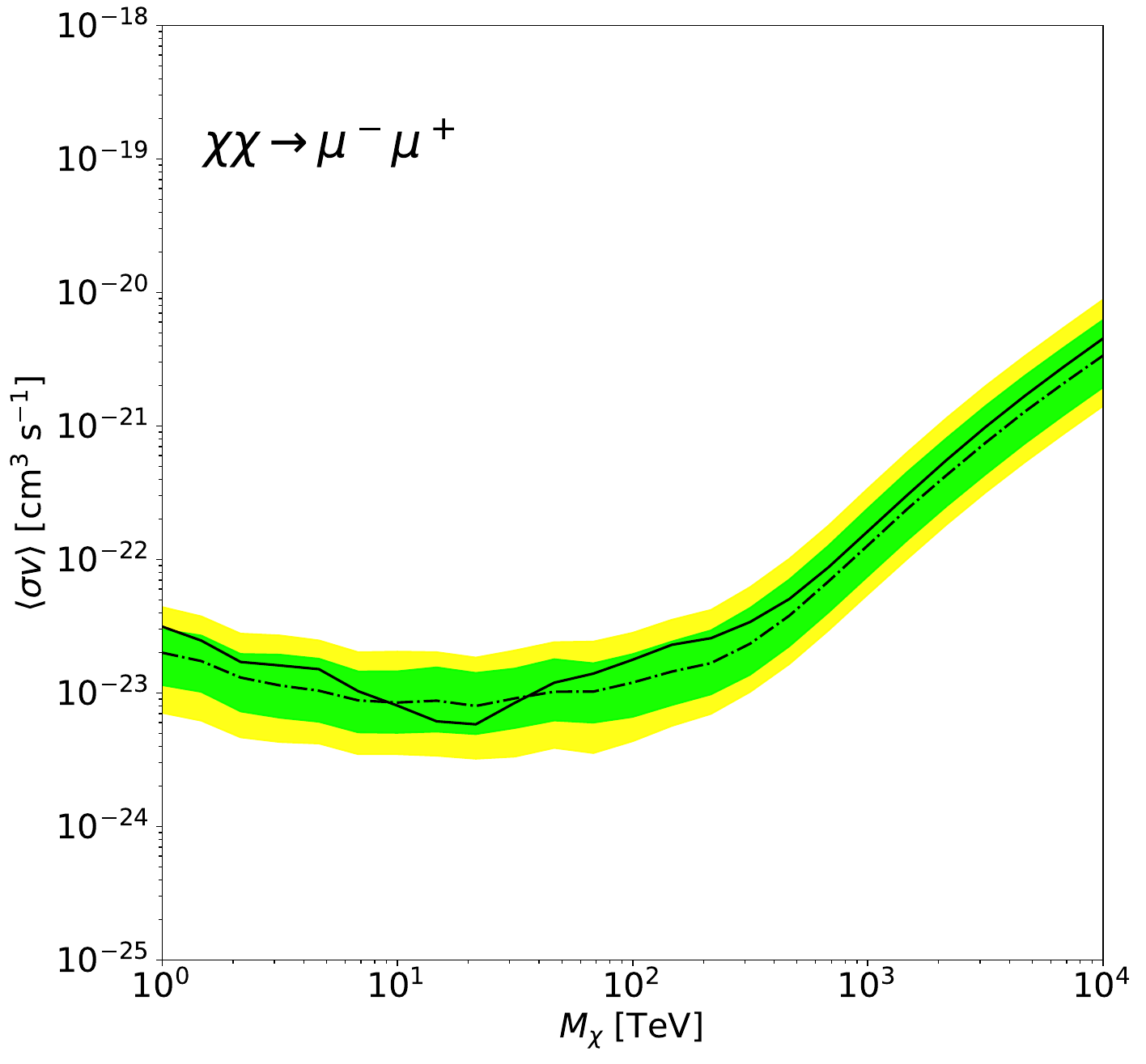} &
   	\includegraphics[width=0.3\textwidth]{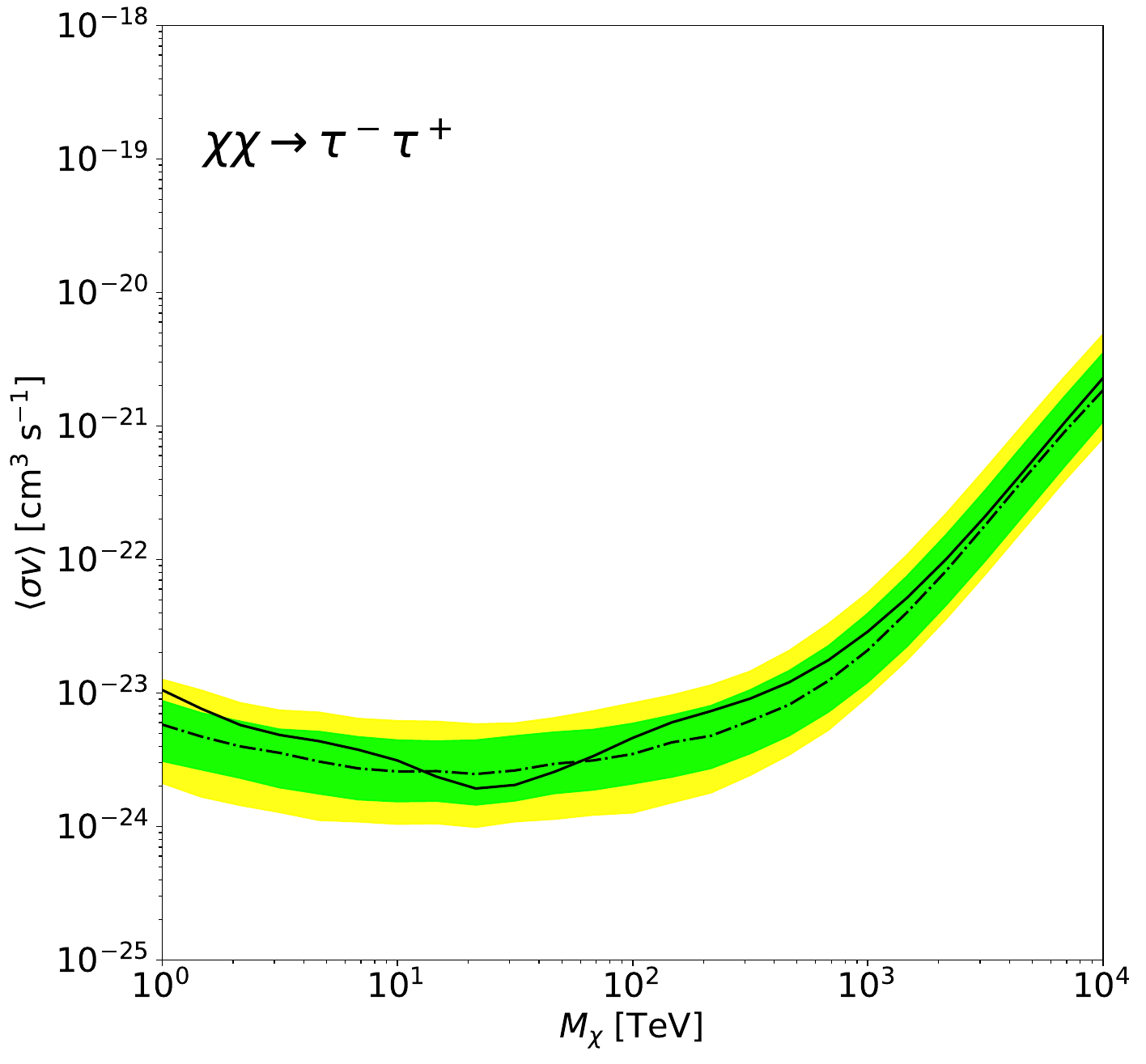} \\
   	 \includegraphics[width=0.3\textwidth]{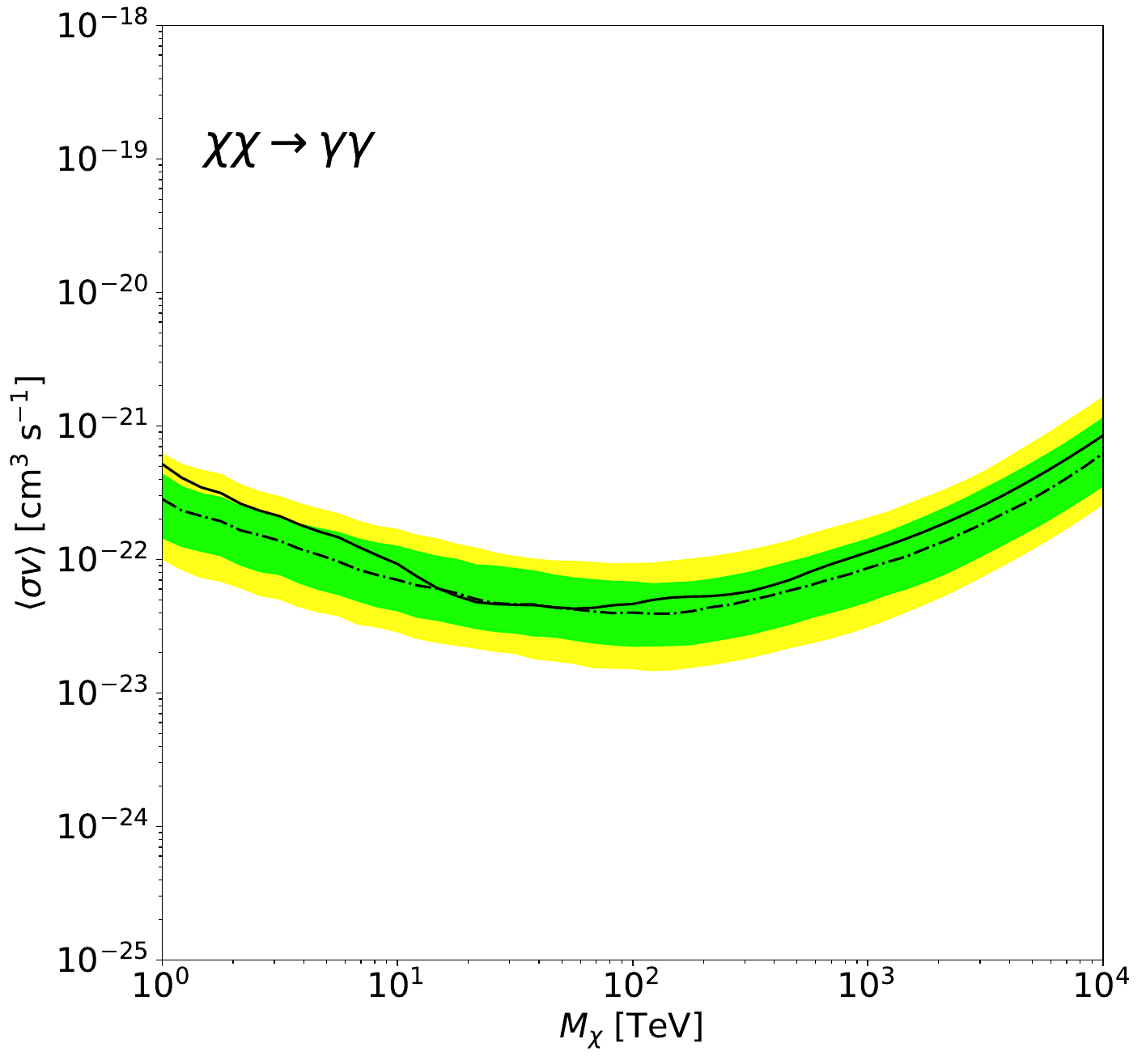} &
   	\includegraphics[width=0.3\textwidth]{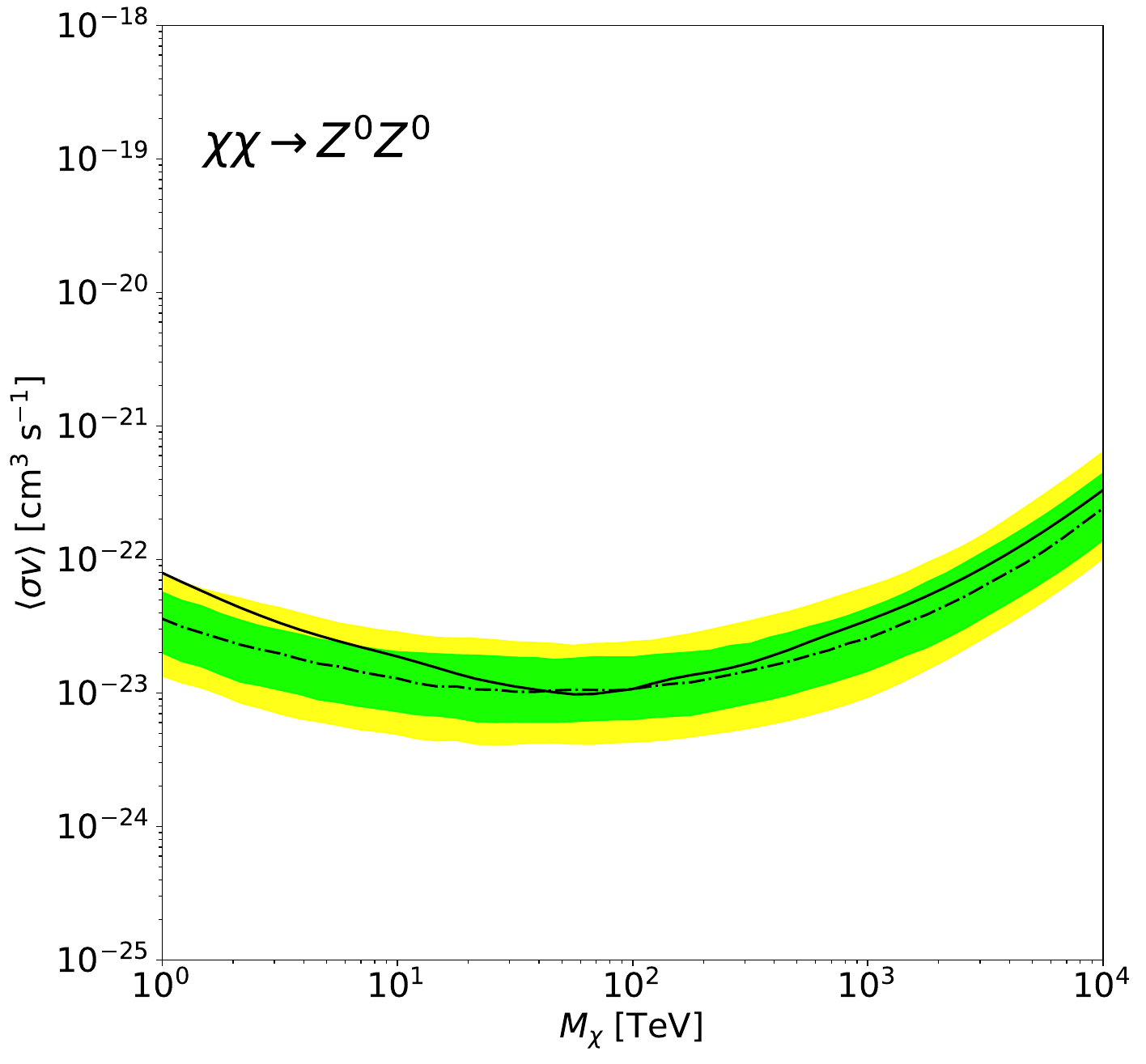} &
	\raisebox{0.75\height}{\includegraphics[width=0.3\textwidth]{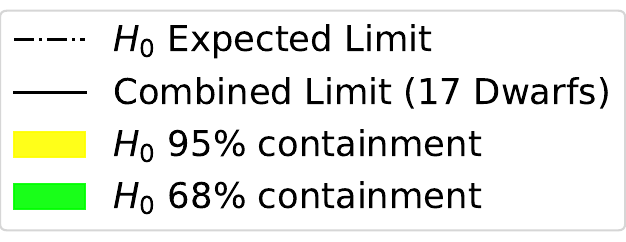}}
	\end{tabular}
    }
    \caption{Brazil bands on \sv~versus $m_\chi$ for $\chi\chi \rightarrow b\overline{b}$, $t\overline{t}$, $u\overline{u}$, $d\overline{d}$, $W^-W^+$, $\nu_e\overline{\nu}_e$, $e^-e^+$, $\mu^-\mu^+$, $\tau^-\tau^+$, $\gamma\gamma$ and $Z^0Z^0$. Limits are with \LS{} \J-factors \cite{DM_Strigari20}. The solid line represents the observed combined limit. The dashed line represents the expected annihilation cross-section from HAWC background. The green and yellow bands show 68\% and 95\% containment respectively.}
\label{fig:LSmtd_bb_1of2}
\end{figure}

\begin{figure}[htbp]
\centering{
	\begin{tabular}{ccc}
   	\includegraphics[width=0.3\textwidth]{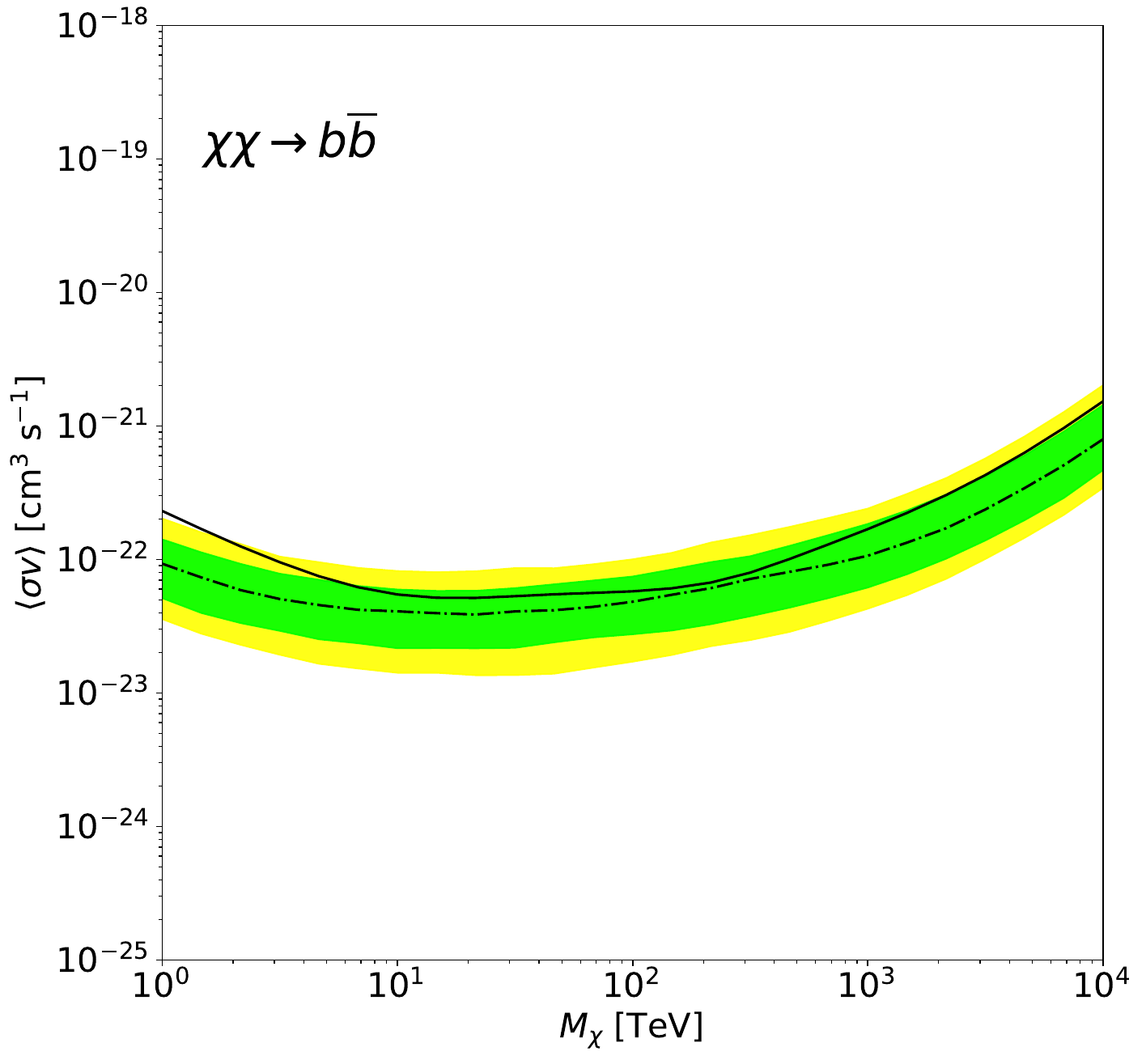} &
   	\includegraphics[width=0.3\textwidth]{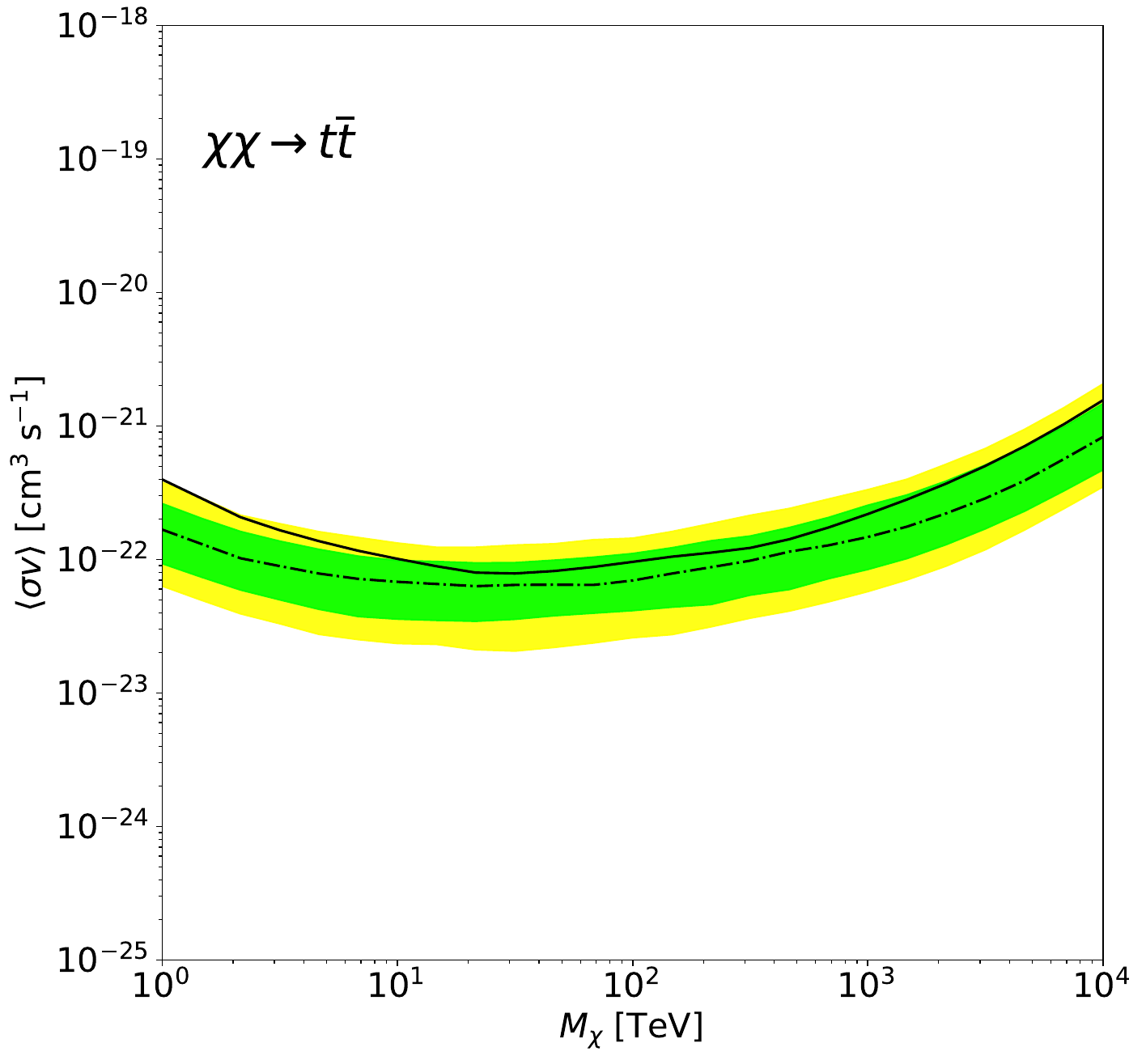} &
   	\includegraphics[width=0.3\textwidth]{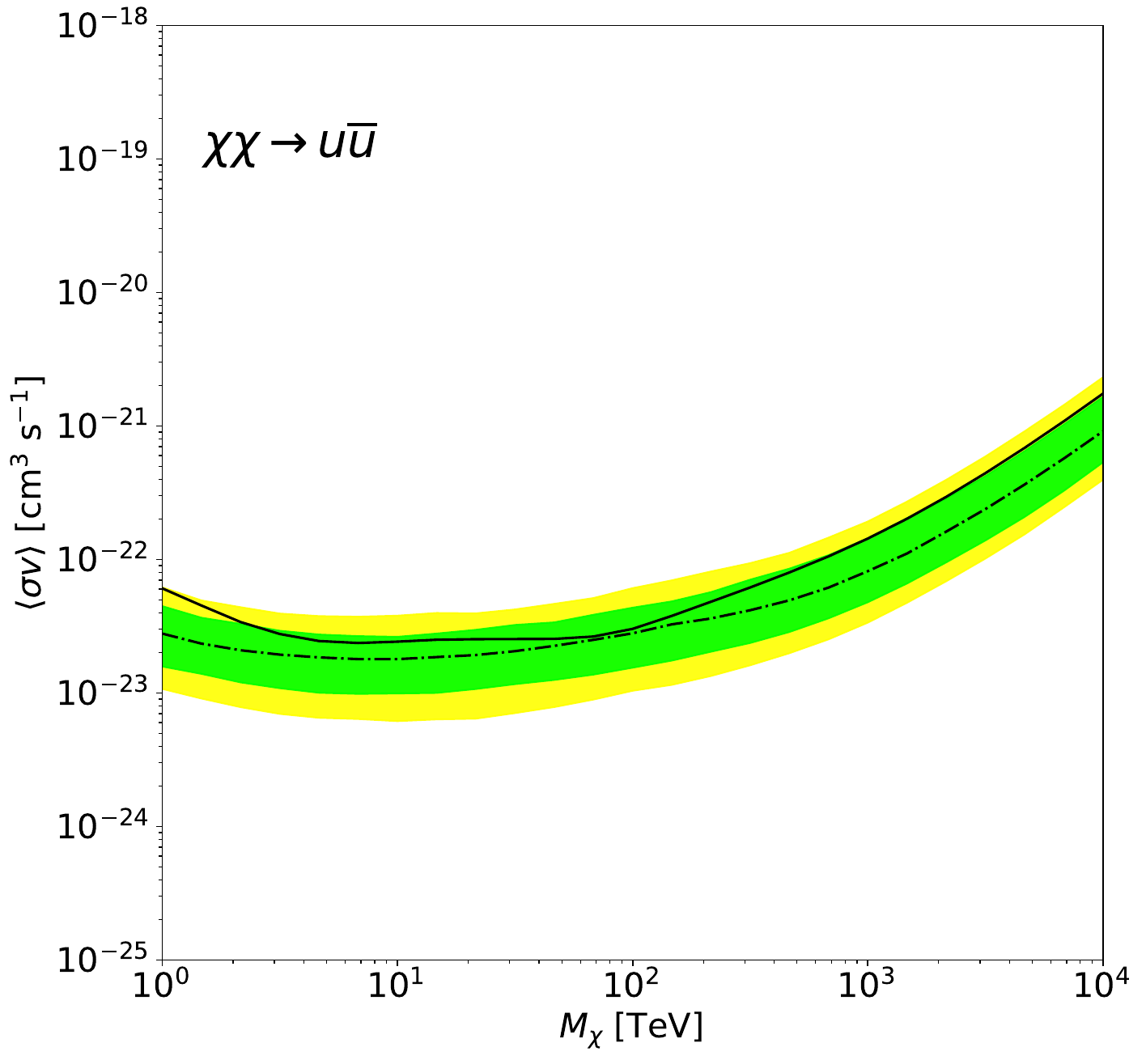} \\
   	\includegraphics[width=0.3\textwidth]{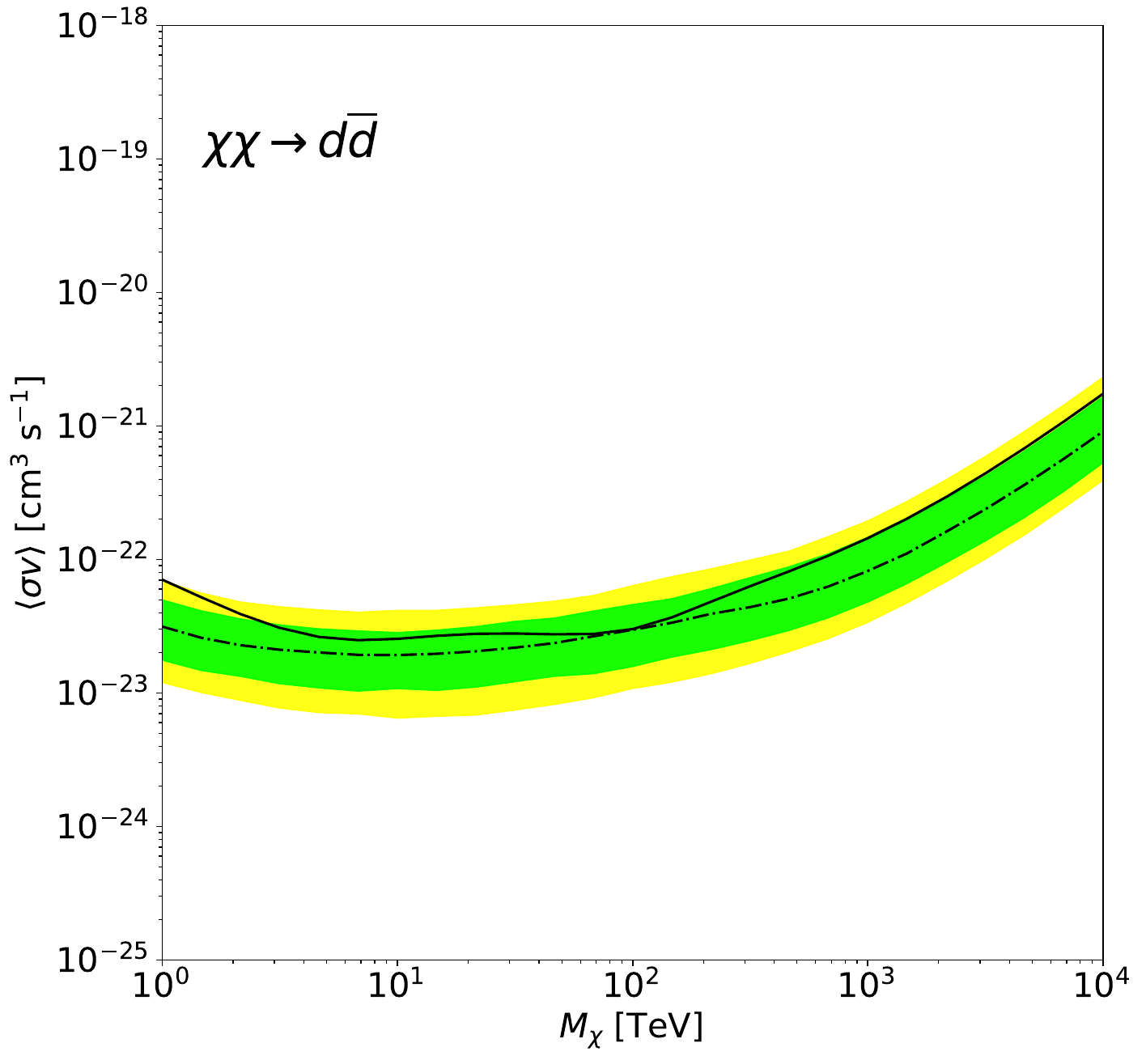} &
   	\includegraphics[width=0.3\textwidth]{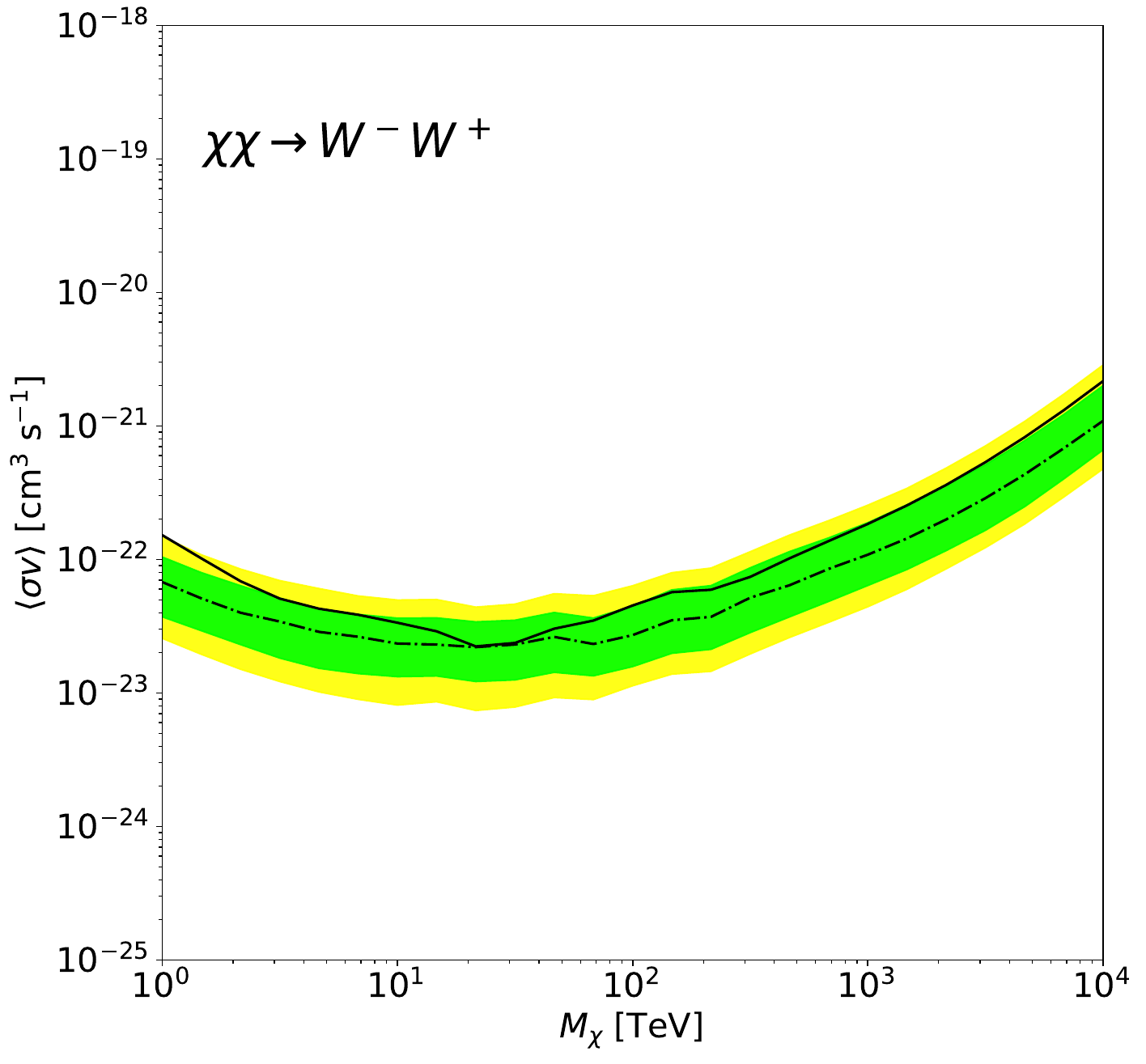} &
 	\includegraphics[width=0.3\textwidth]{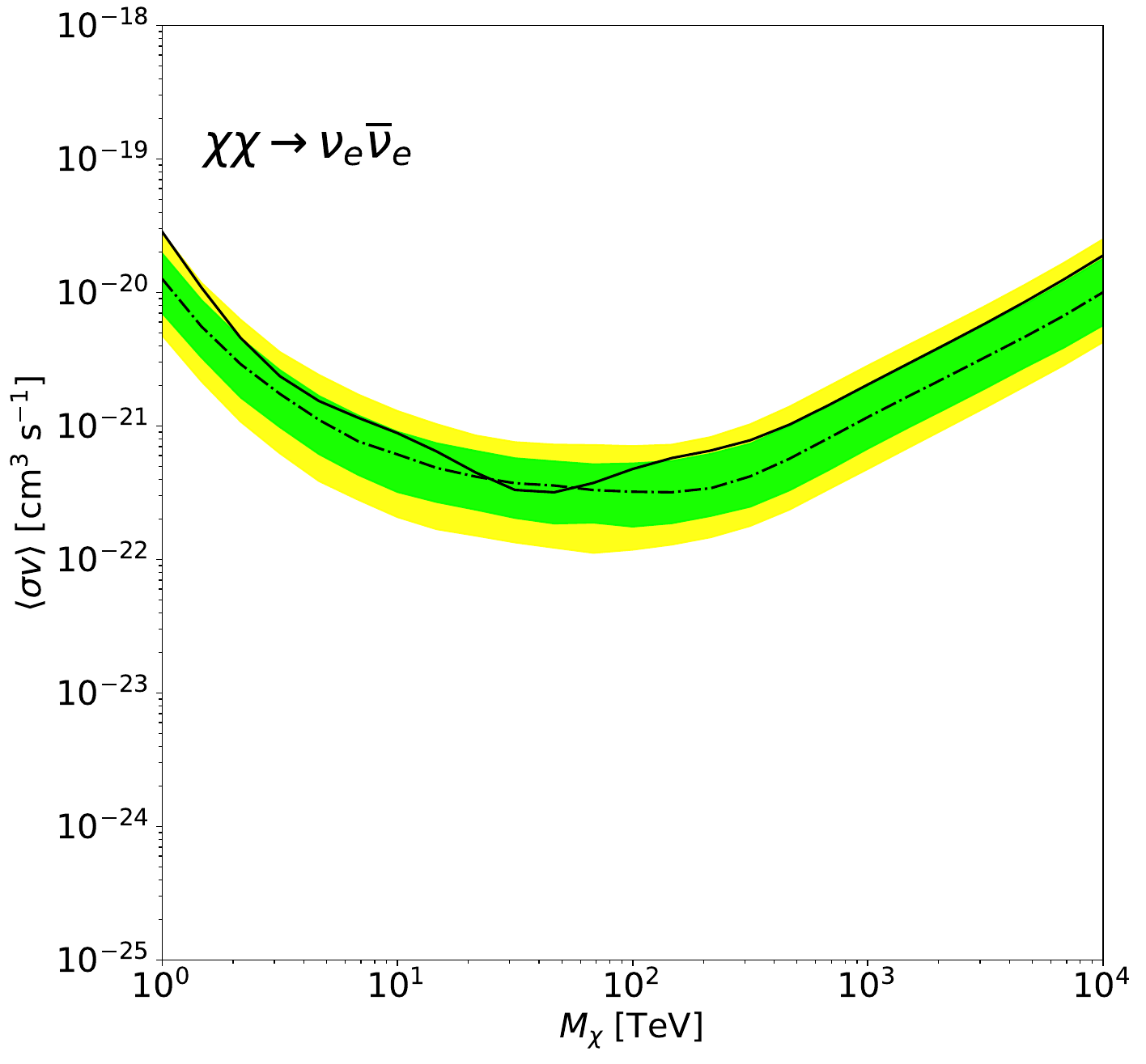} \\
  	\includegraphics[width=0.3\textwidth]{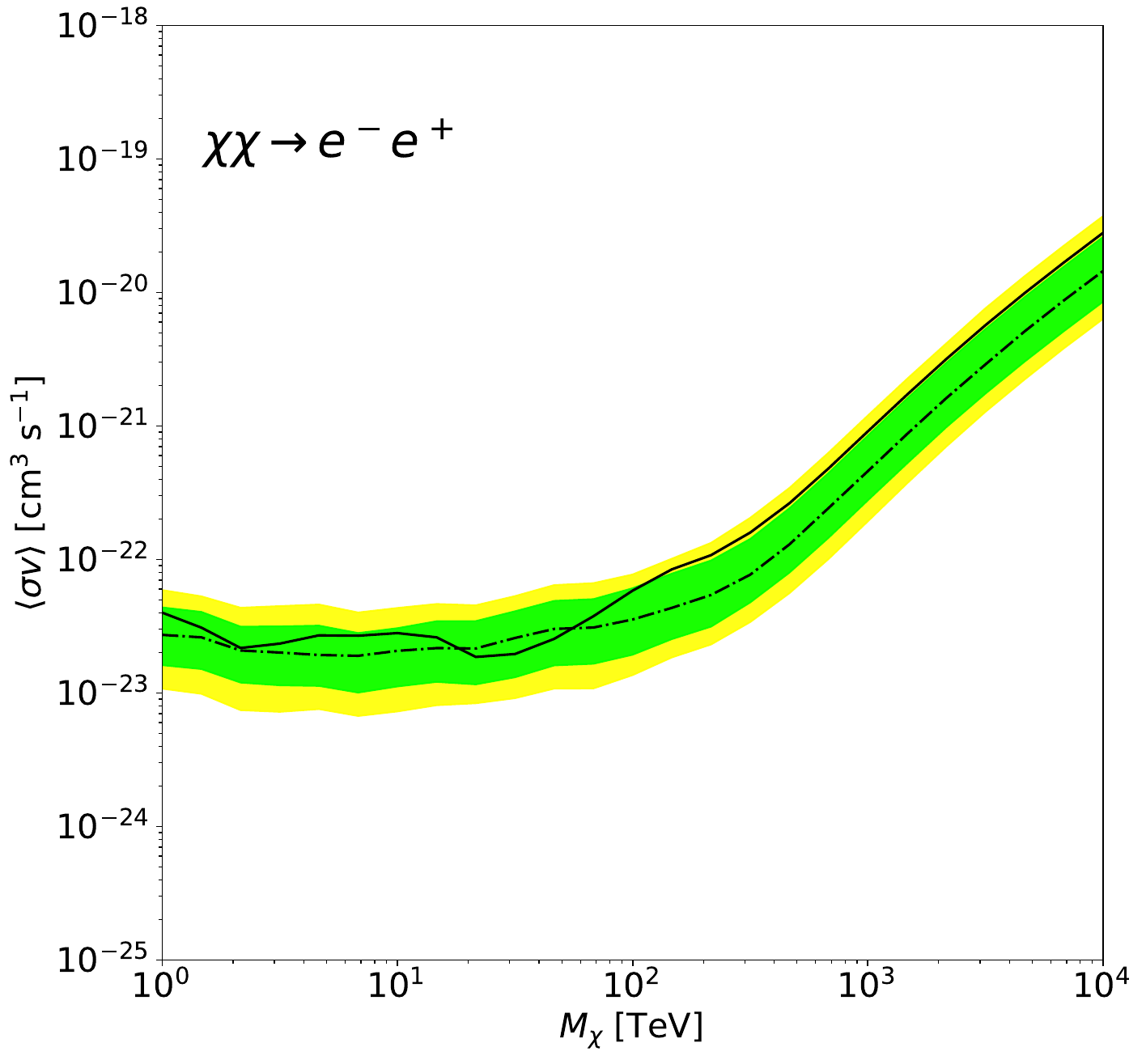} &
   	\includegraphics[width=0.3\textwidth]{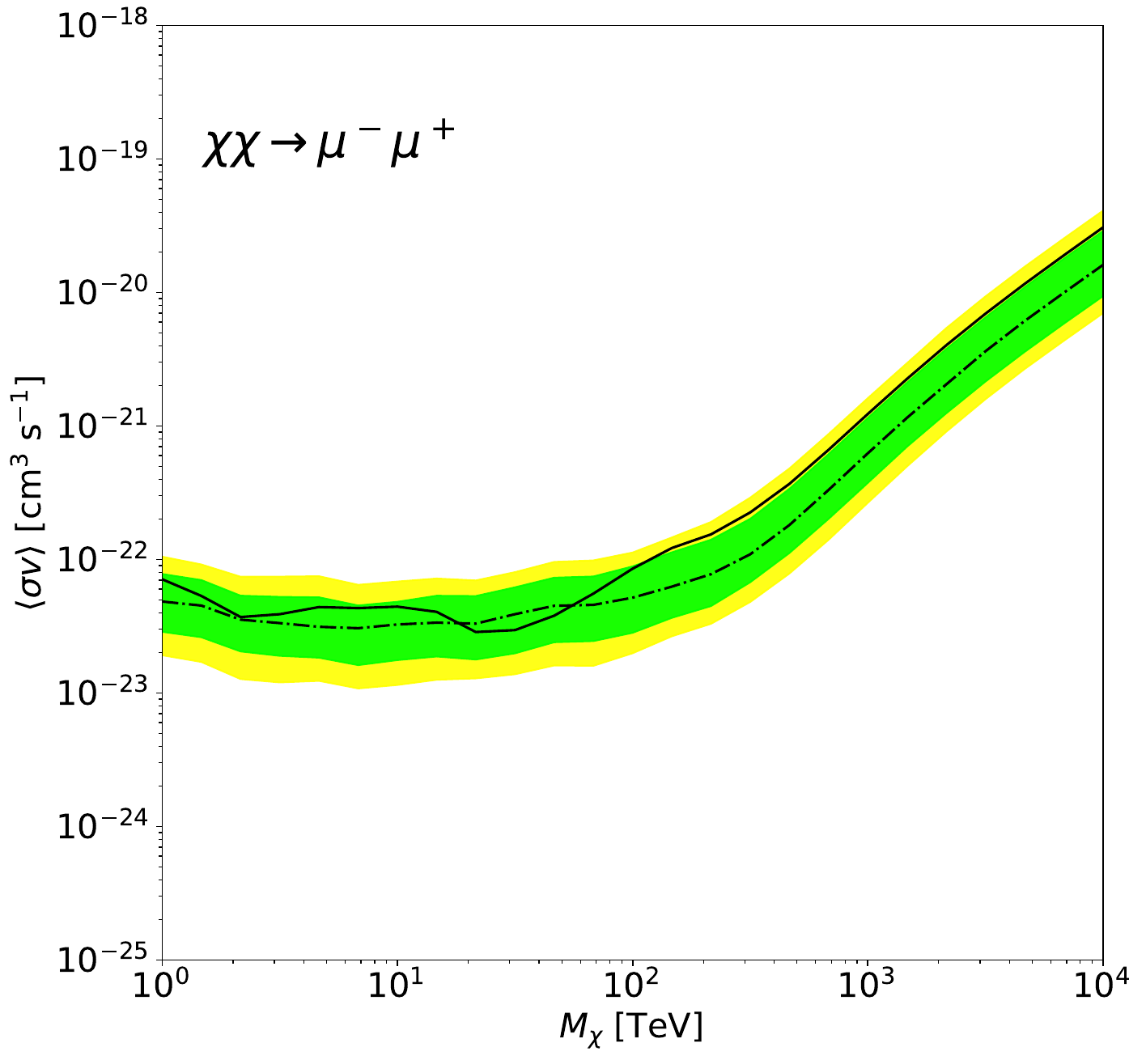} &
   	\includegraphics[width=0.3\textwidth]{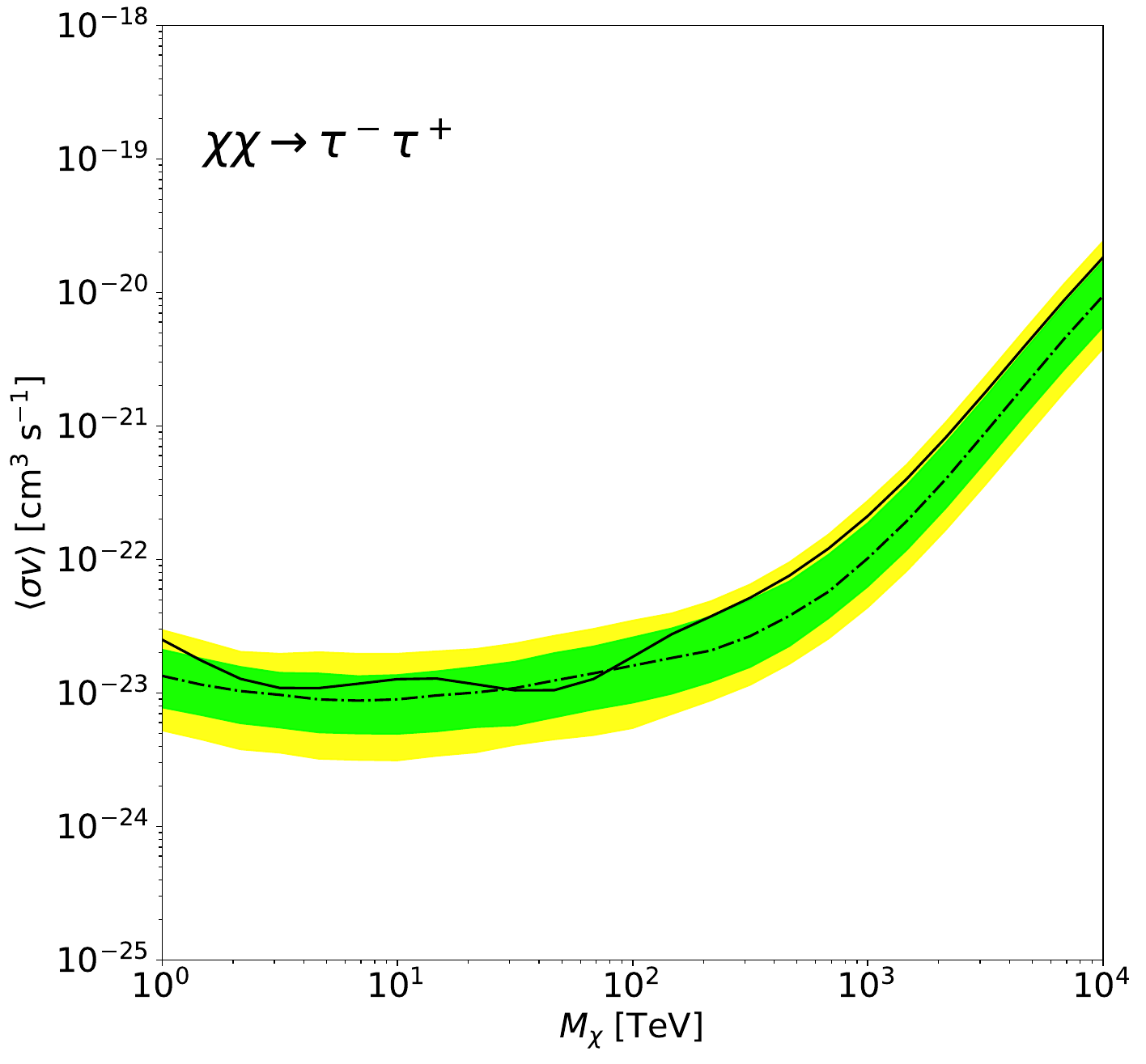} \\
   	\includegraphics[width=0.3\textwidth]{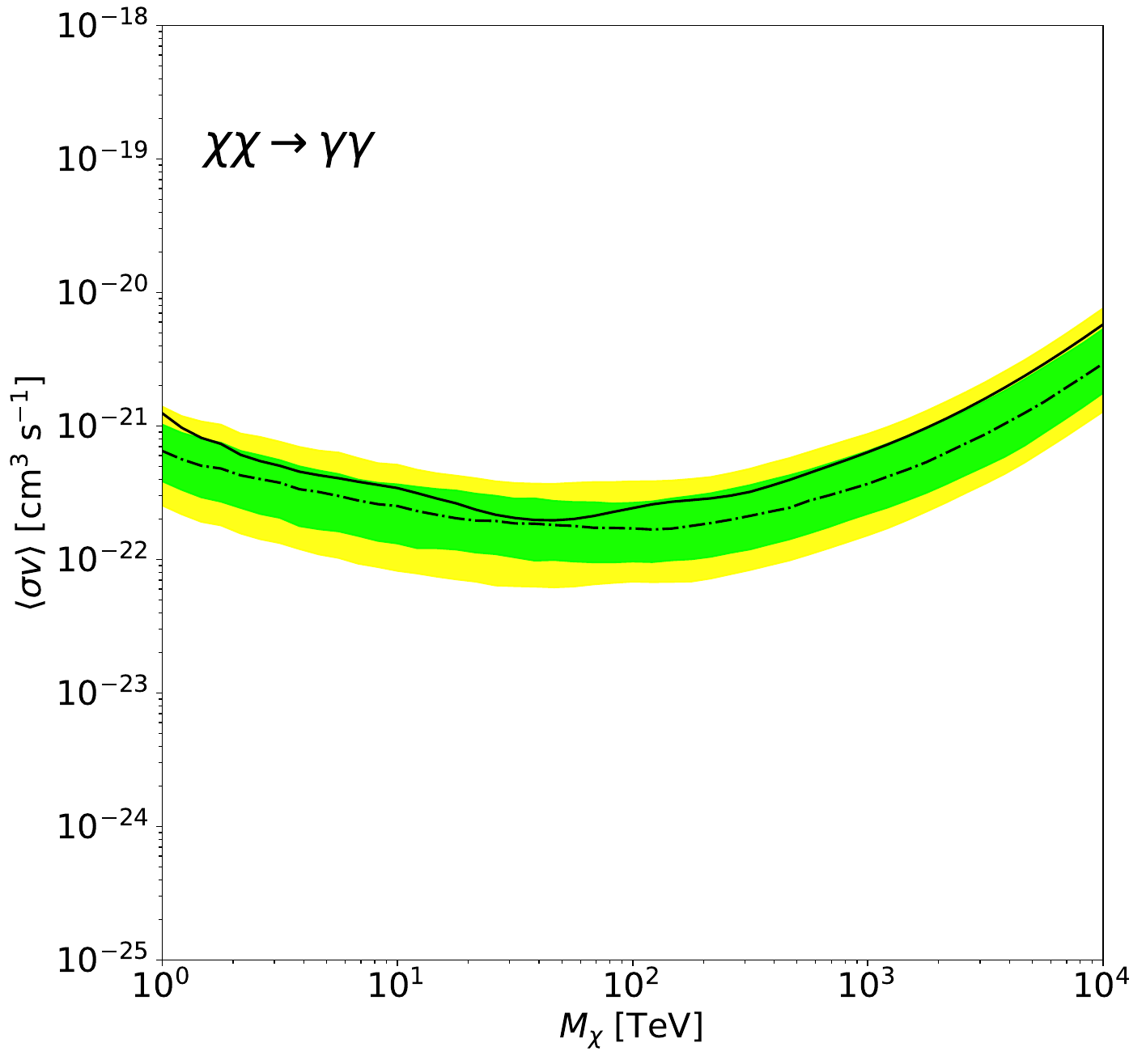} &
   	\includegraphics[width=0.3\textwidth]{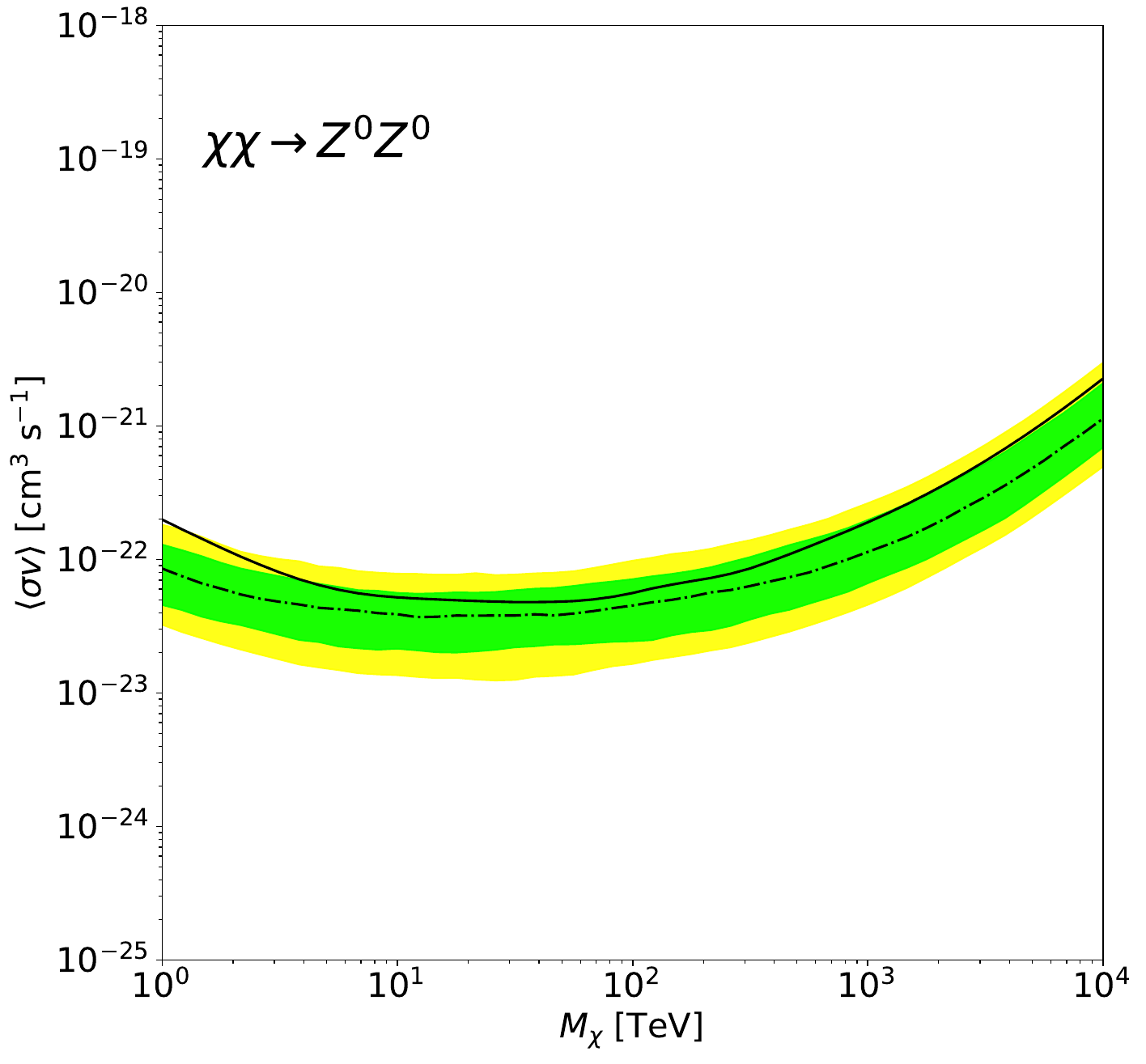} &
    	\raisebox{0.75\height}{\includegraphics[width=0.3\textwidth]{figures/BrazilBands/LEGEND_BrazilBand.pdf}}
	\end{tabular}
    }
    \caption{Same as \cref{fig:LSmtd_bb_1of2} but with \GS{} \J-factors \cite{Ando_2020}.}
\label{fig:GSmtd_bb_1of2}
\end{figure}

\begin{figure}[htbp]
\centering{
	\begin{tabular}{ccc}
    	\includegraphics[width=0.3\textwidth]{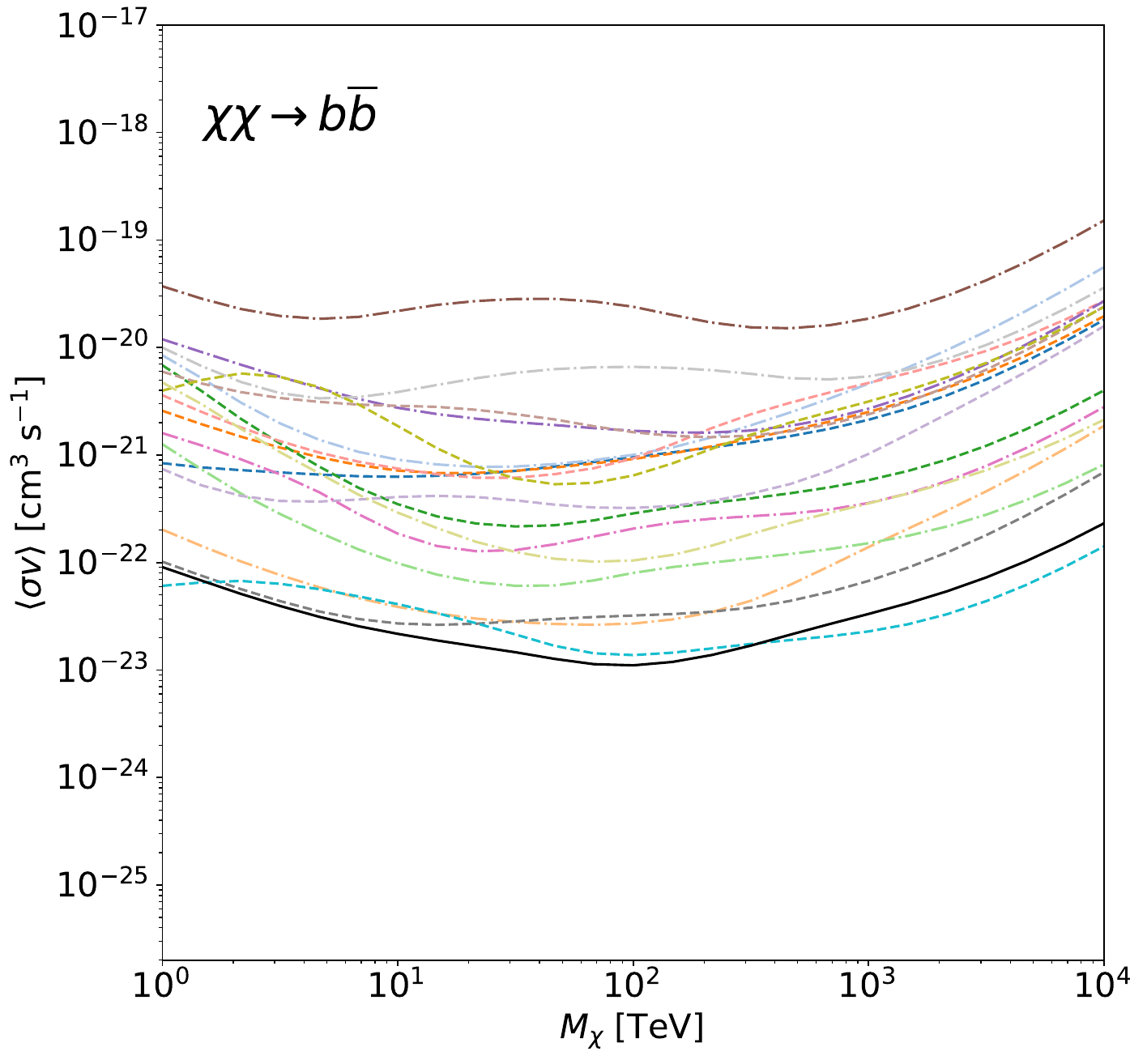} &
    	\includegraphics[width=0.3\textwidth]{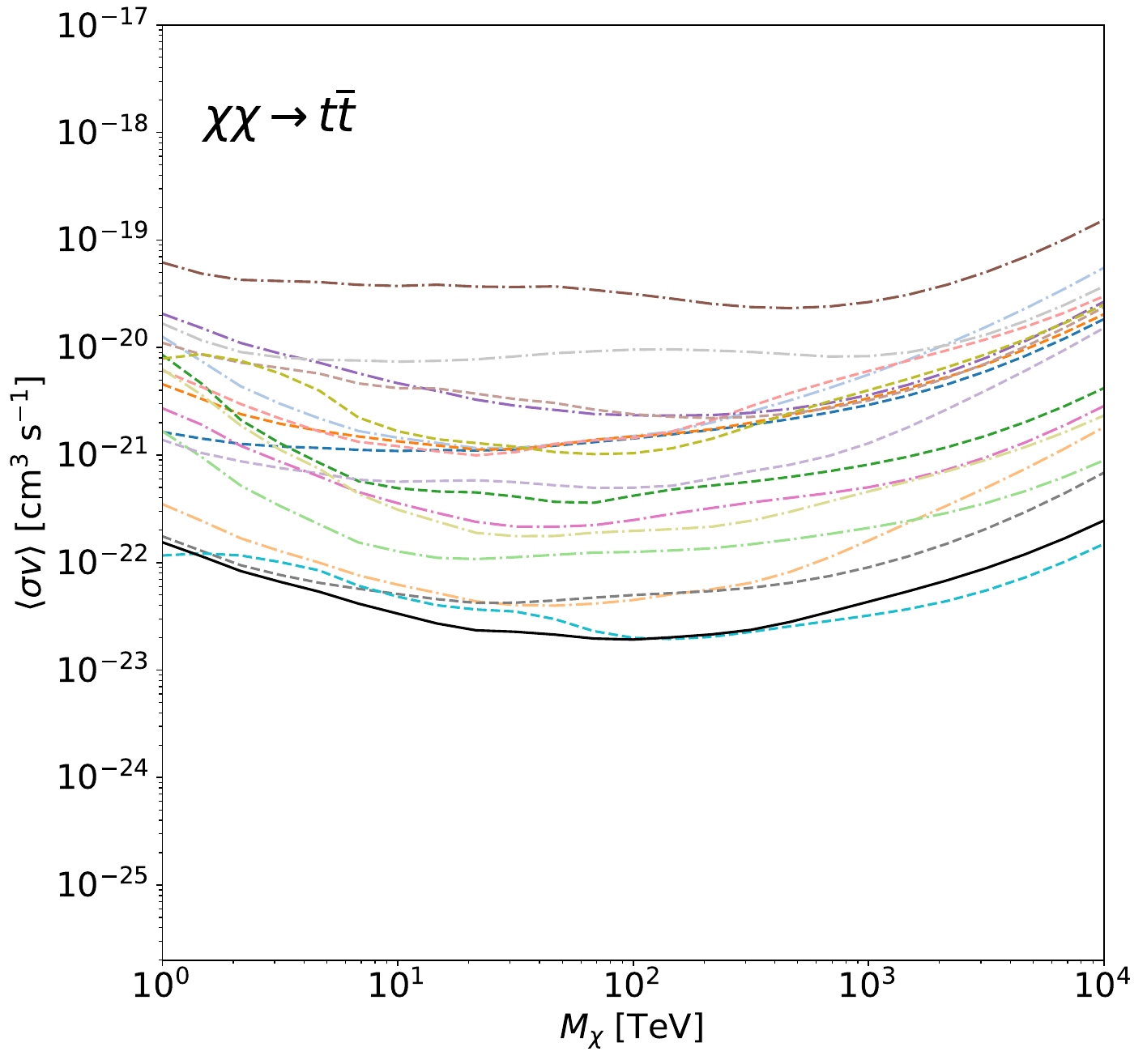} &
   	\includegraphics[width=0.3\textwidth]{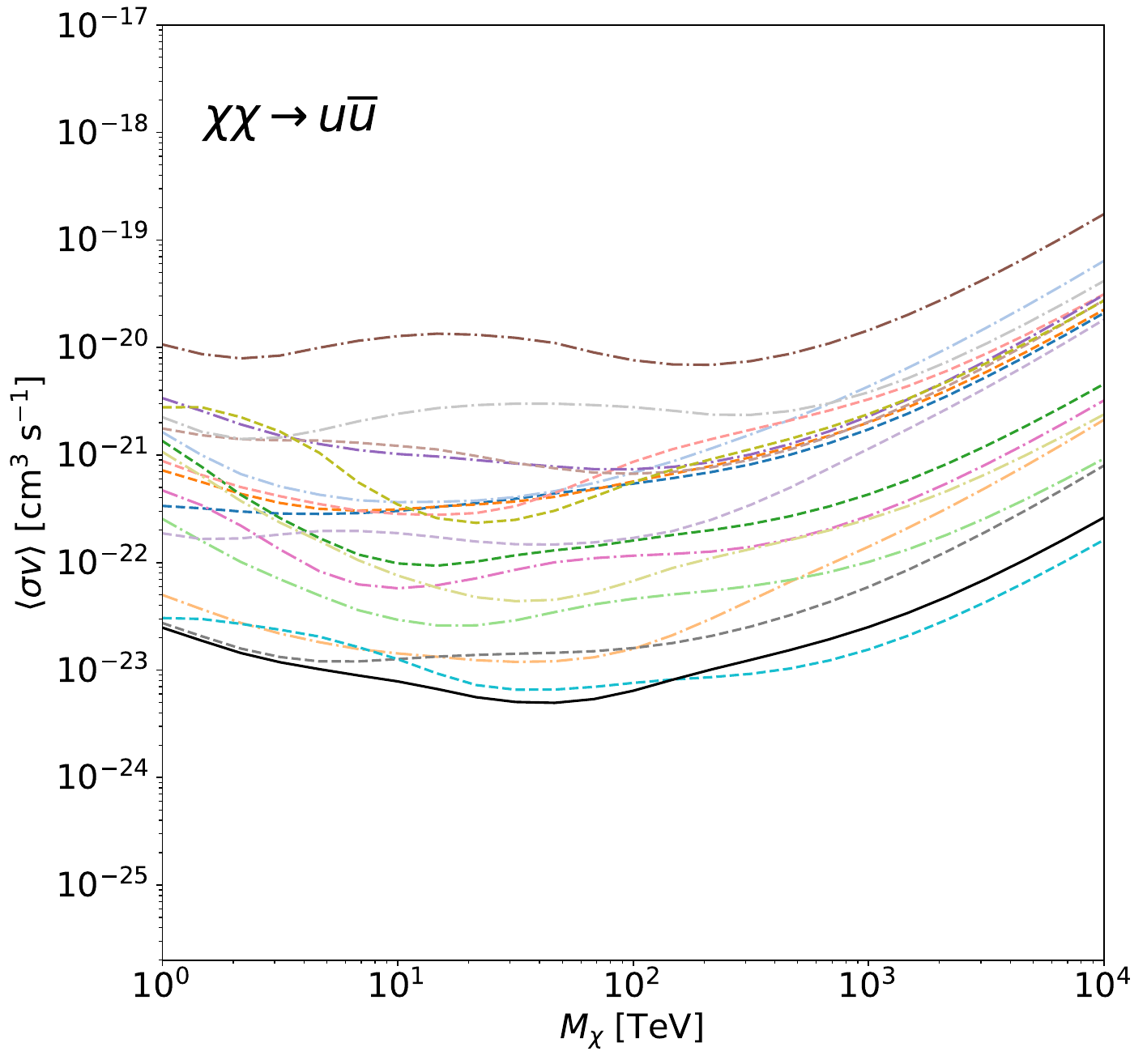} \\
    	\includegraphics[width=0.3\textwidth]{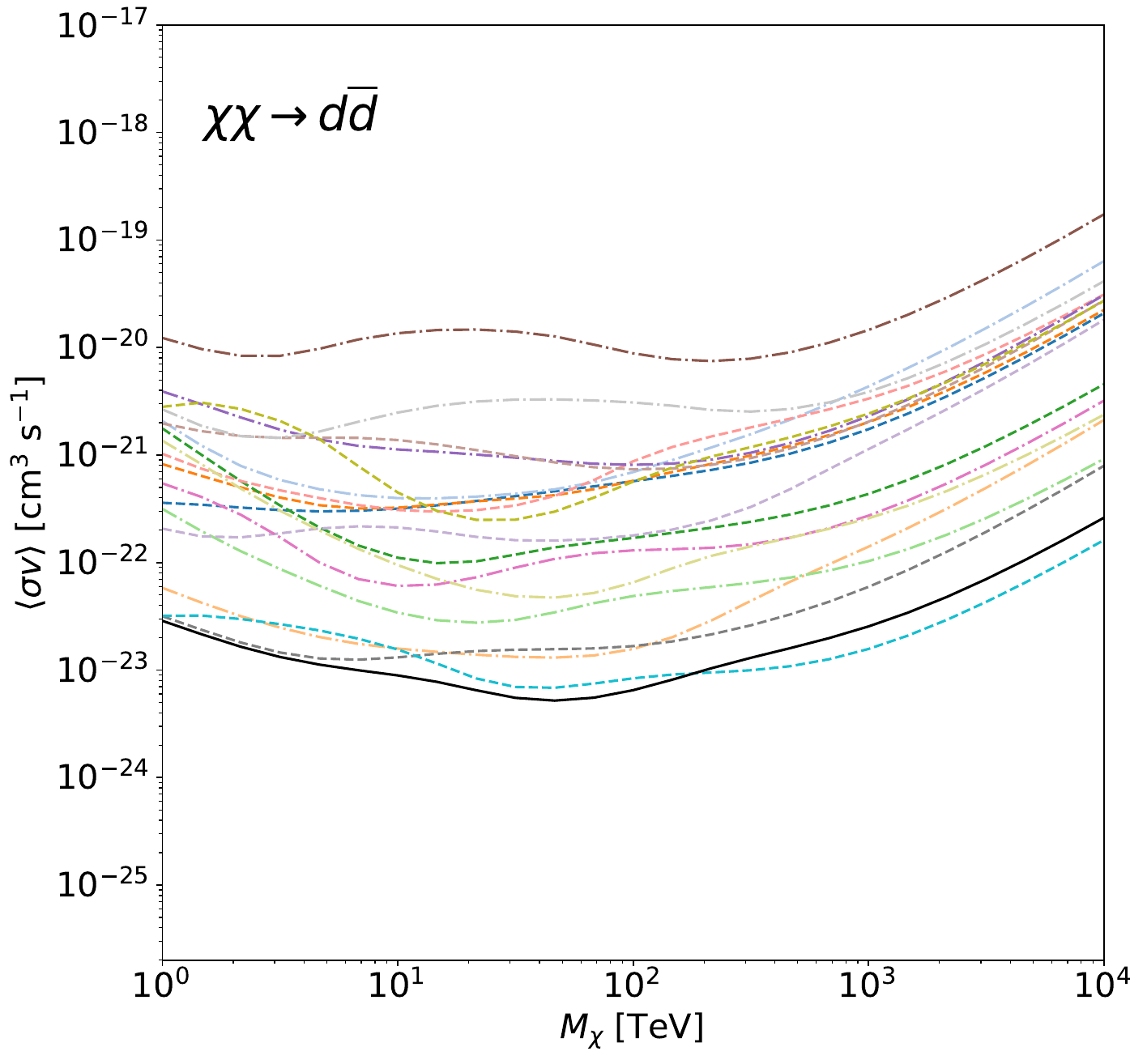} &
    	\includegraphics[width=0.3\textwidth]{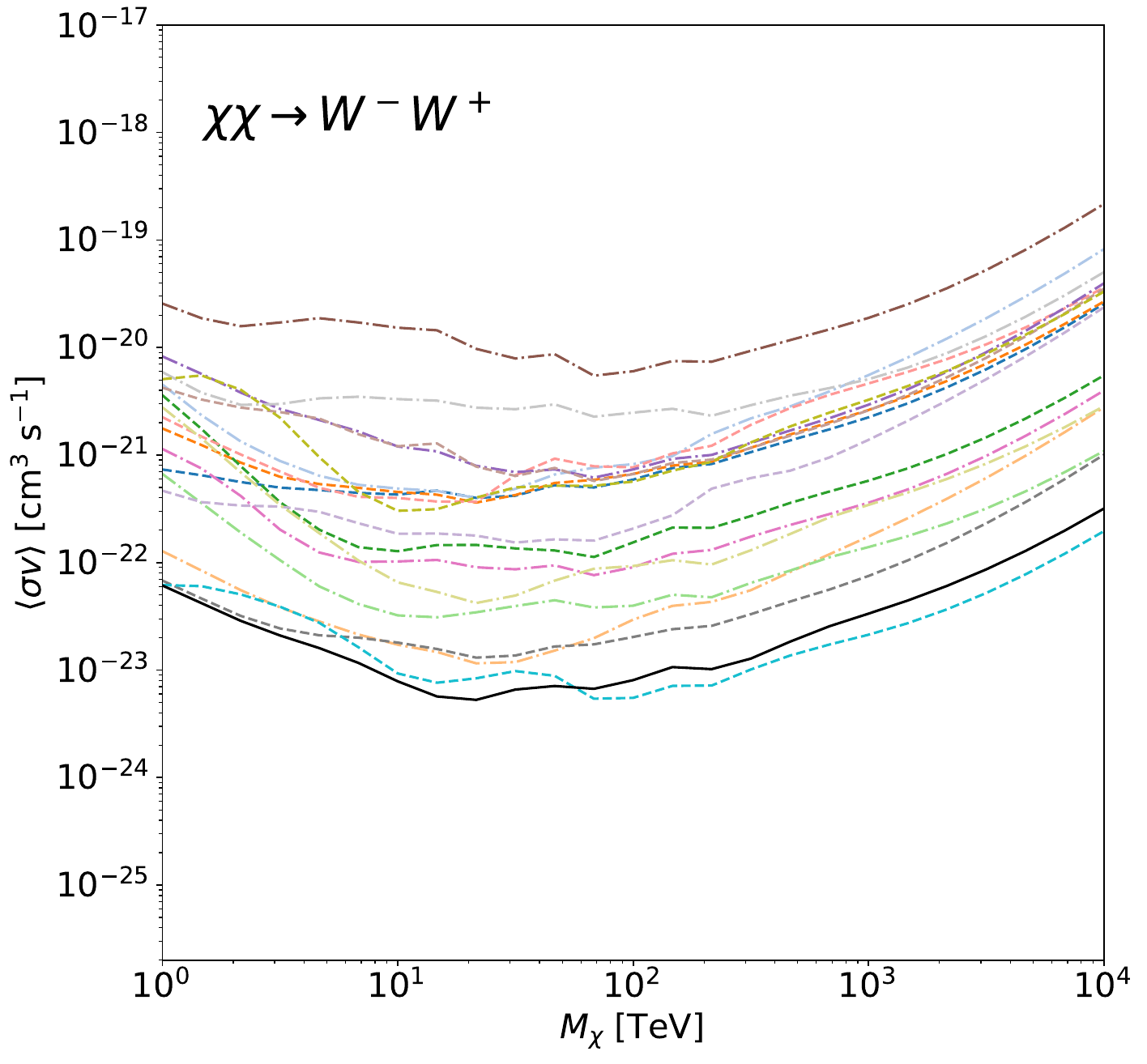} &
   	\includegraphics[width=0.3\textwidth]{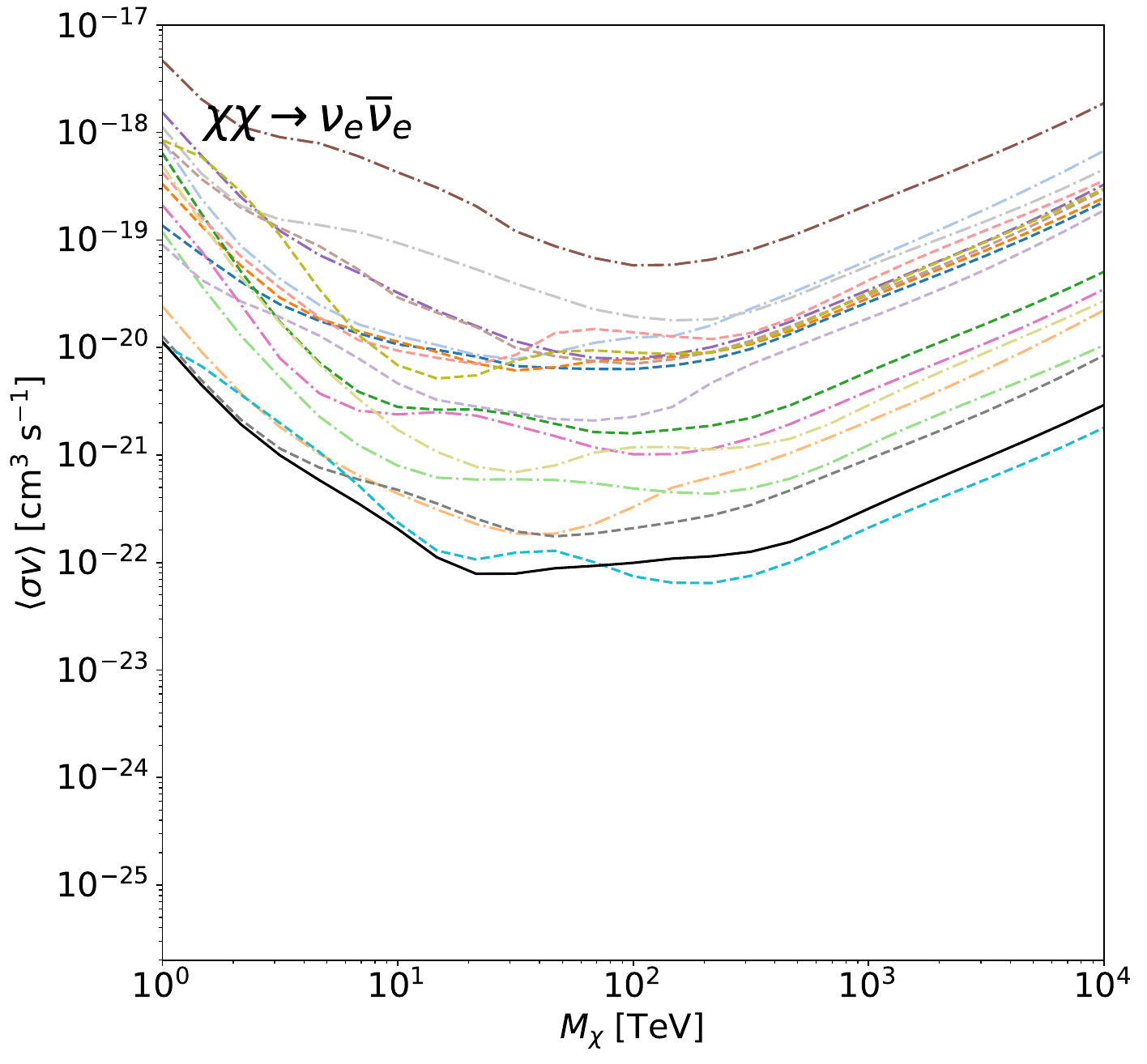} \\
    	\includegraphics[width=0.3\textwidth]{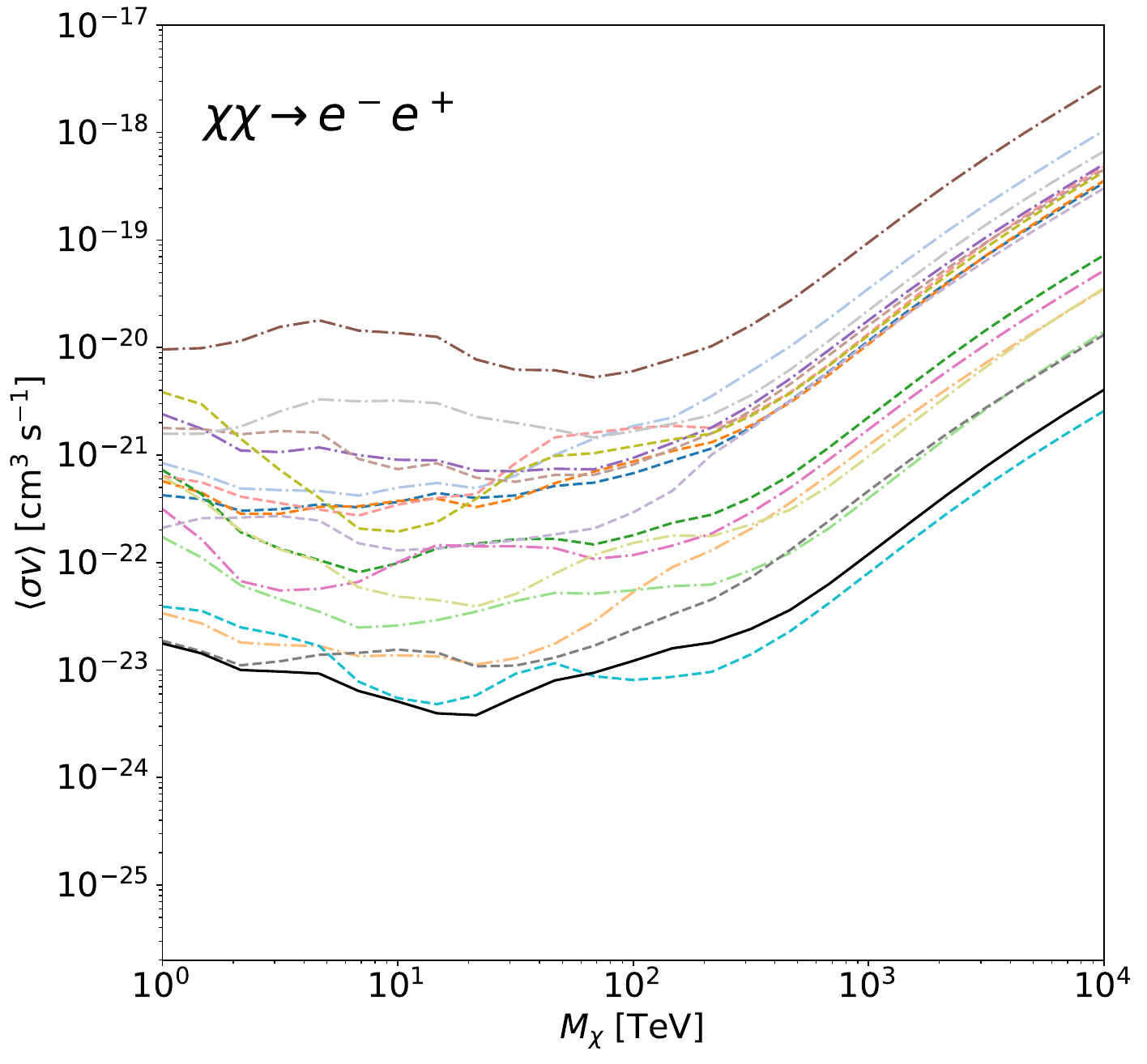} &
    	\includegraphics[width=0.3\textwidth]{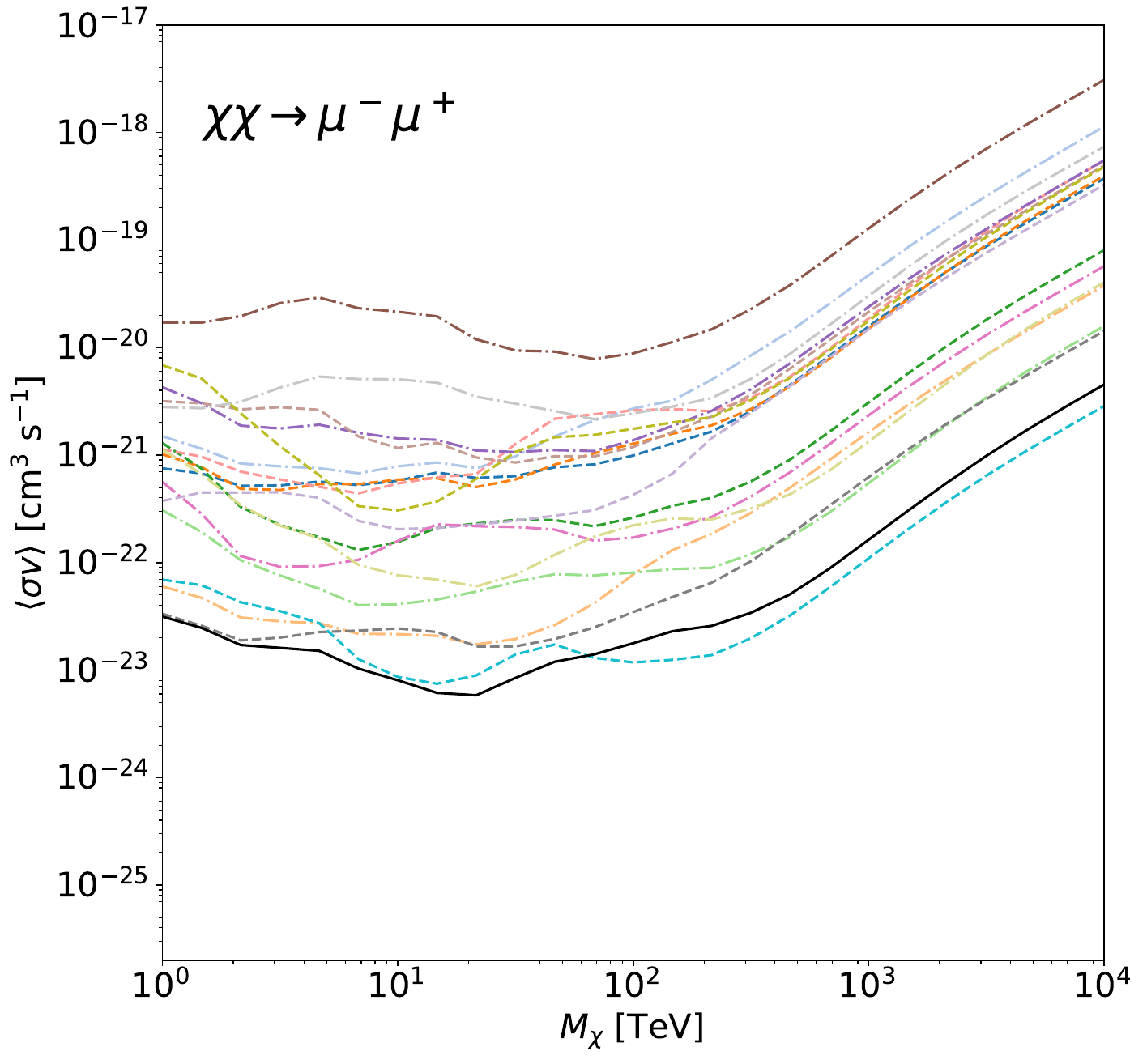} &
    	\includegraphics[width=0.3\textwidth]{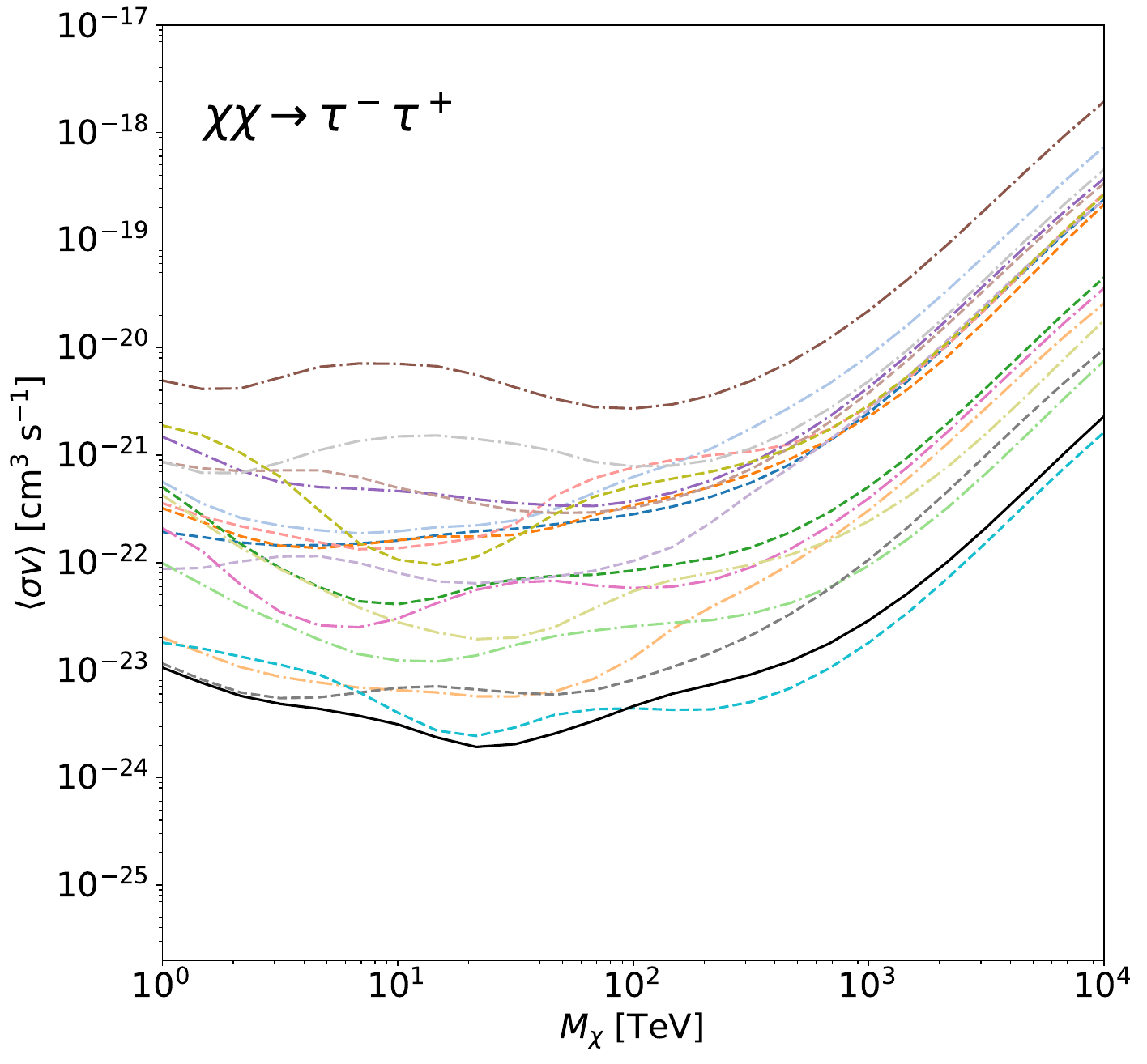} \\
    	\includegraphics[width=0.3\textwidth]{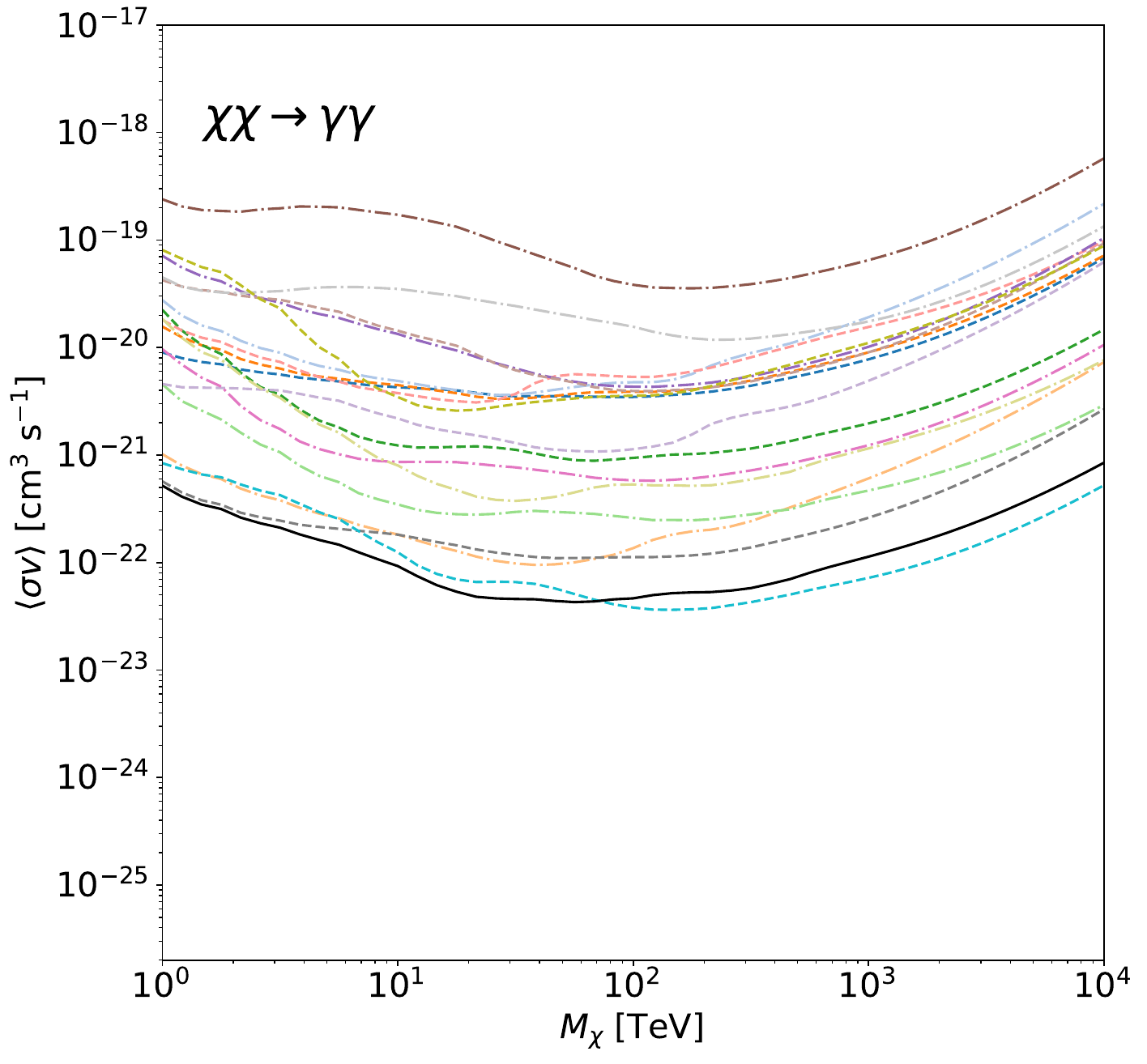} &
    	\includegraphics[width=0.3\textwidth]{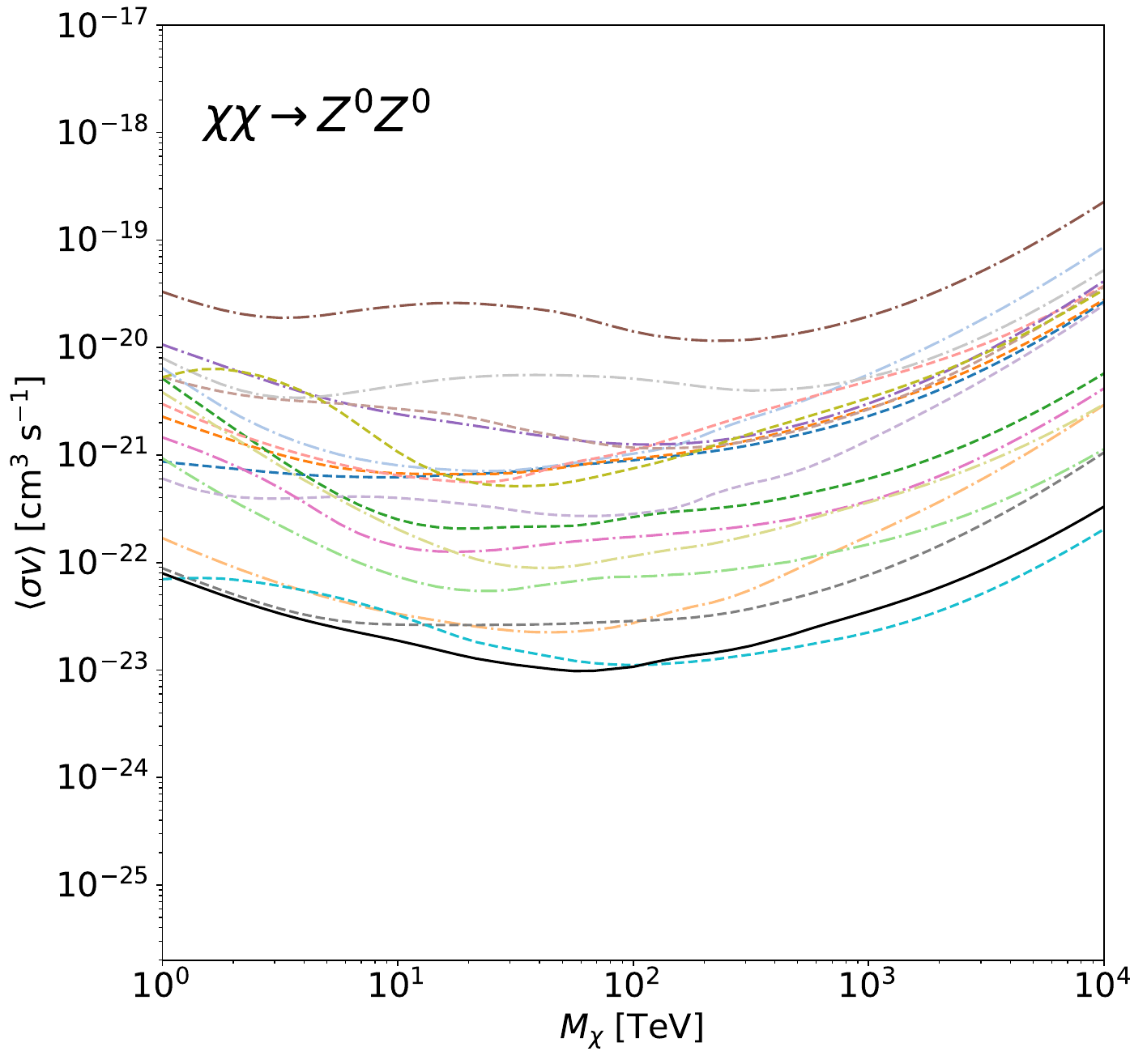} &
    	\raisebox{0.3\height}{\includegraphics[width=0.3\textwidth]{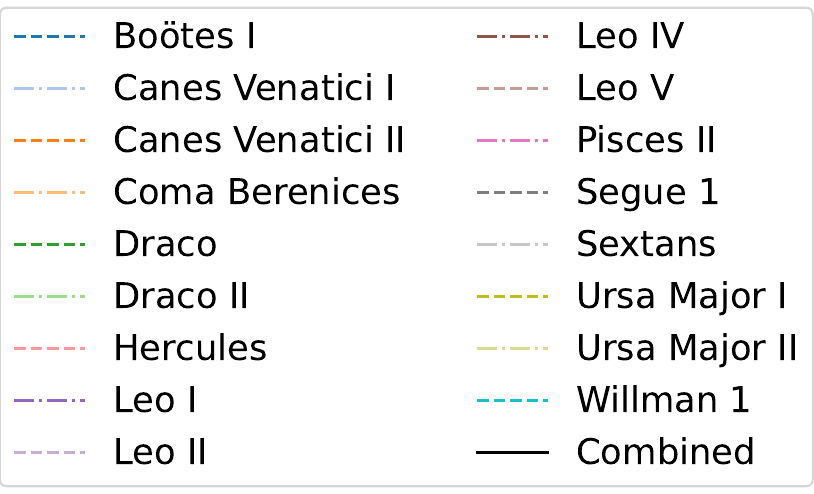}}
	\end{tabular}
    }
    \caption{HAWC upper limits at 95\% confidence level on \sv~versus $m_\chi$ for $\chi\chi \rightarrow b\overline{b}$, $t\overline{t}$, $u\overline{u}$, $d\overline{d}$, $W^-W^+$, $\nu_e\overline{\nu}_e$, $e^-e^+$, $\mu^-\mu^+$, $\tau^-\tau^+$, $\gamma\gamma$ and $Z^0Z^0$. Limits are with \LS{} \J-factors \cite{DM_Strigari20}. The solid line represents the observed combined limit. Dashed lines represent limits from individual dSphs.}
\label{fig:LSmtd_limits_1of2}
\end{figure}

\begin{figure}[htbp]
\centering{
	\begin{tabular}{ccc}
   	\includegraphics[width=0.3\textwidth]{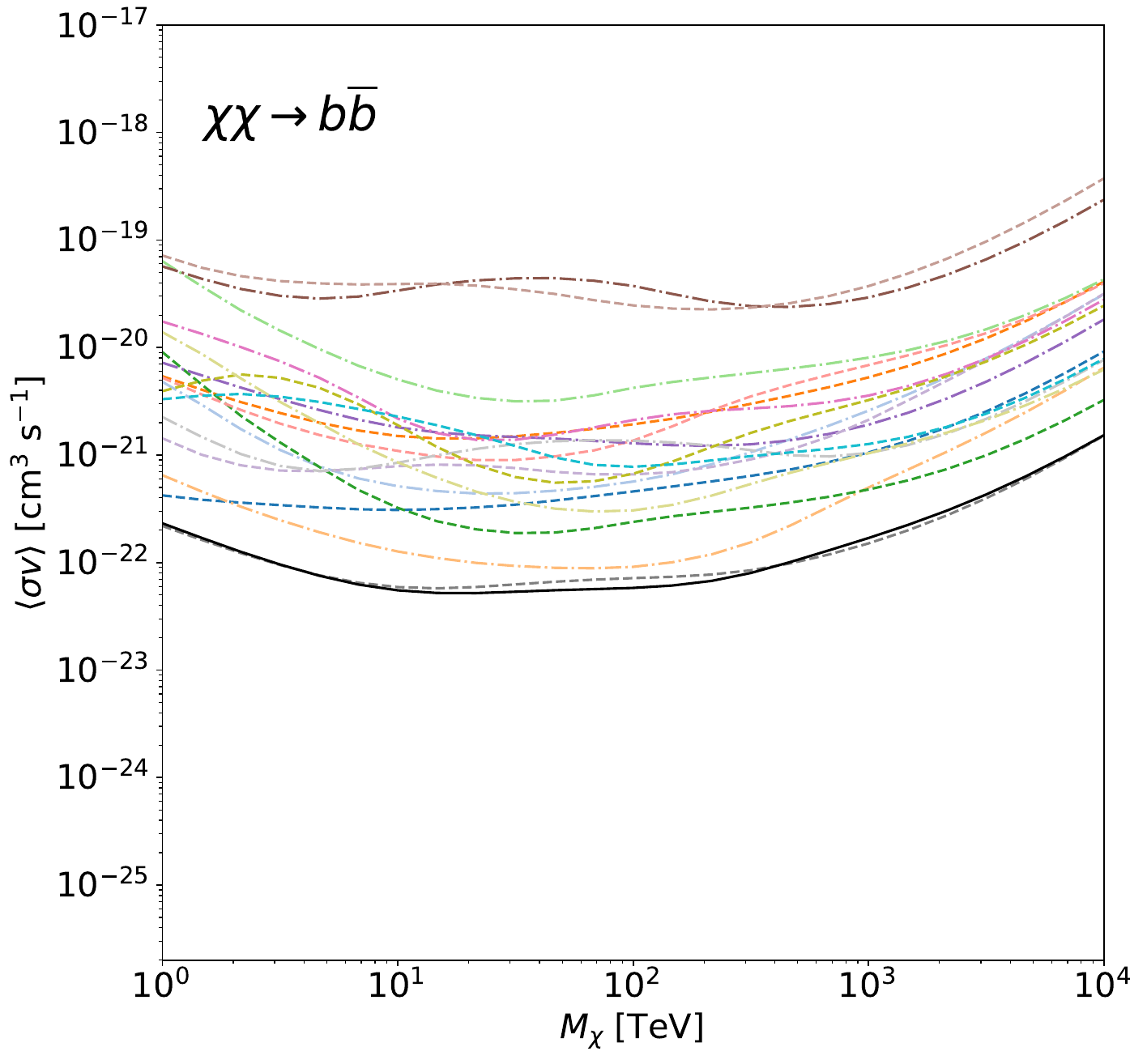} &
    	\includegraphics[width=0.3\textwidth]{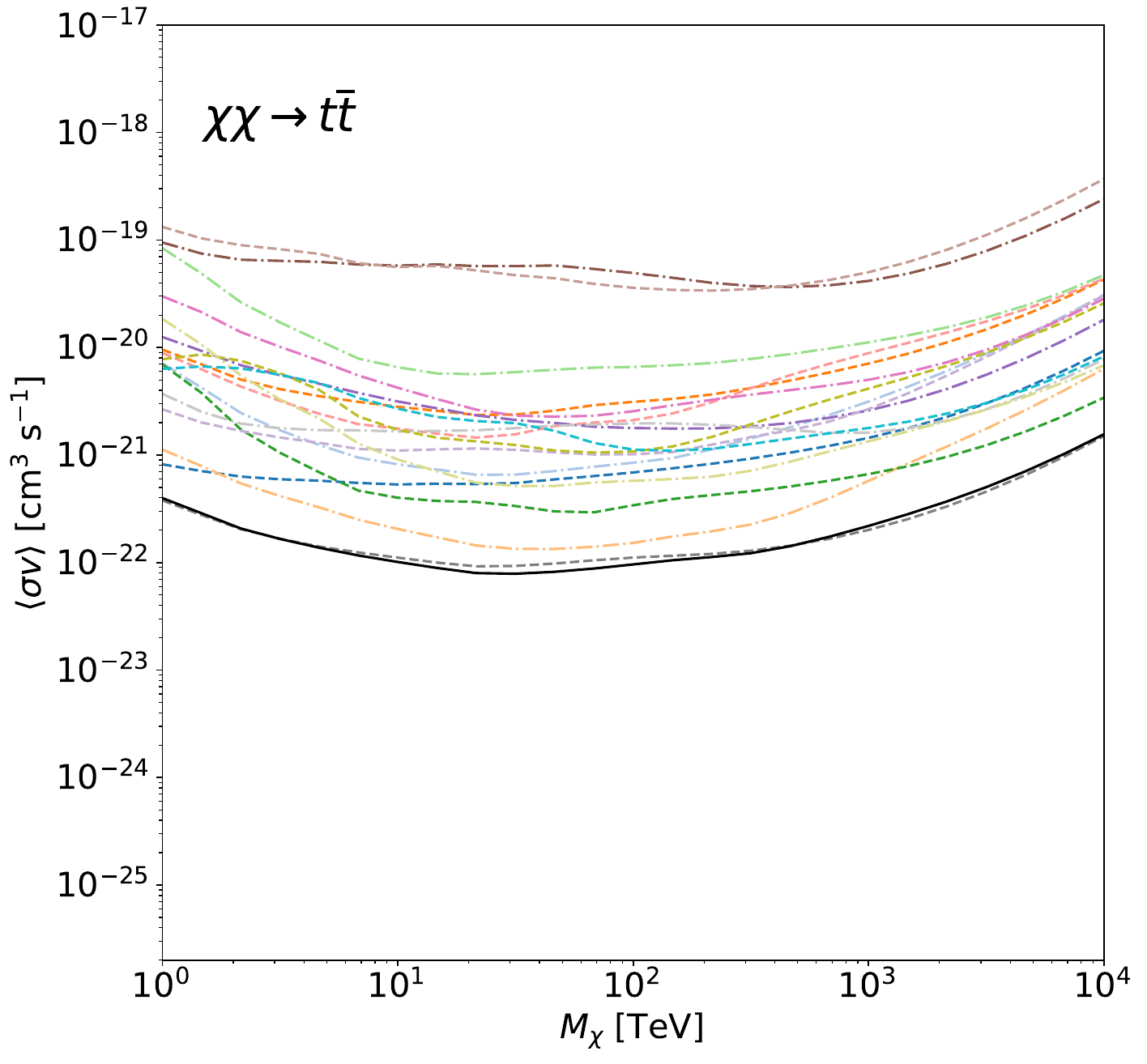} &
    	\includegraphics[width=0.3\textwidth]{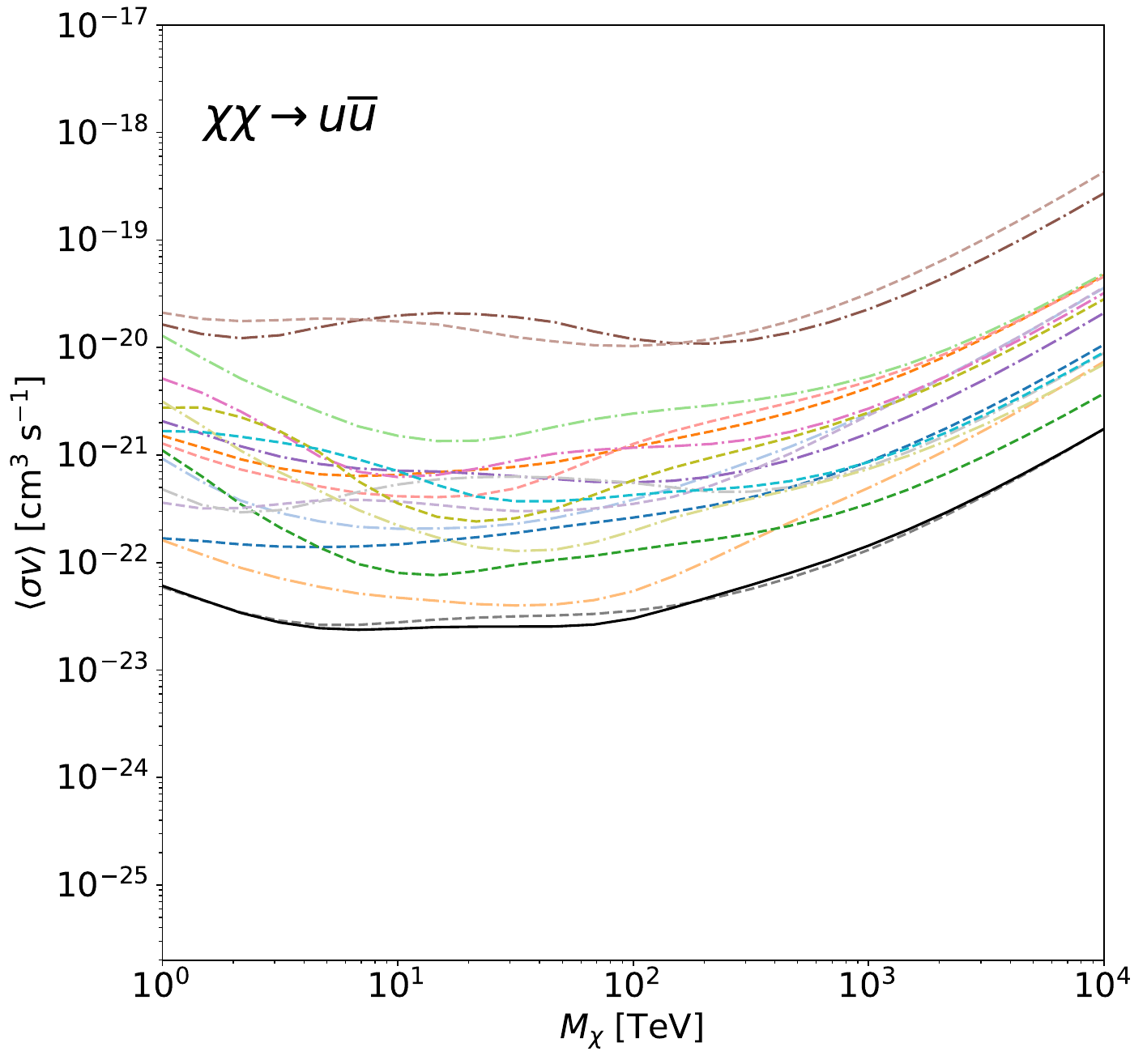} \\
    	\includegraphics[width=0.3\textwidth]{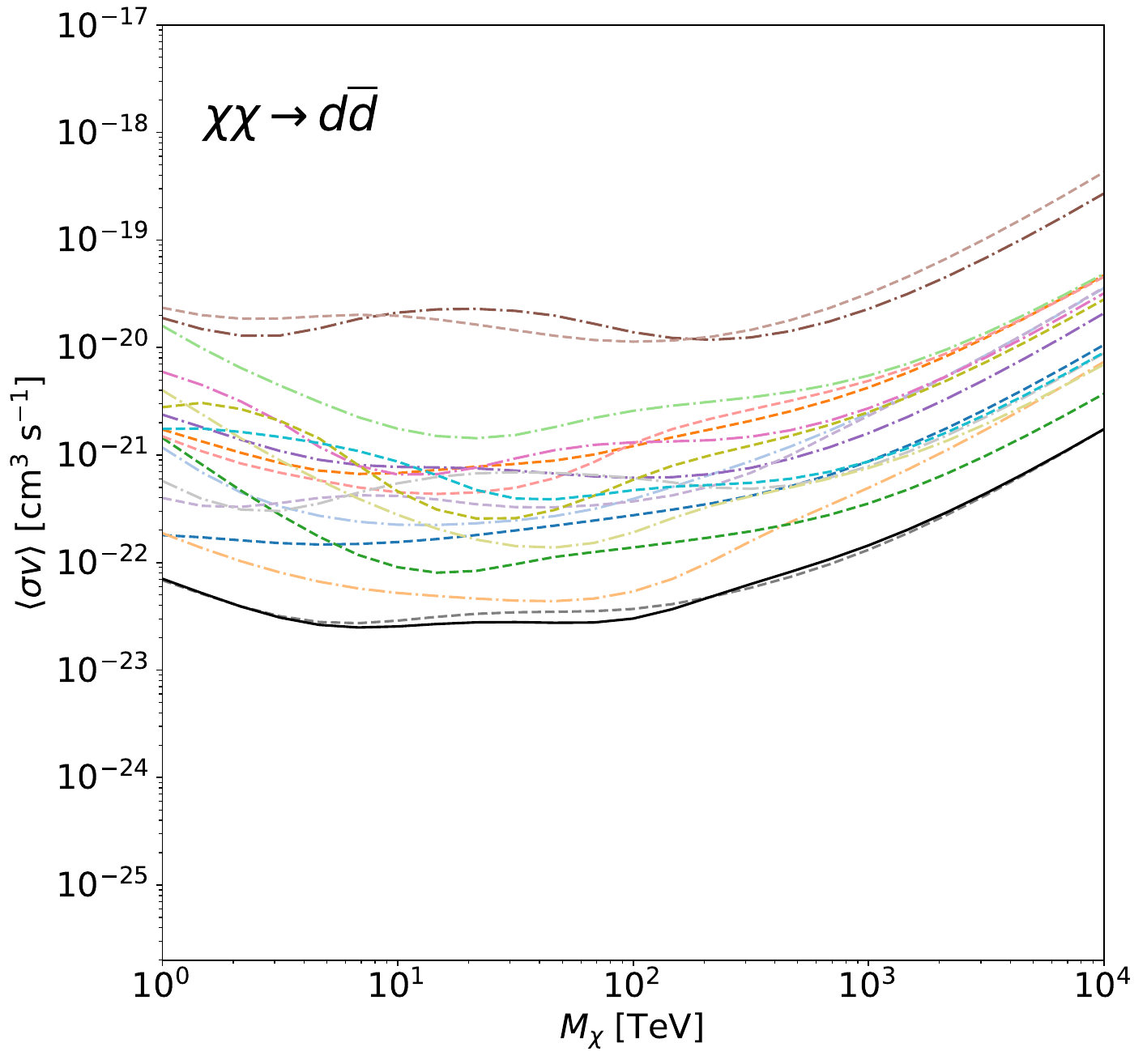} &
    	\includegraphics[width=0.3\textwidth]{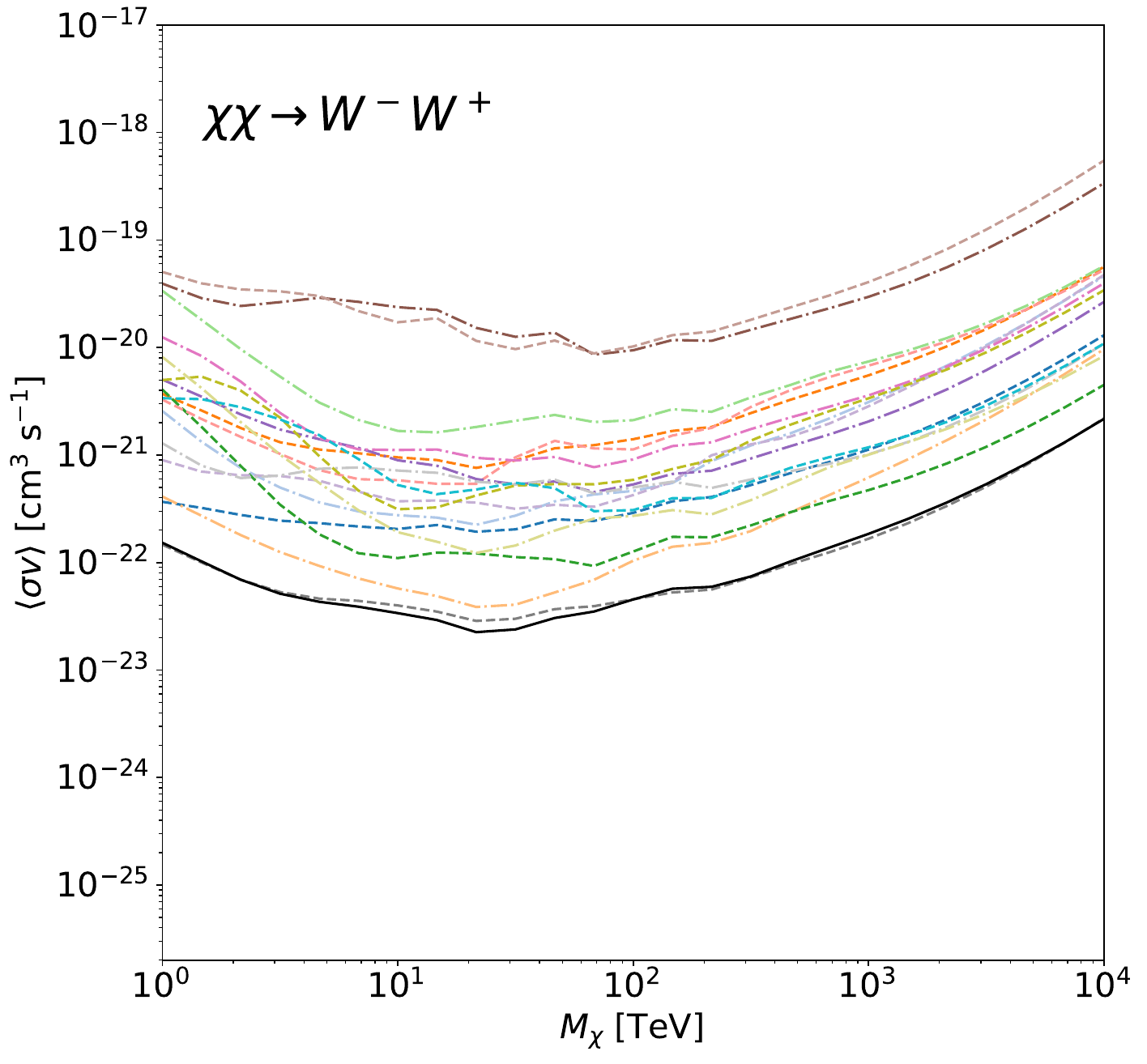} &
     	\includegraphics[width=0.3\textwidth]{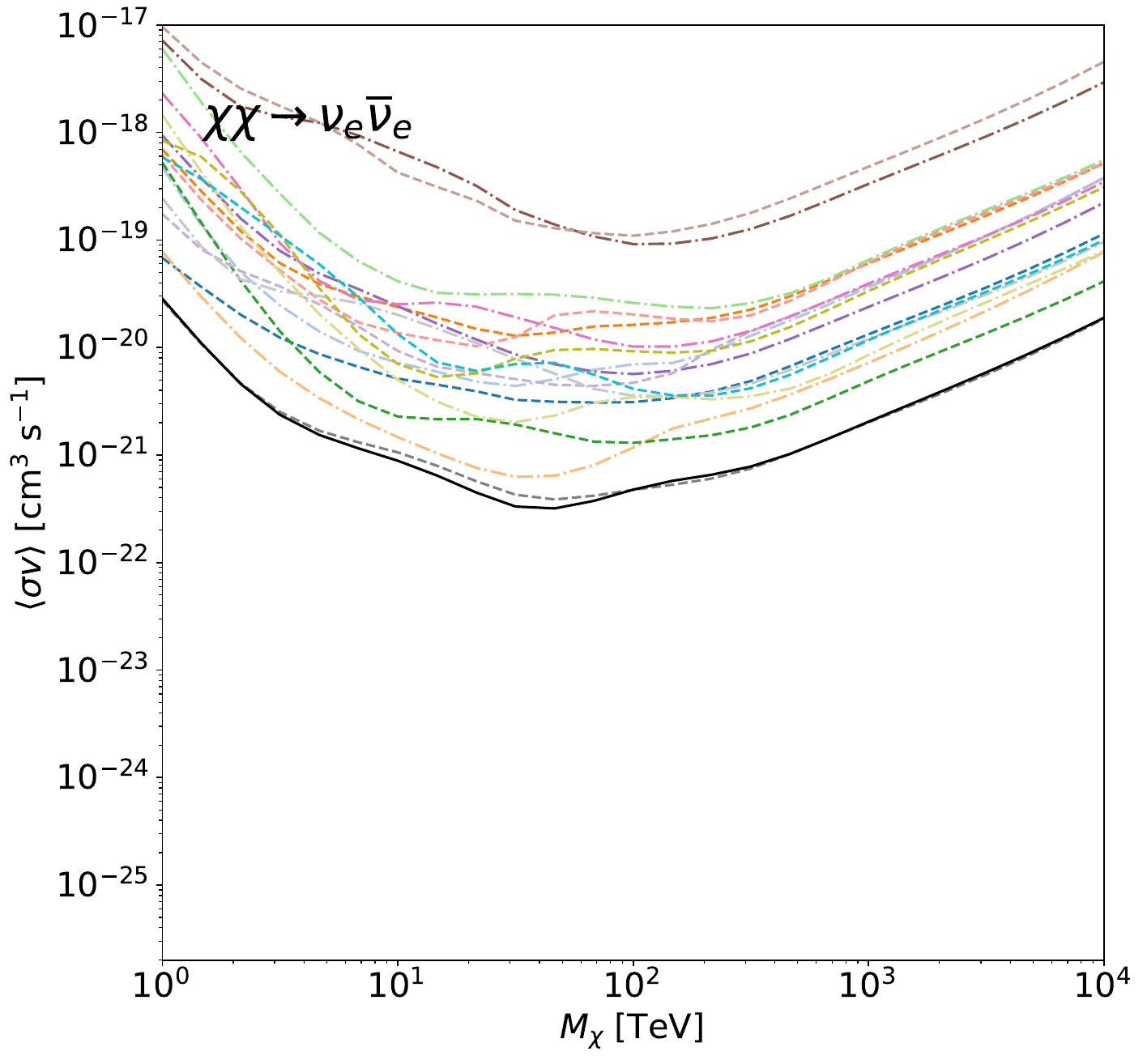} \\
	\includegraphics[width=0.3\textwidth]{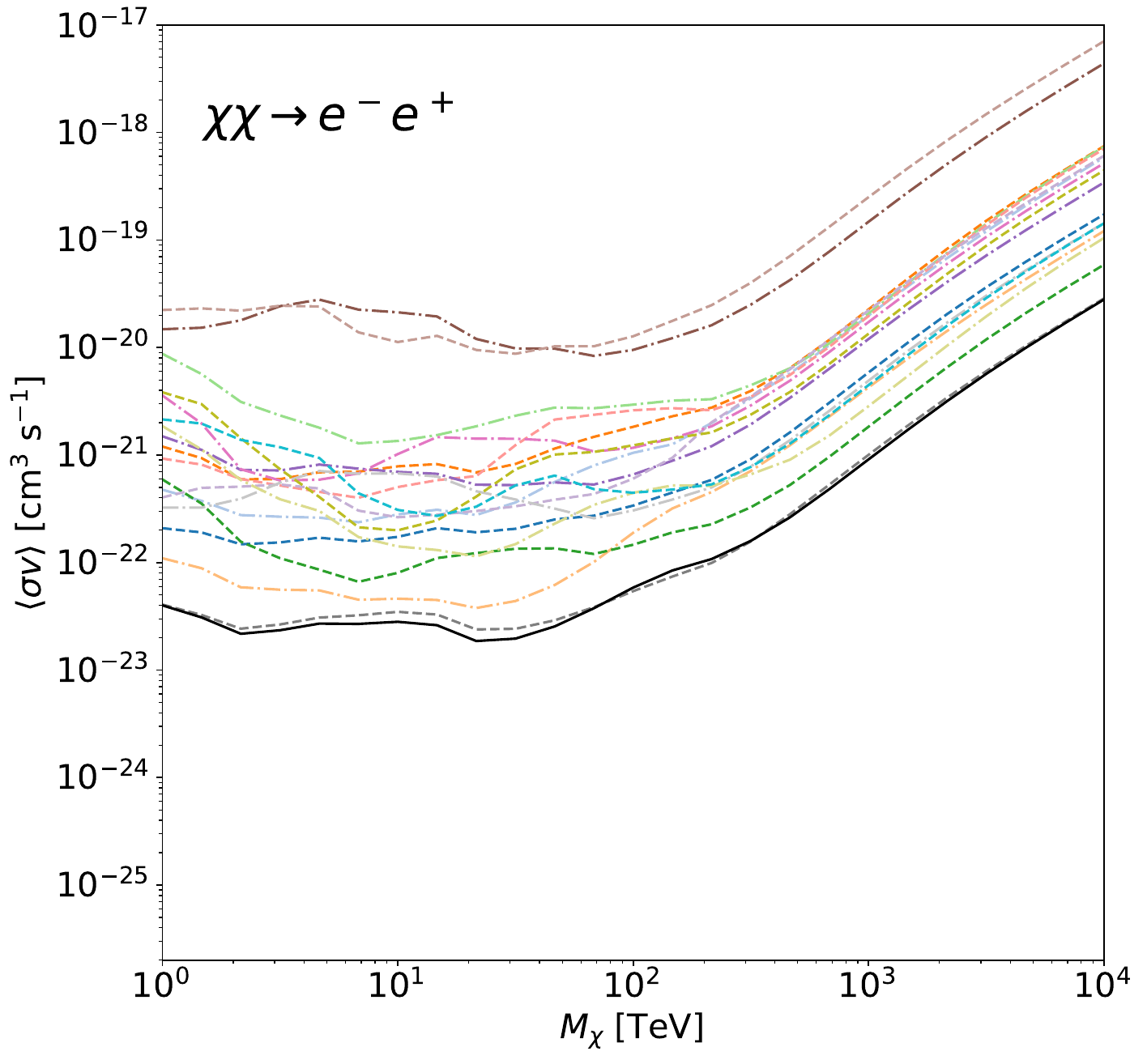} &
    	\includegraphics[width=0.3\textwidth]{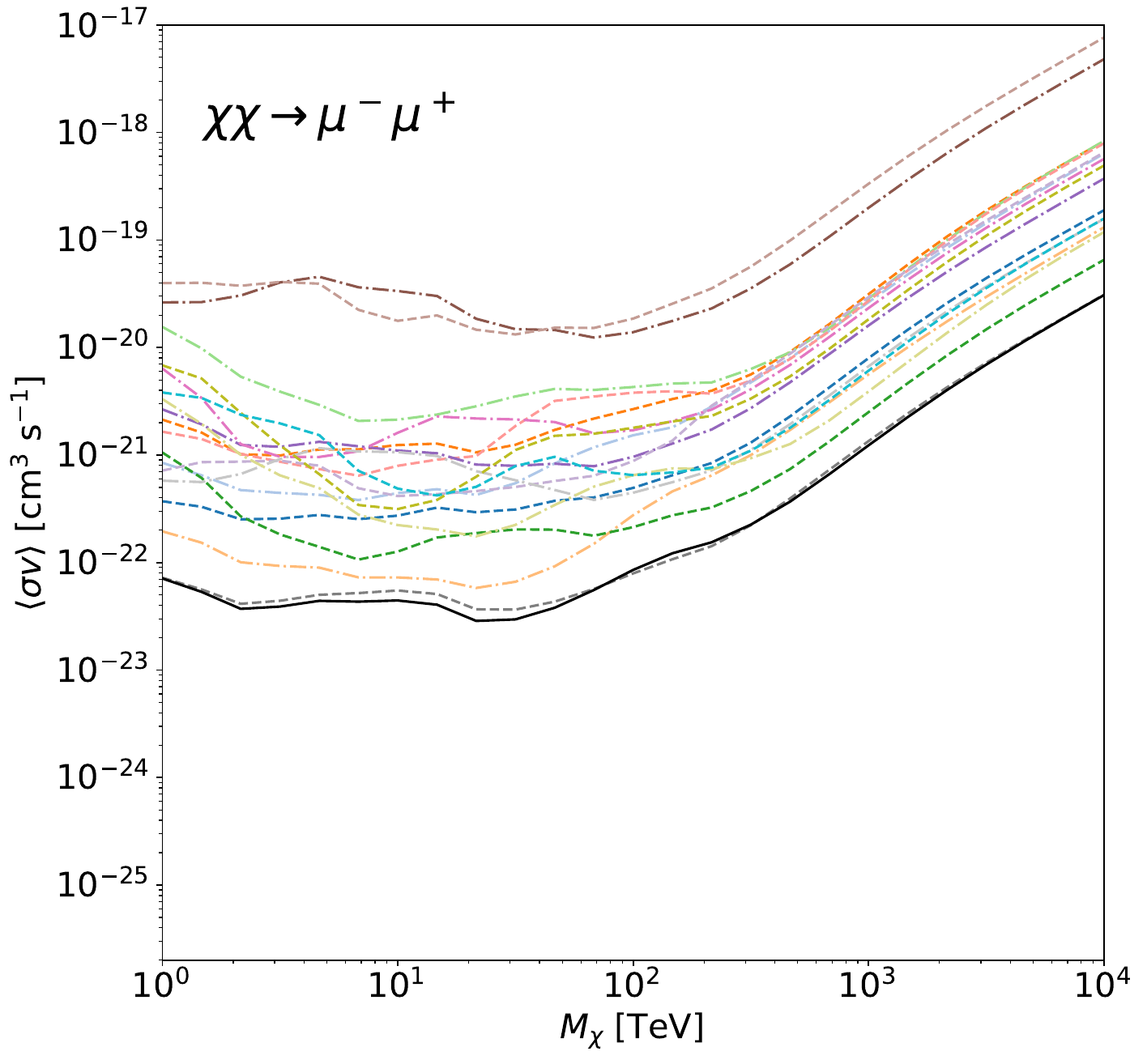} &
    	\includegraphics[width=0.3\textwidth]{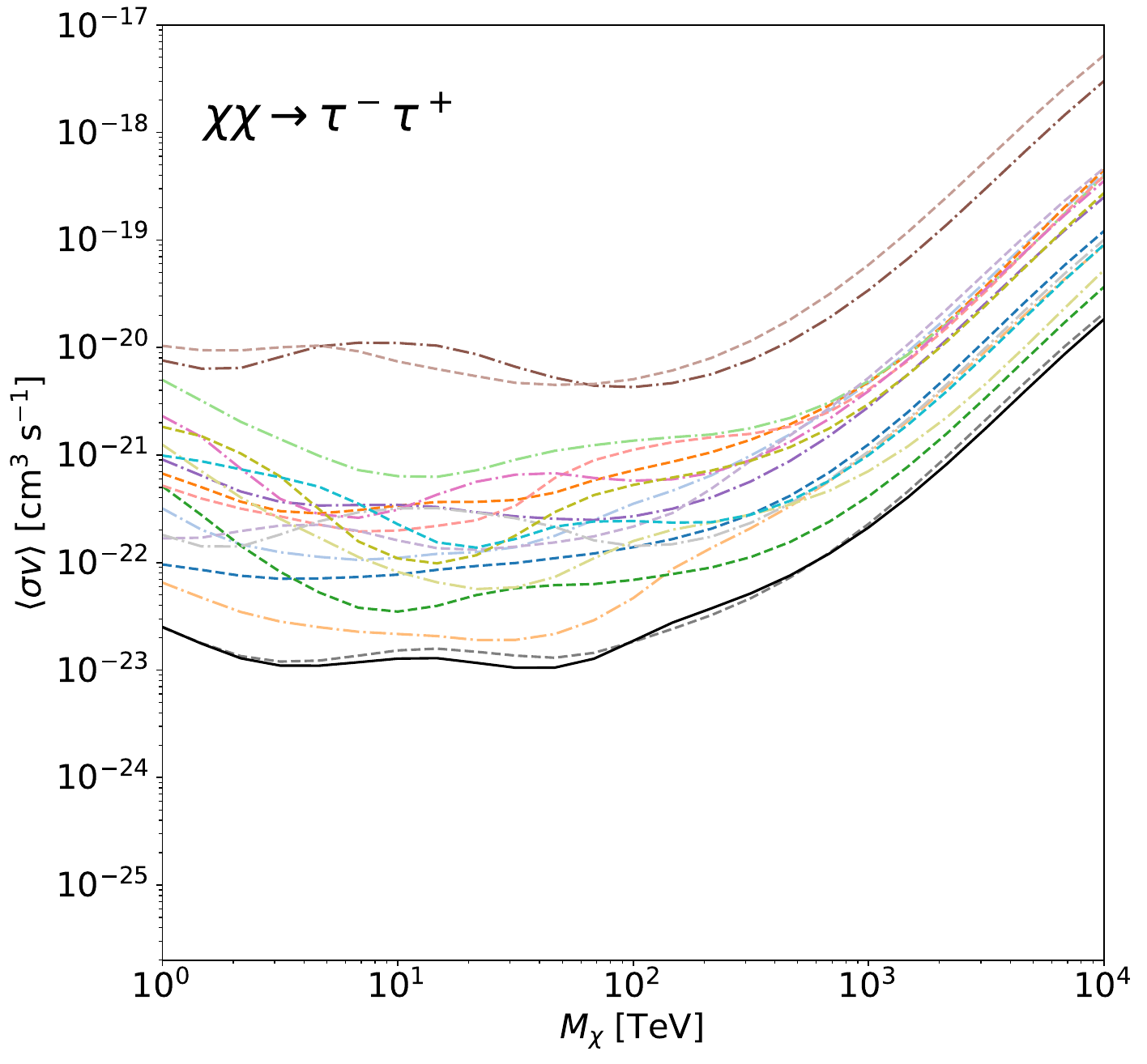} \\
	\includegraphics[width=0.3\textwidth]{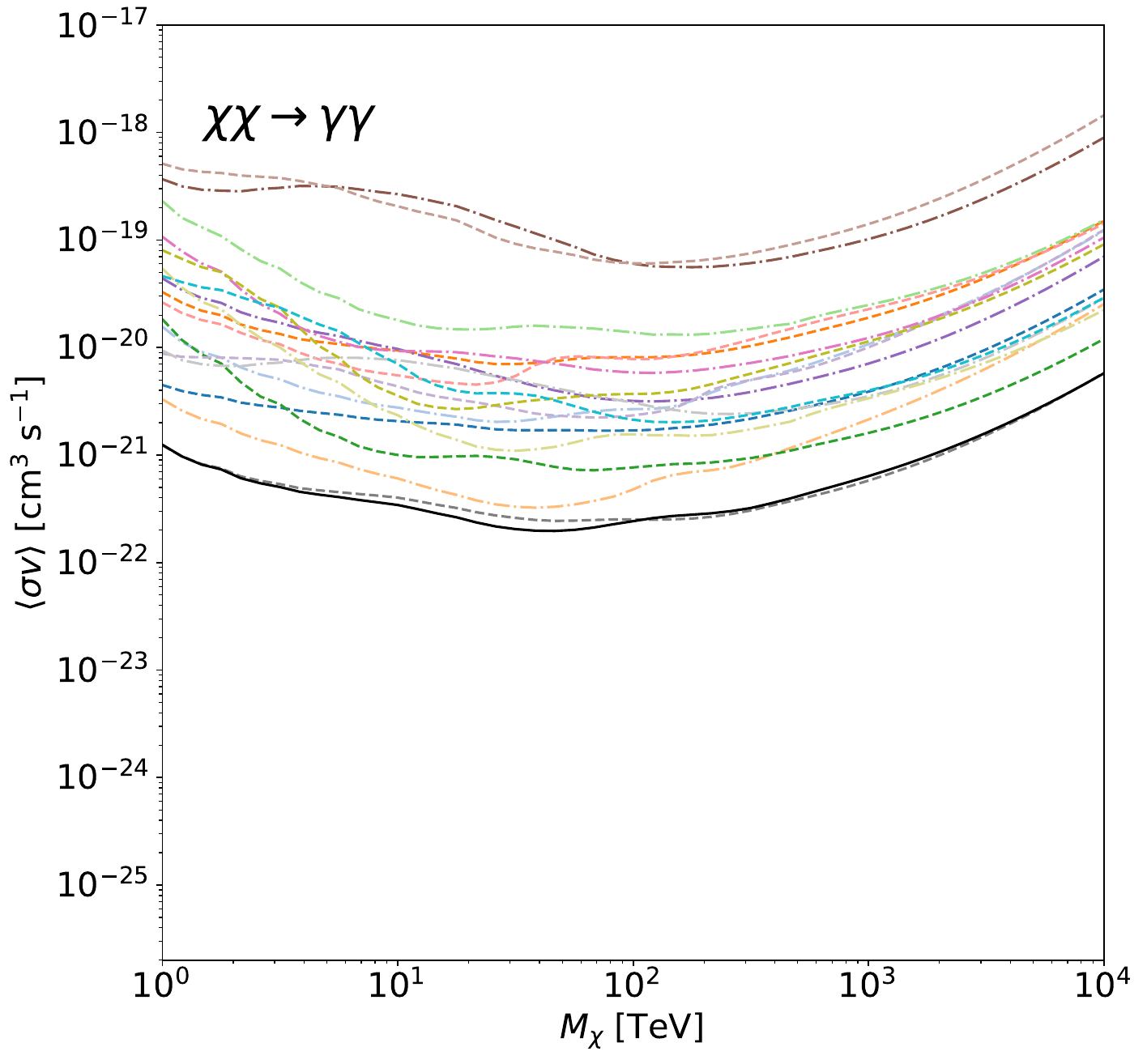} &
    	\includegraphics[width=0.3\textwidth]{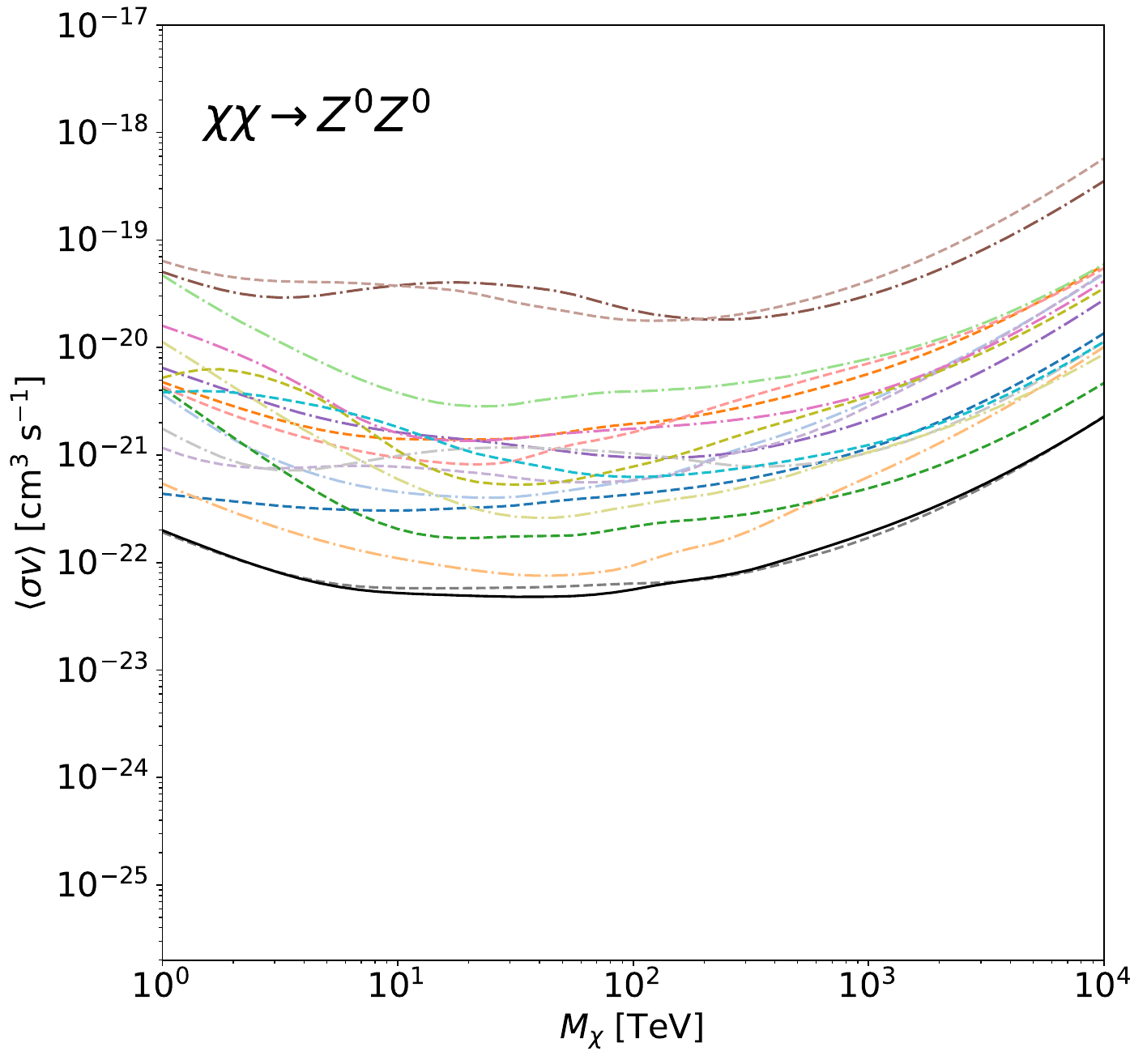} &
    	\raisebox{0.3\height}{\includegraphics[width=0.3\textwidth]{figures/NEWresults/LEGEND_Combined_results.pdf}}
	\end{tabular}
    }
    \caption{Same as \cref{fig:LSmtd_limits_1of2} but with \GS{} \J-factors \cite{Ando_2020}.}
\label{fig:GSmtd_limits_1of2}
\end{figure}

\begin{figure}[htbp]
\centering{
	\begin{tabular}{ccc}
    \includegraphics[width=0.3\textwidth]{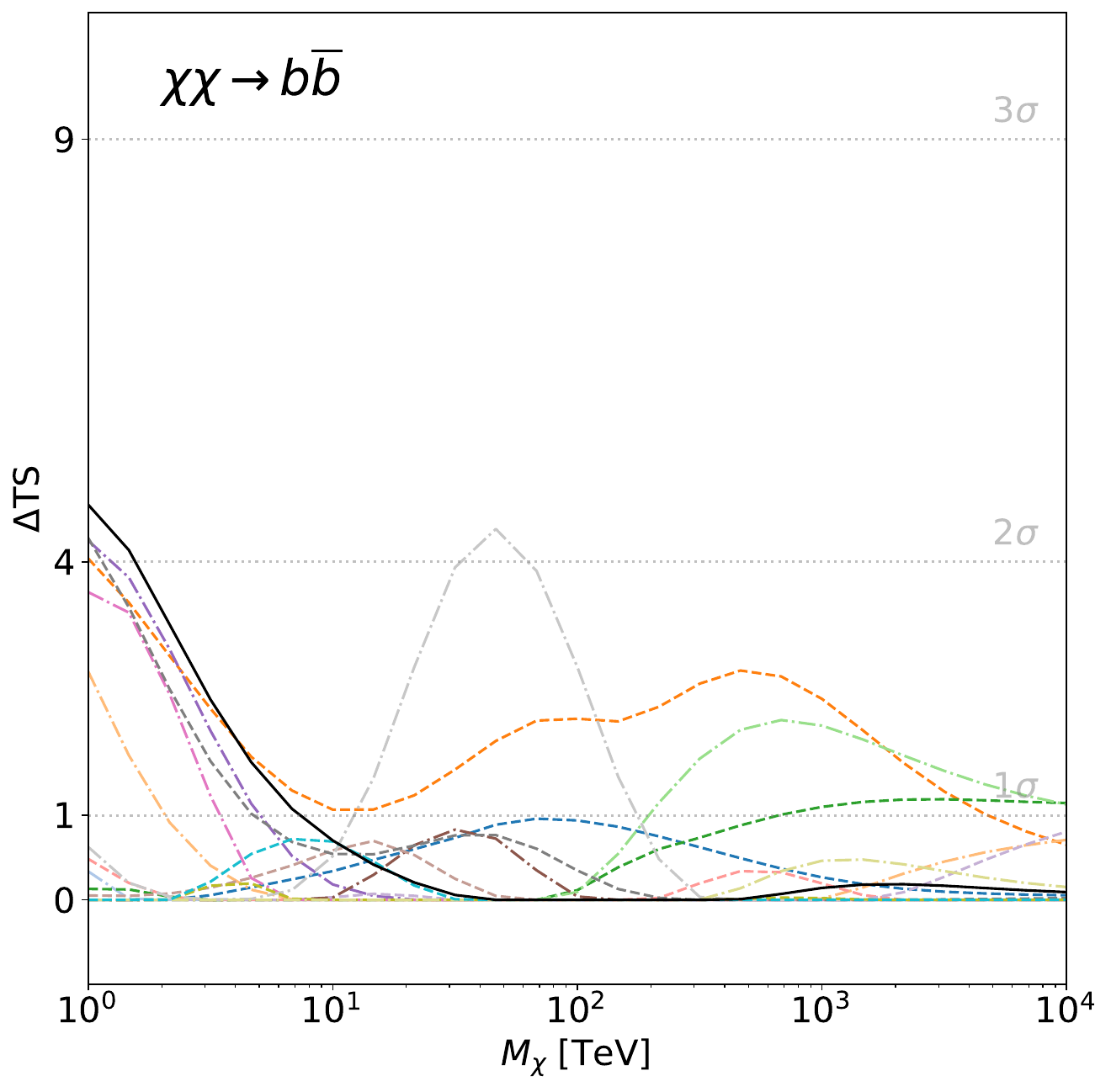} &
    \includegraphics[width=0.3\textwidth]{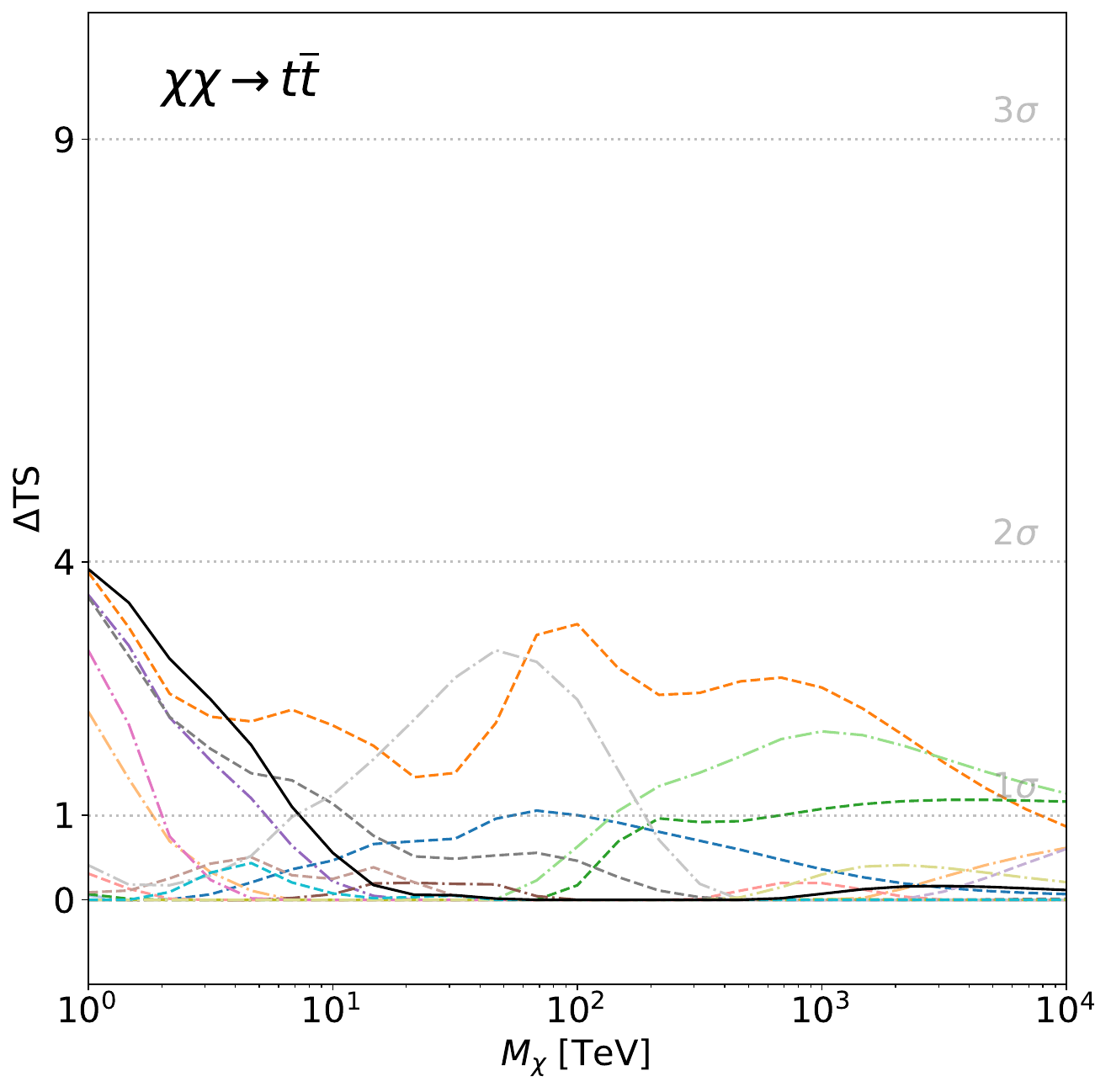} &
    \includegraphics[width=0.3\textwidth]{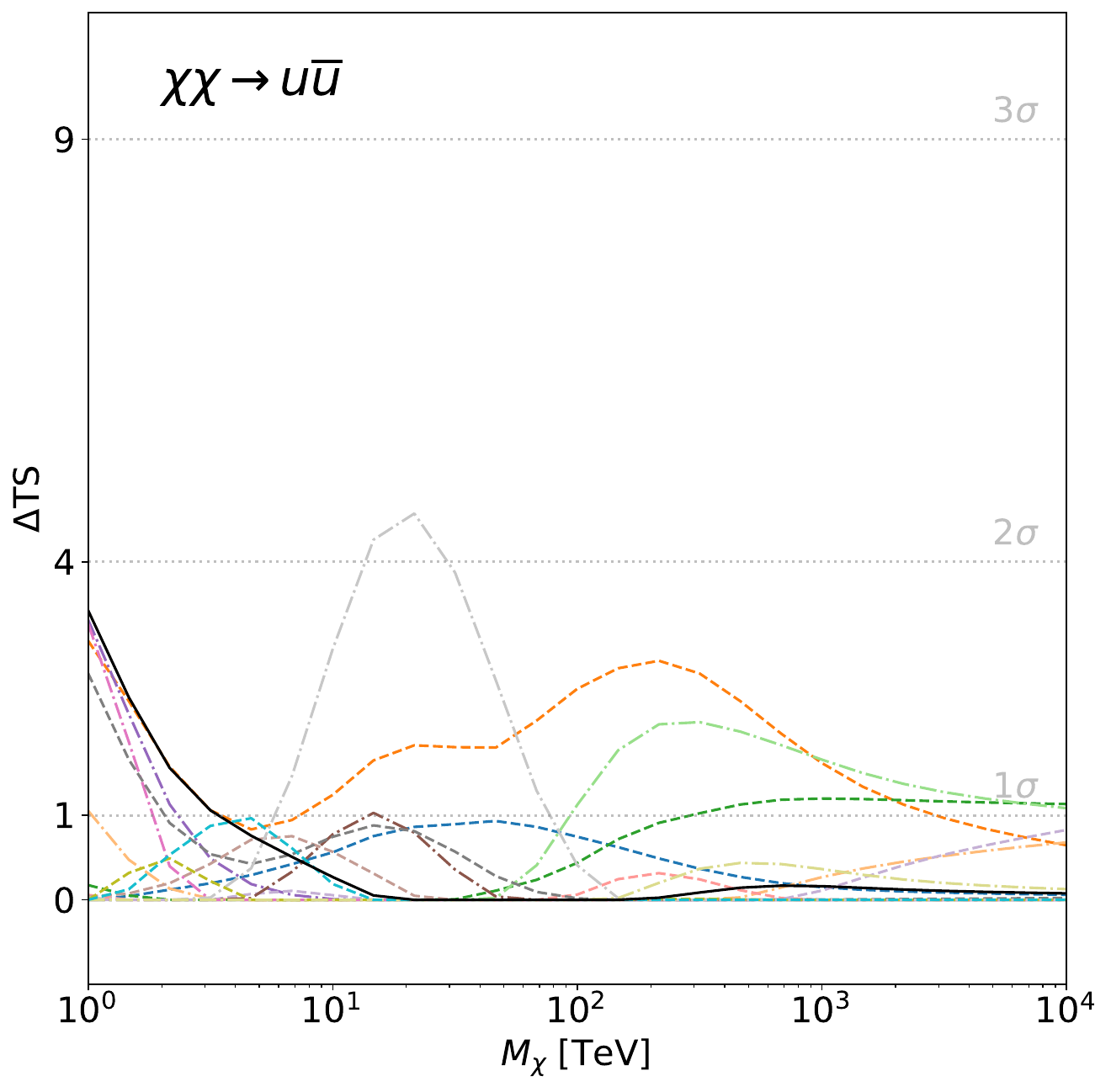} \\
    \includegraphics[width=0.3\textwidth]{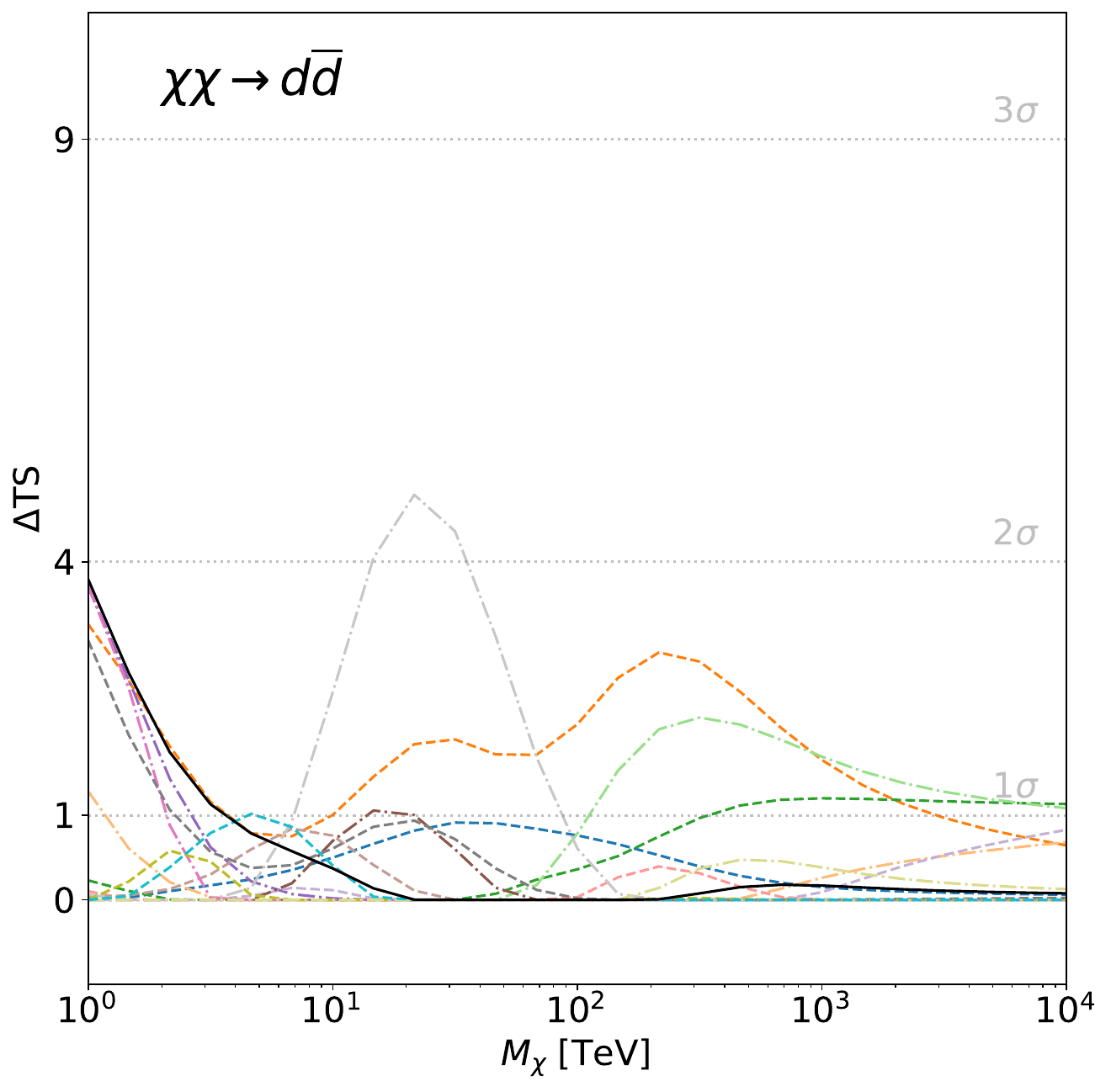} &
    \includegraphics[width=0.3\textwidth]{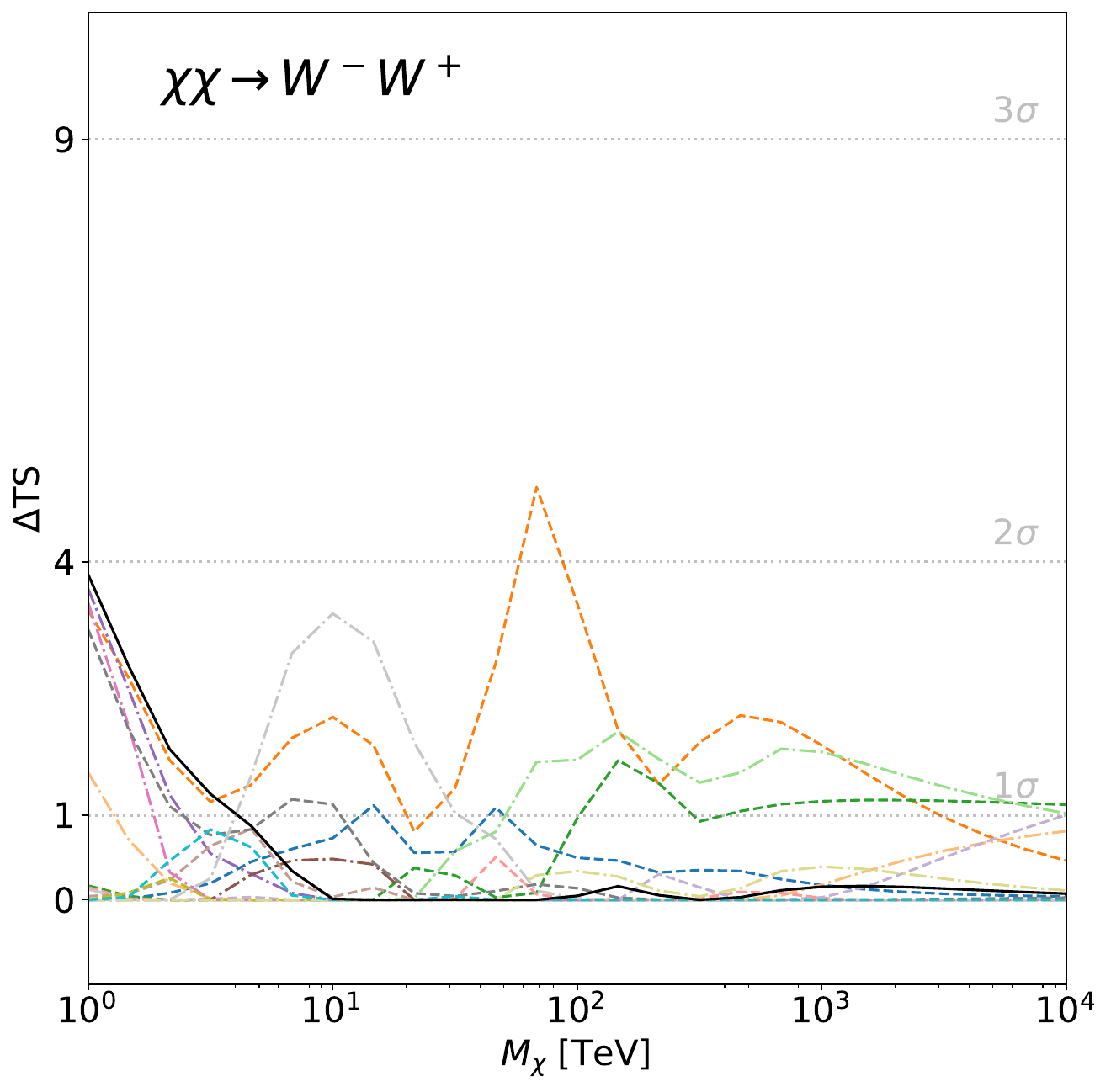} &
     \includegraphics[width=0.3\textwidth]{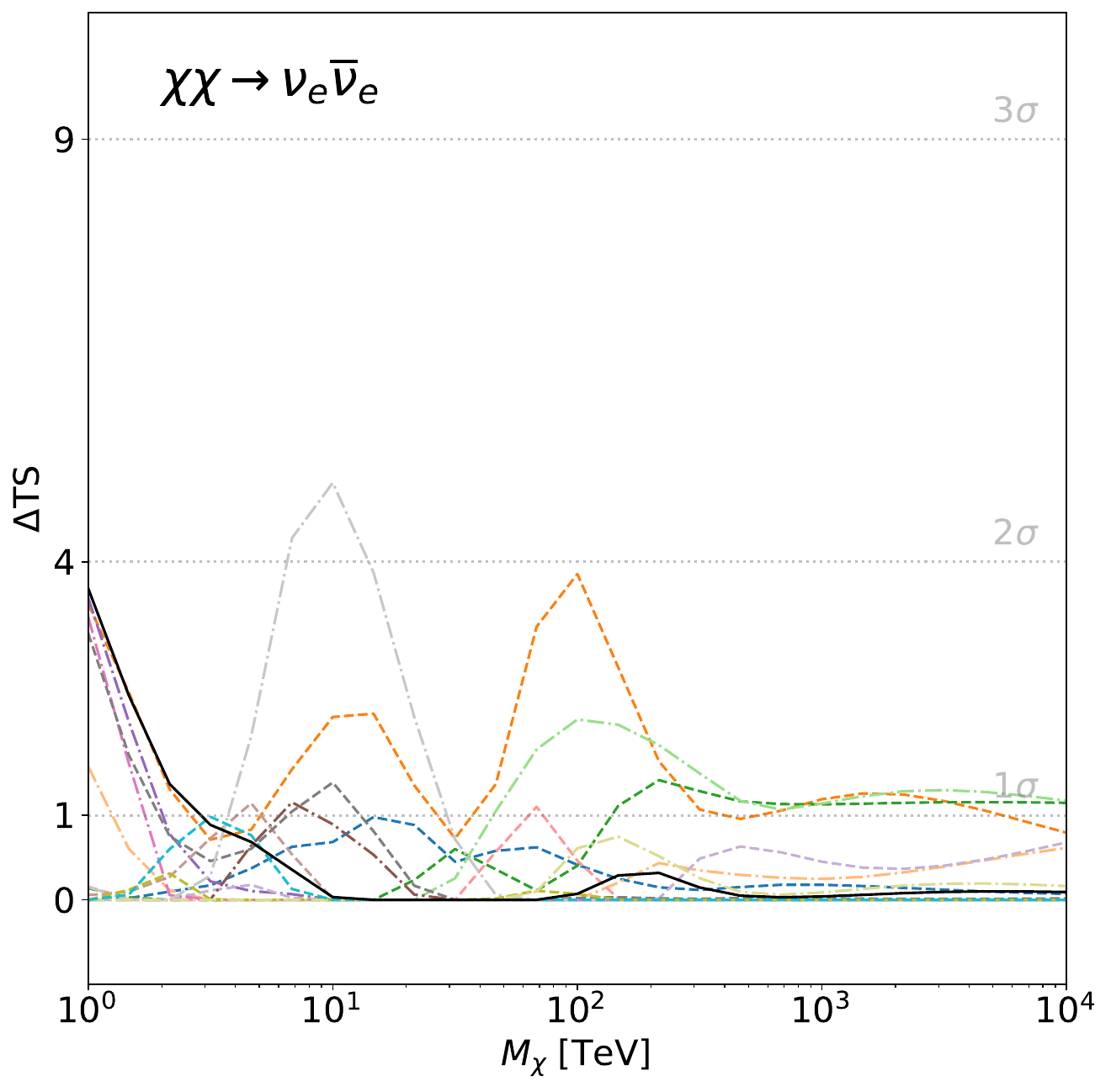} \\
    \includegraphics[width=0.3\textwidth]{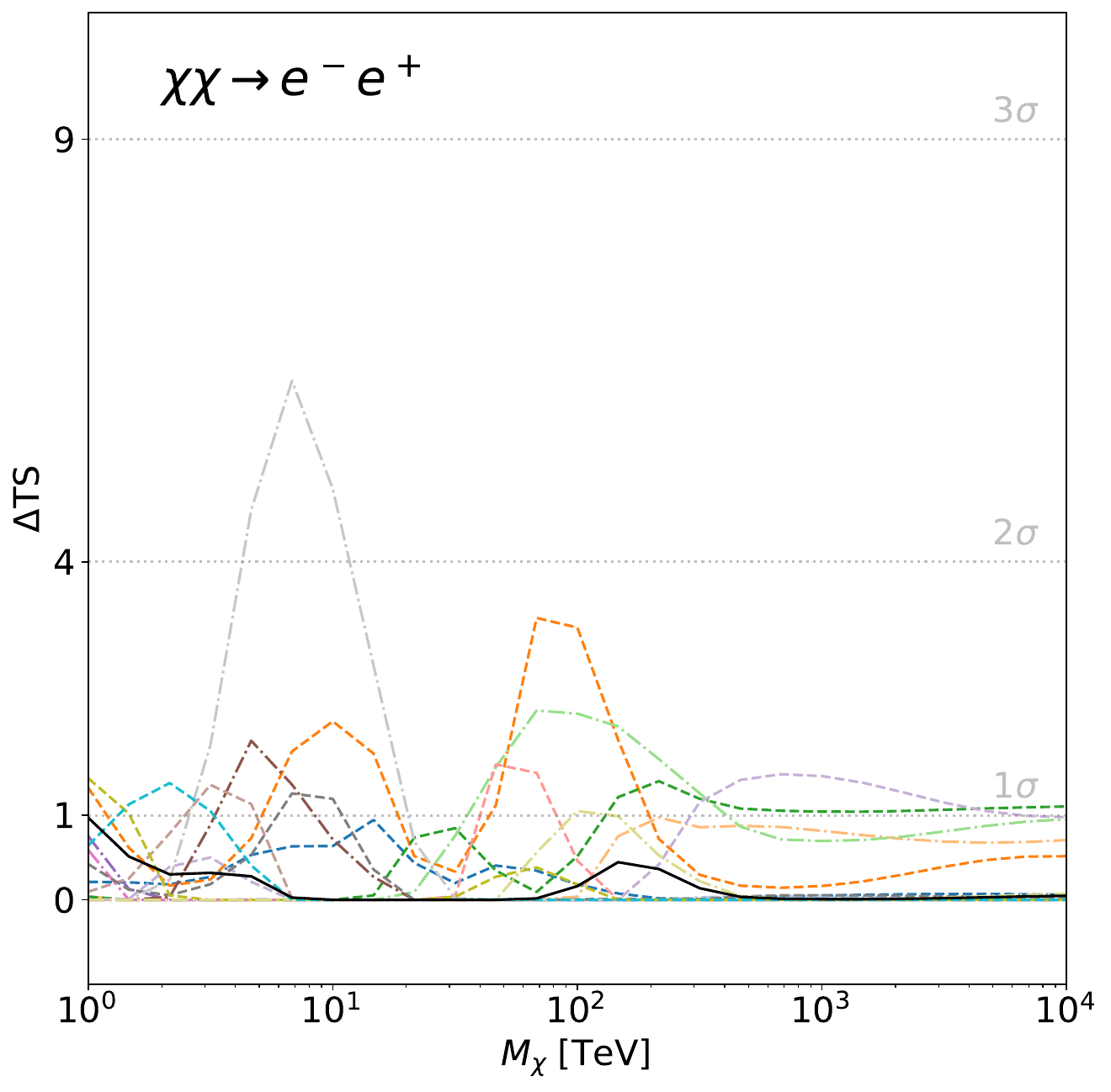} &
    \includegraphics[width=0.3\textwidth]{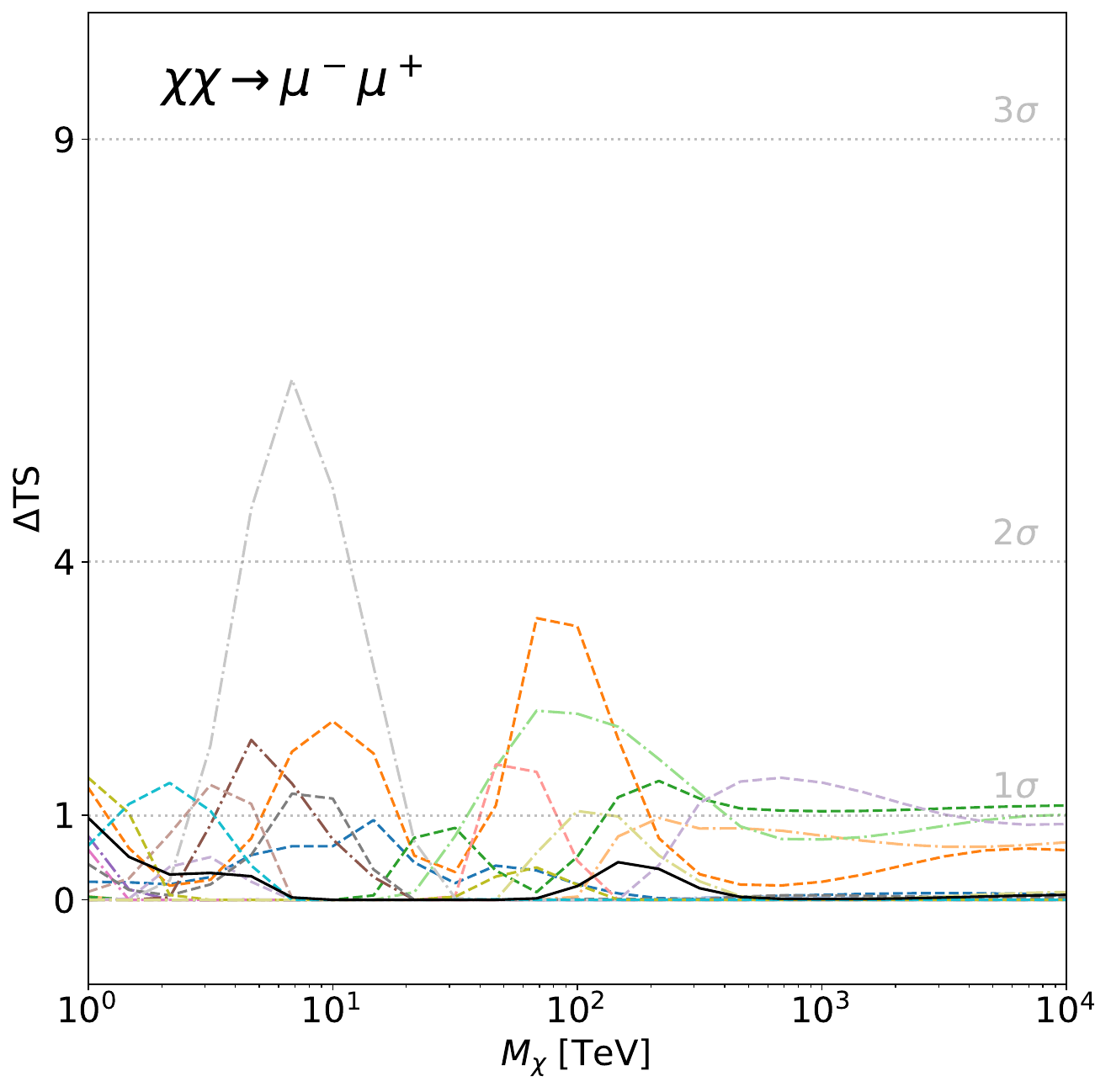} &
    \includegraphics[width=0.3\textwidth]{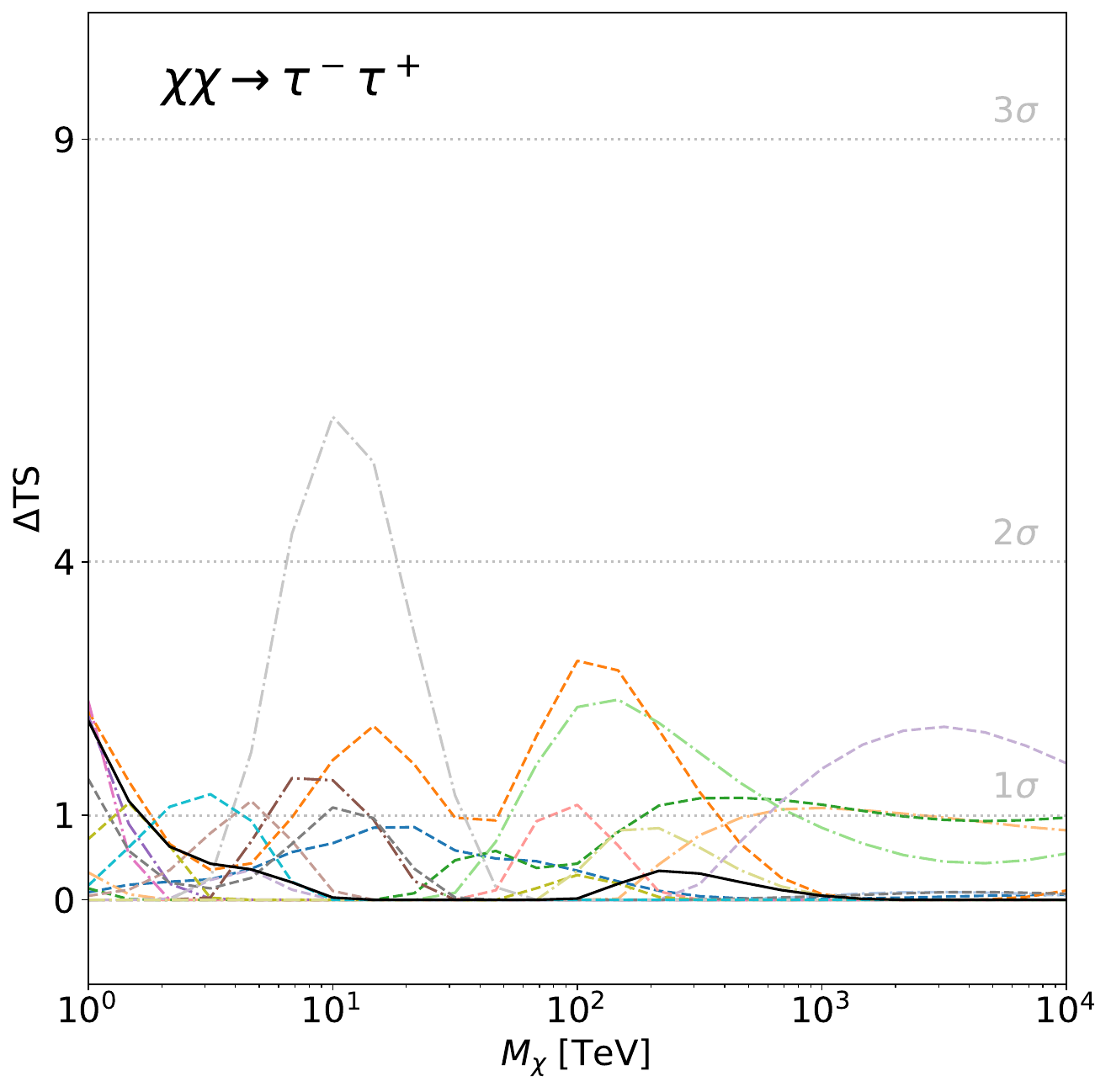} \\
    \includegraphics[width=0.3\textwidth]{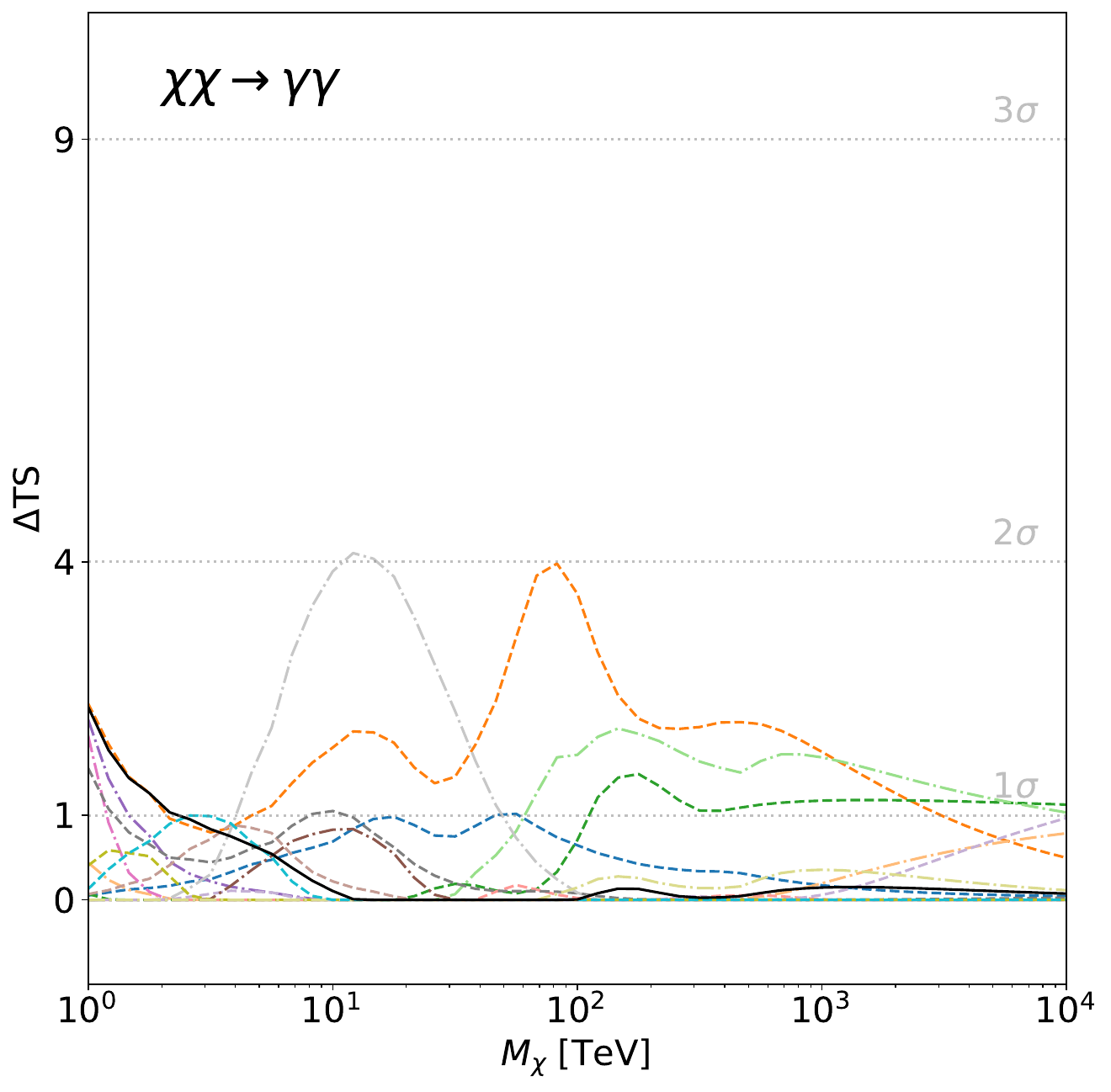} &
    \includegraphics[width=0.3\textwidth]{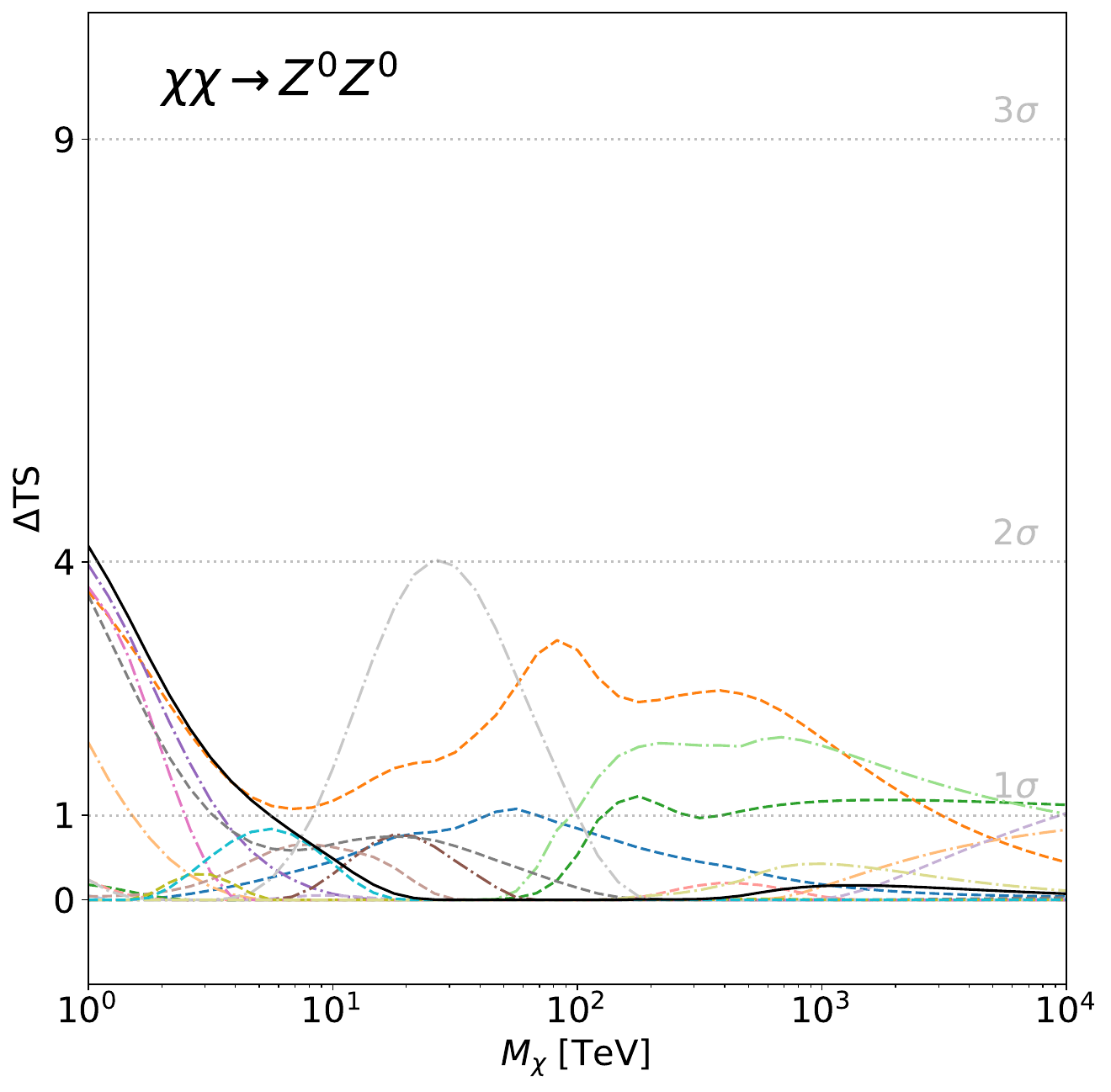} &
    	\raisebox{0.4\height}{\includegraphics[width=0.3\textwidth]{figures/NEWresults/LEGEND_Combined_results.pdf}}
	\end{tabular}
    }
    \caption{HAWC TS values for best fit \sv~versus $m_\chi$ for SM annihilation channels: $\chi\chi \rightarrow b\overline{b}$, $t\overline{t}$, $u\overline{u}$, $d\overline{d}$, $W^-W^+$, $\nu_e\overline{\nu}_e$, $e^-e^+$, $\mu^-\mu^+$, $\tau^-\tau^+$, $\gamma\gamma$ and $Z^0Z^0$. Limits use \LS{} \J-factors. The solid black line shows the combined best fit TS values. The colored, dashed lines are the TS values from each dSph.}
\label{fig:LSmtd_TS_1of2}
\end{figure}

\begin{figure}[htbp]
\centering{
	\begin{tabular}{ccc}
    	\includegraphics[width=0.3\textwidth]{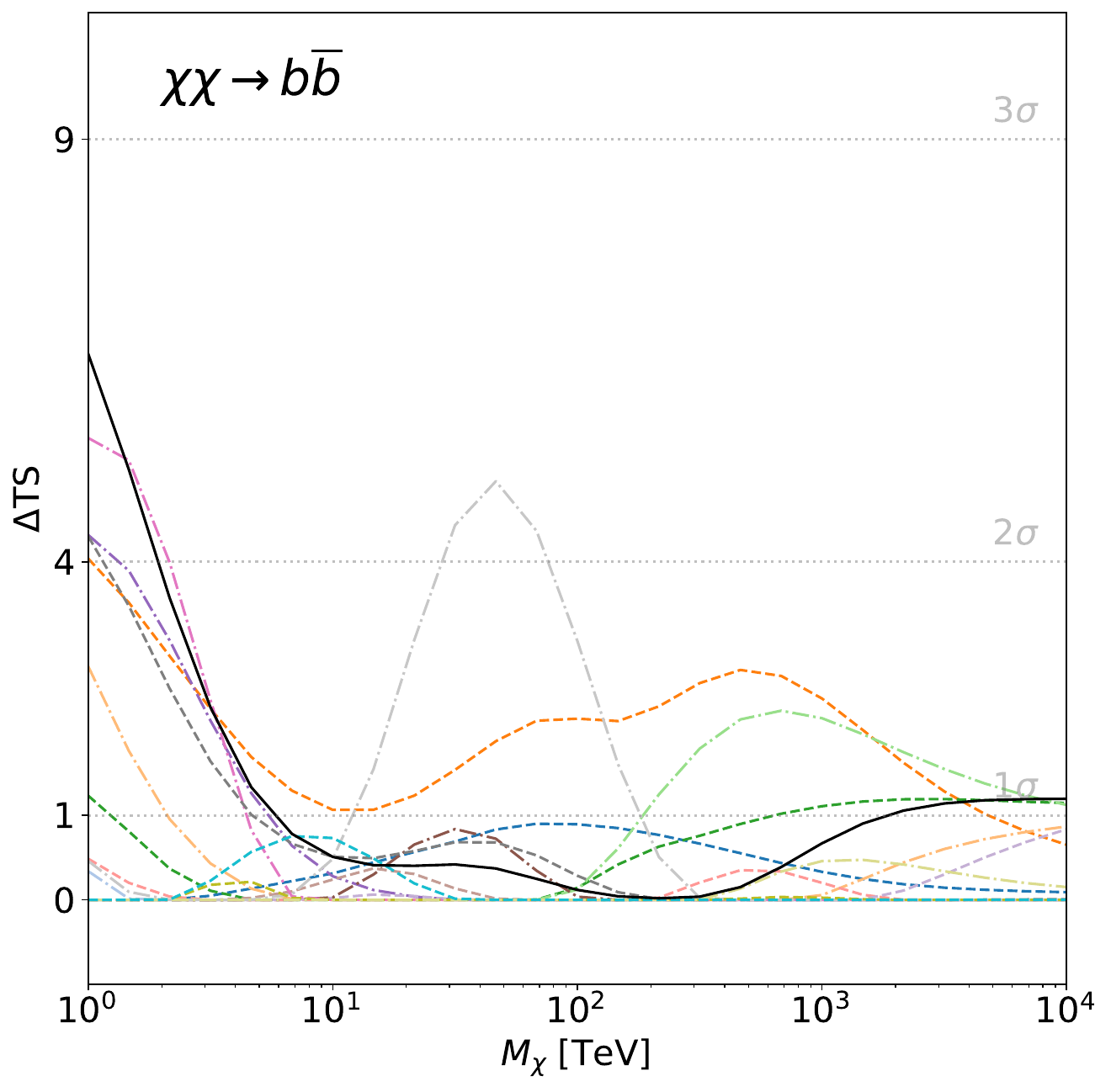} &
   	\includegraphics[width=0.3\textwidth]{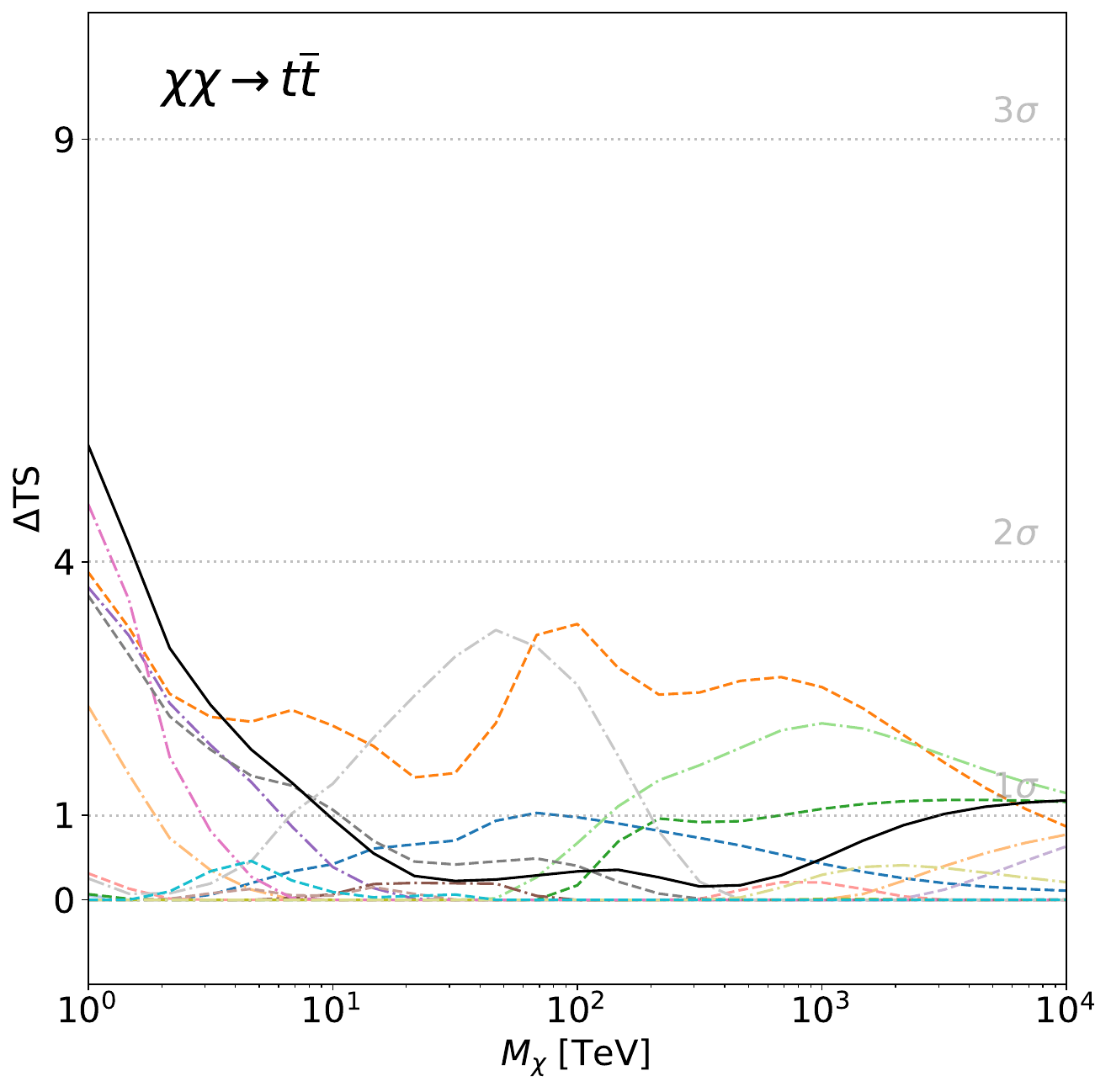} &
    	\includegraphics[width=0.3\textwidth]{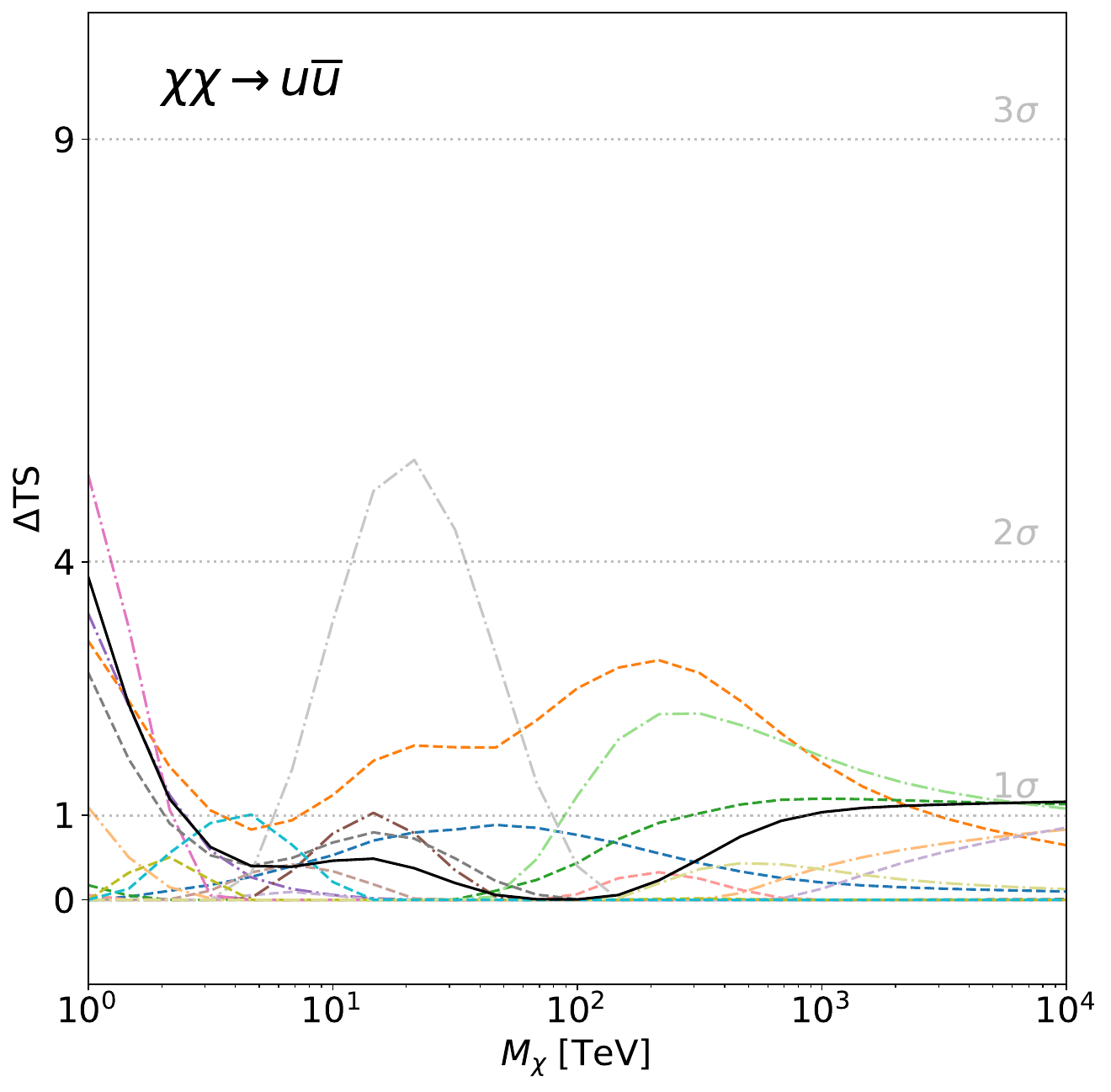} \\
    	\includegraphics[width=0.3\textwidth]{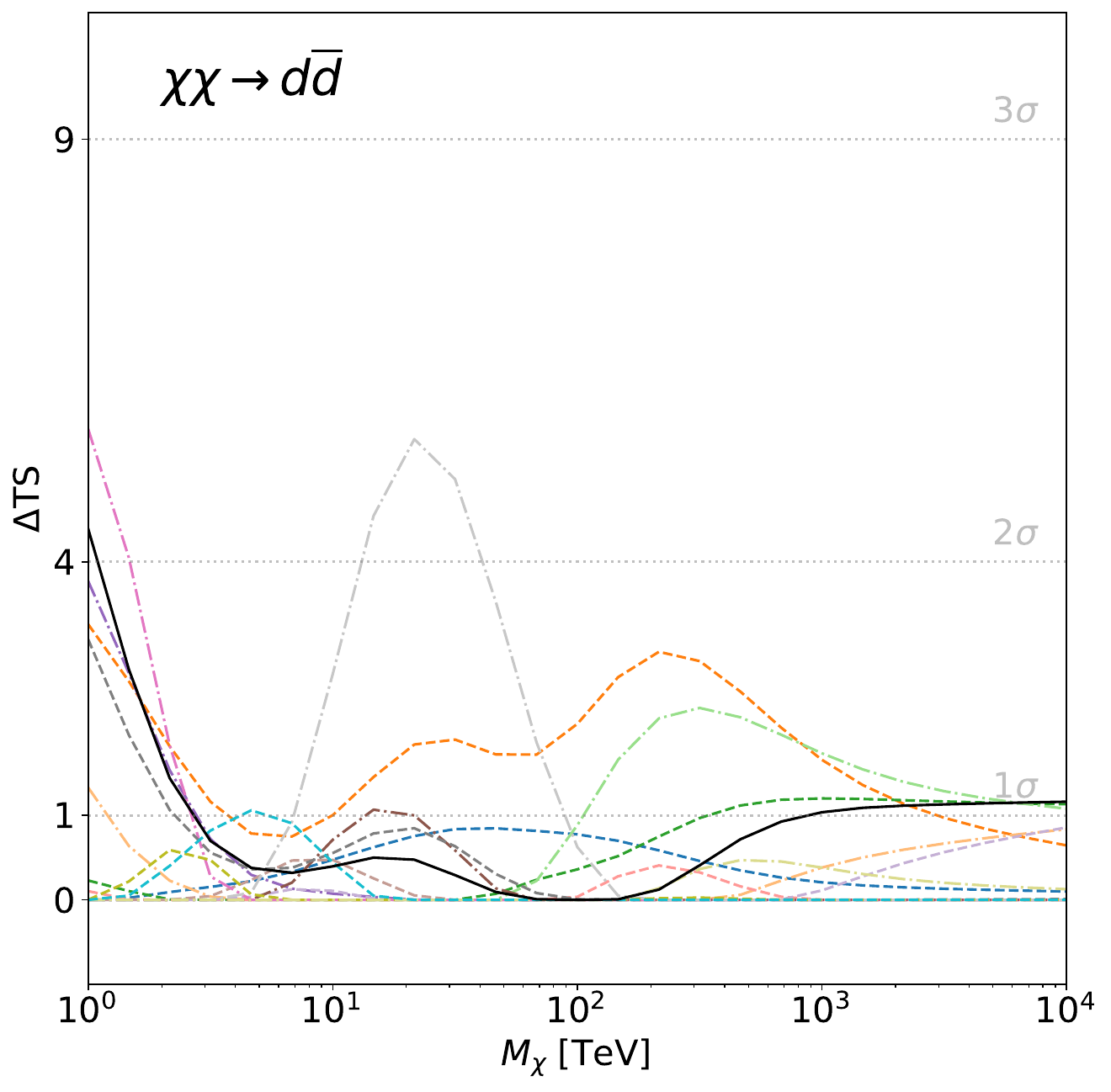} &
    	\includegraphics[width=0.3\textwidth]{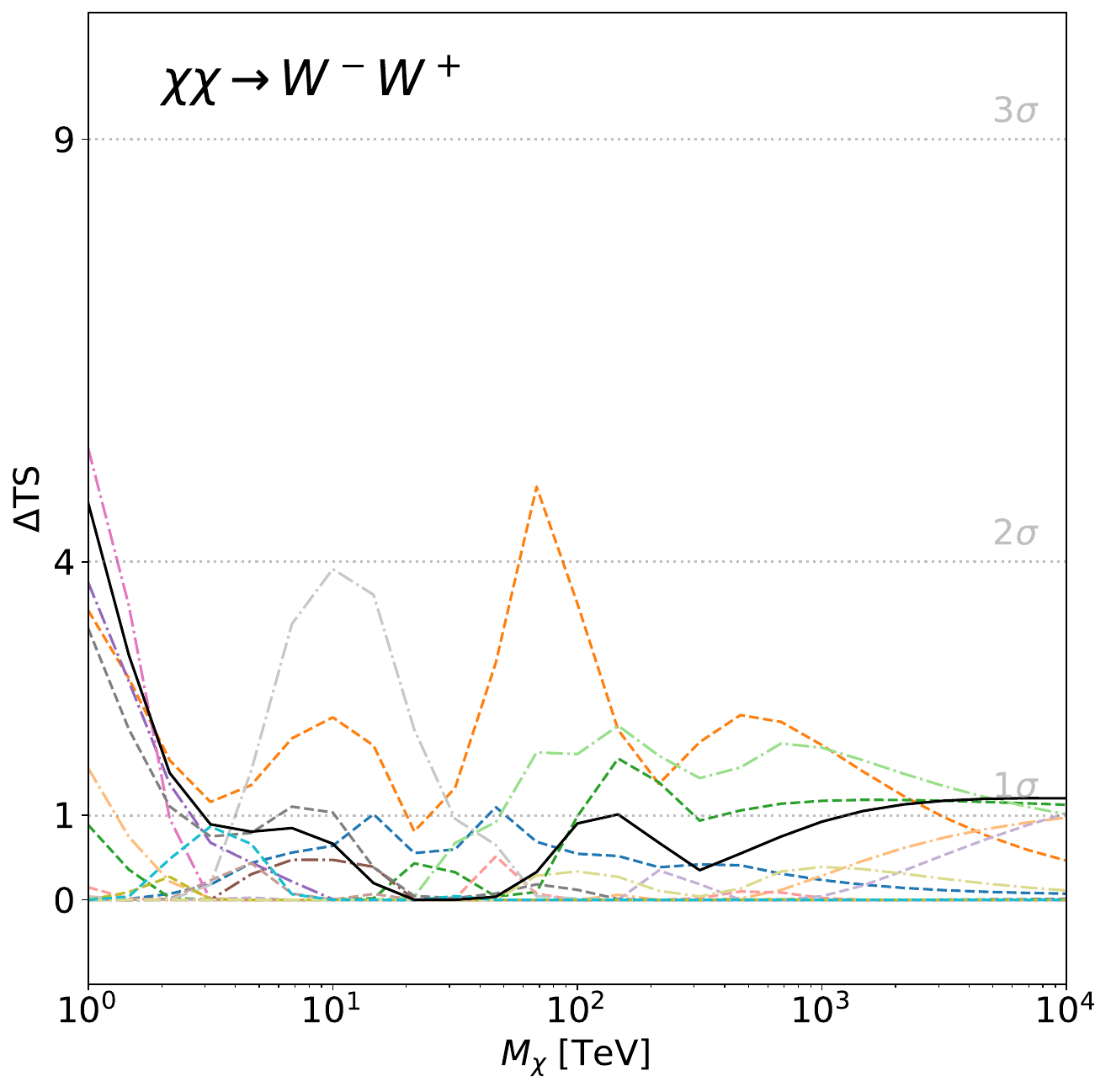} &
    	\includegraphics[width=0.3\textwidth]{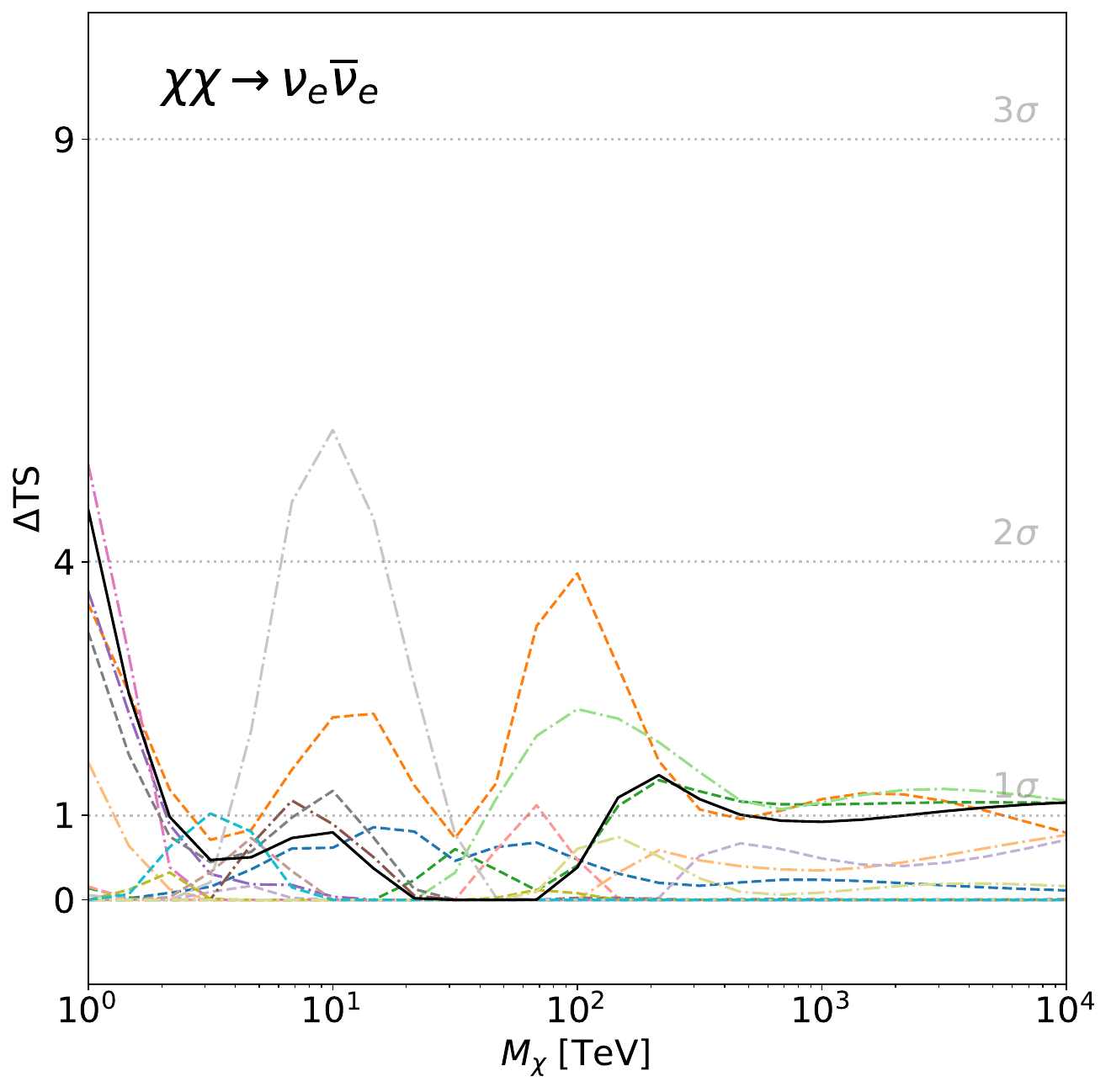} \\
    	\includegraphics[width=0.3\textwidth]{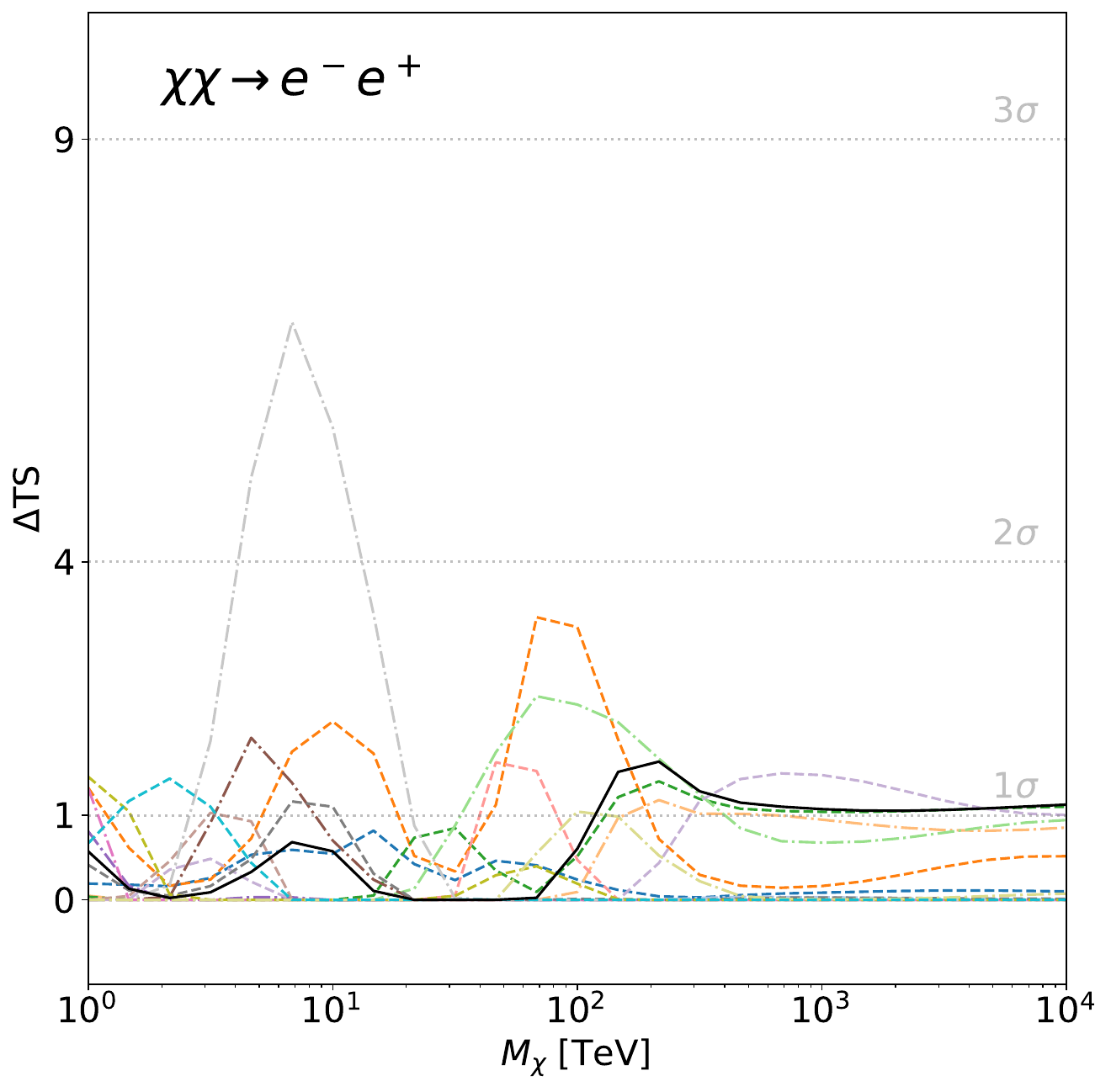} &
    	\includegraphics[width=0.3\textwidth]{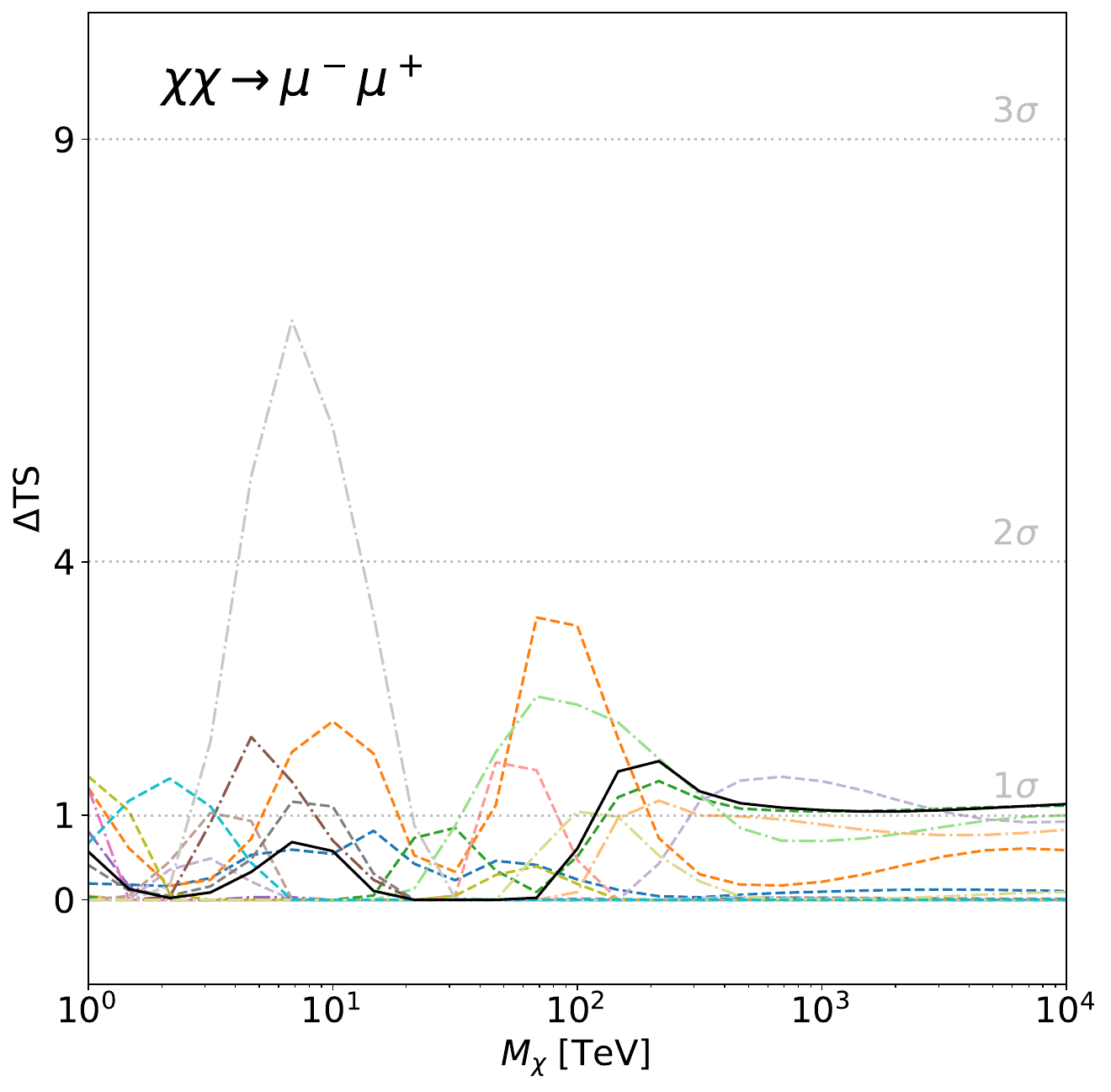} &
    	\includegraphics[width=0.3\textwidth]{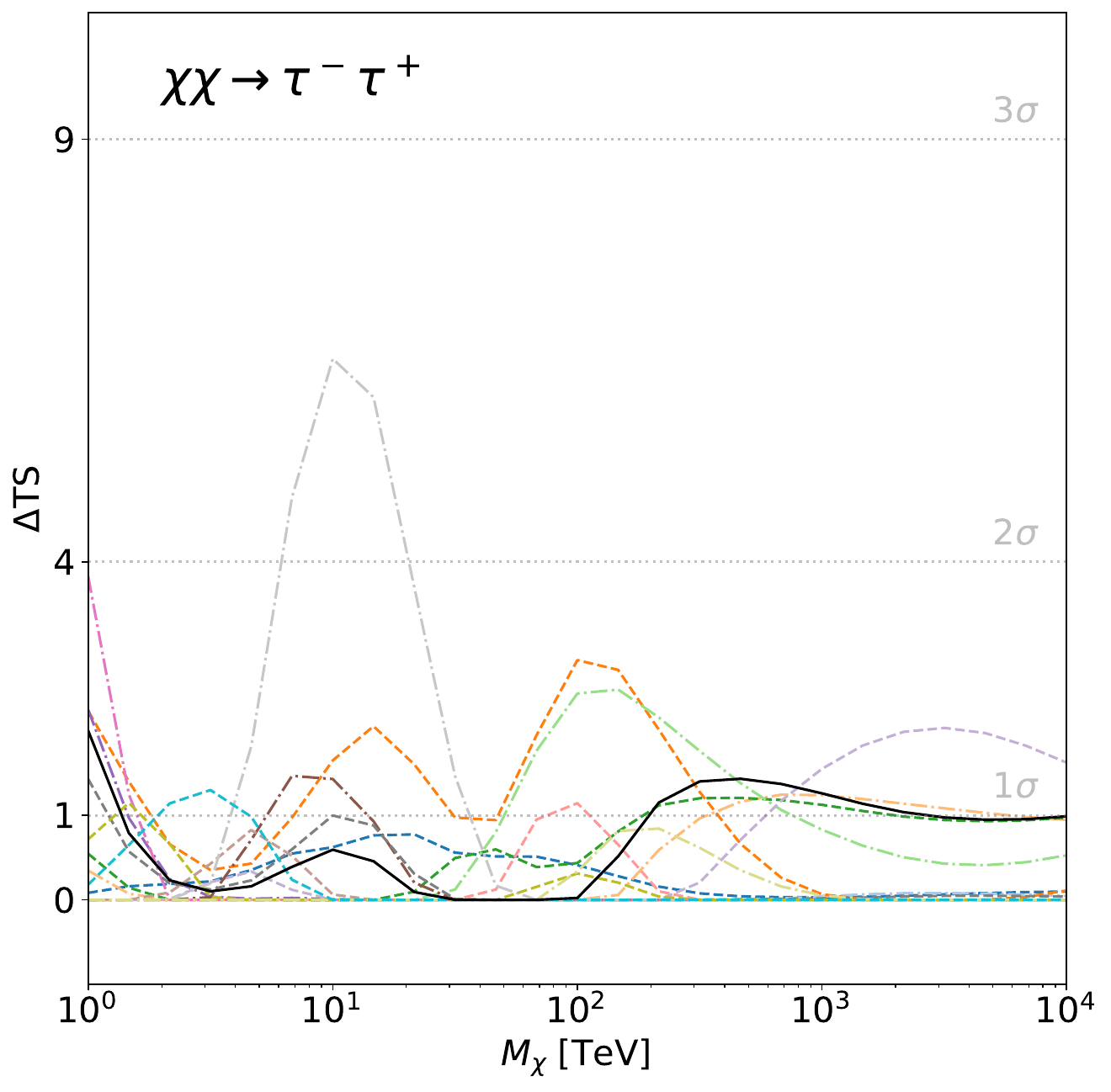} \\
    	\includegraphics[width=0.3\textwidth]{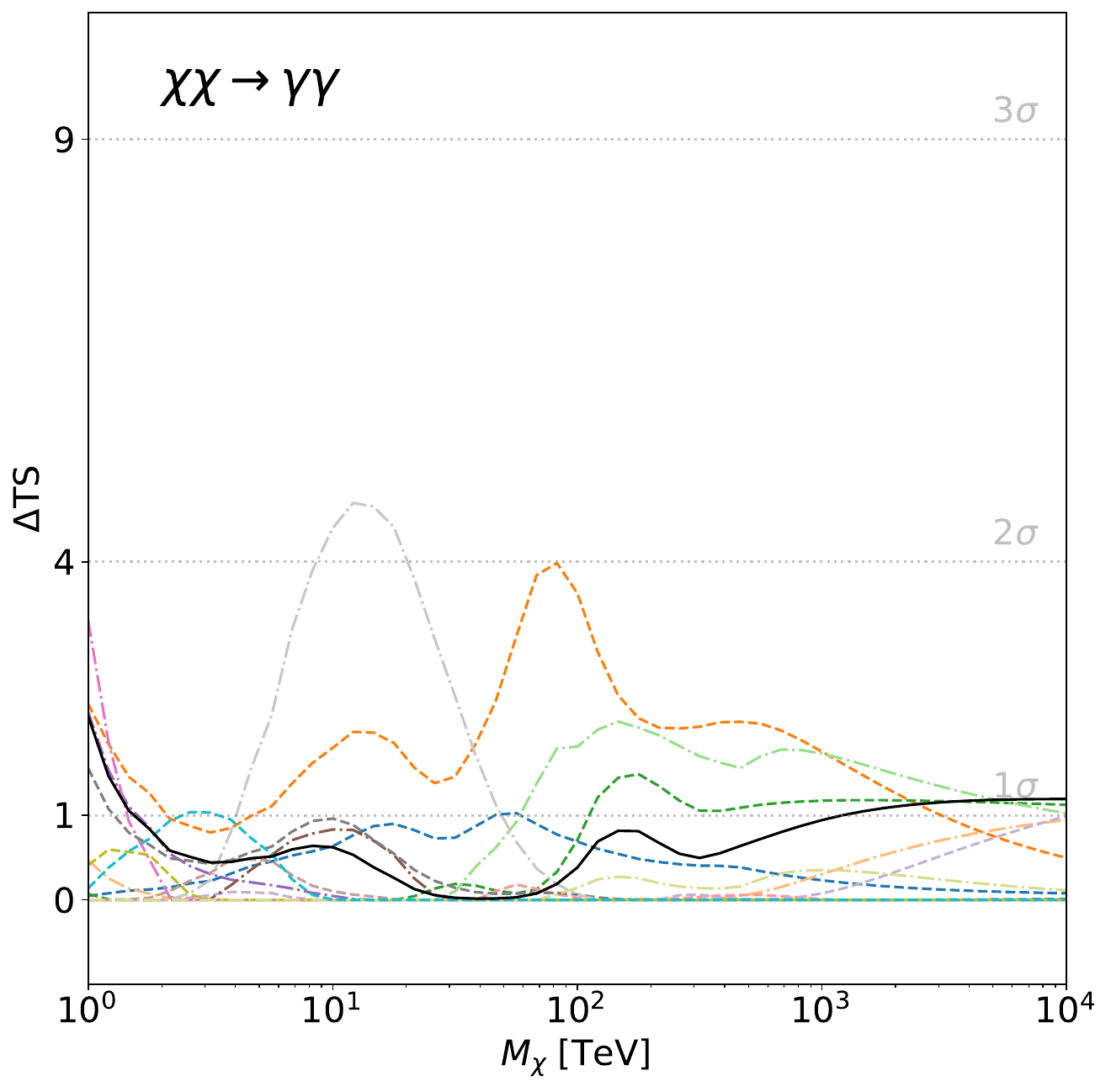} &
    	\includegraphics[width=0.3\textwidth]{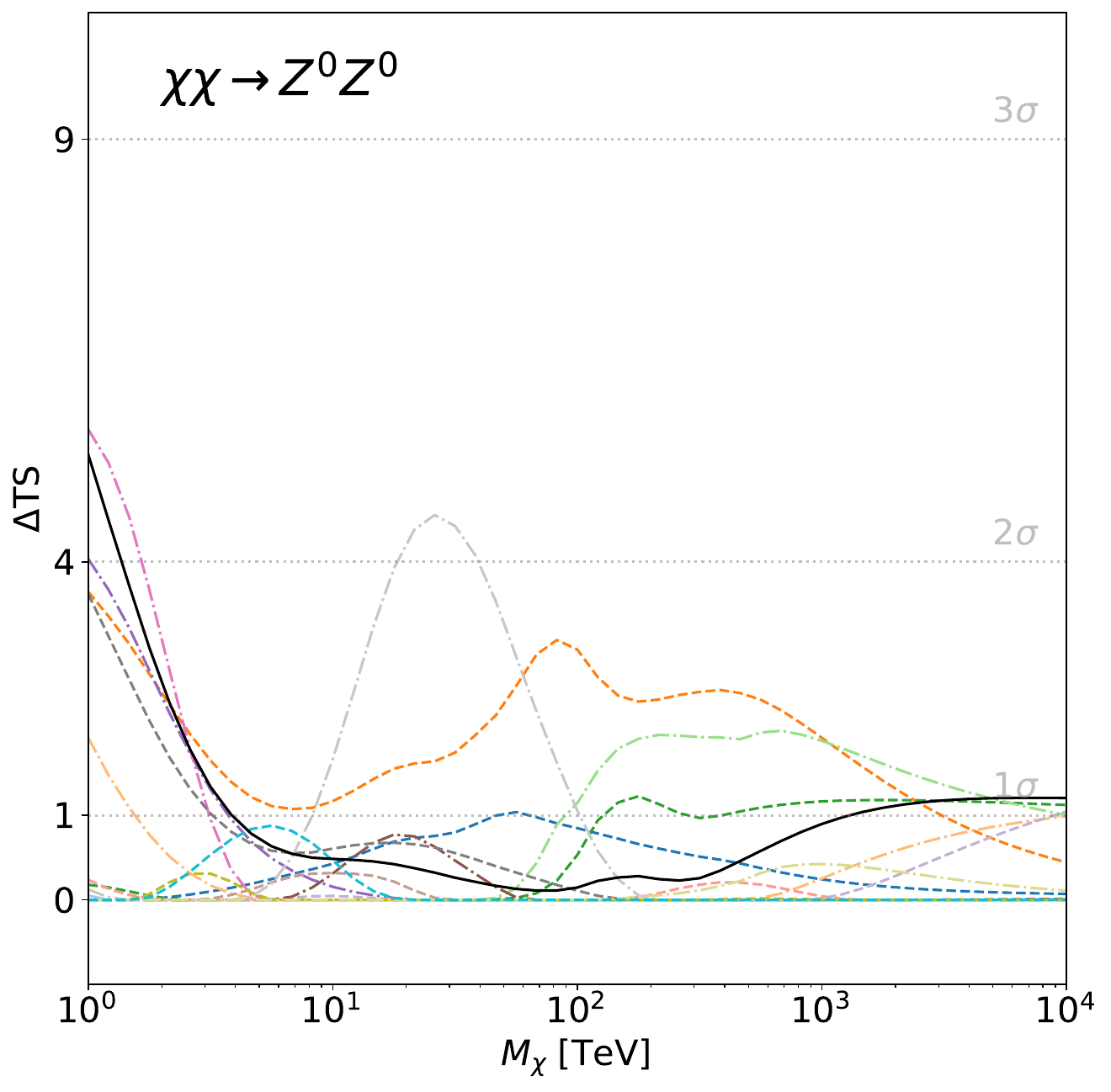} &
    	\raisebox{0.4\height}{\includegraphics[width=0.3\textwidth]{figures/NEWresults/LEGEND_Combined_results.pdf}}
	\end{tabular}
    }
    \caption{Same as \cref{fig:LSmtd_TS_1of2} but with \GS{} \J-factors.}
\label{fig:GSmtd_TS_1of2}
\end{figure}

17 dSphs common between \GS{} and \LS{} are considered for the analysis.
These dSphs are analyzed for emission from DM annihilation according to the likelihood method described in \cref{sec:likelihood}.
The likelihood profiles are then stacked to synthesize a combined limit on the dark matter annihilation cross-section, \sv.
This combination is done for each of the 11 SM annihilation channels over a mass range from $1$ TeV to $10$ PeV. The mass range is separated into $25$ log-spaced bins for all channels except $\gamma\gamma$ and $Z^0Z^0$, which have $49$ log-spaced bins.
The combined limits for all annihilation channels with HAWC's observations are shown in \cref{fig:LSmtd_limits_1of2} for the \LS{} catalog and \cref{fig:GSmtd_limits_1of2} for the \GS{} catalog.
Test statistics (TS) of the best fit \sv~values for each $m_\chi$ and SM channel are shown in \cref{fig:LSmtd_TS_1of2} for the \LS{} catalog and \cref{fig:GSmtd_TS_1of2} for the \GS{} catalog.
Here, we show updated limits for $\chi\chi \rightarrow b\overline{b}$, $t\overline{t}$, $u\overline{u}$, $d\overline{d}$, $W^-W^+$, $\nu_e\overline{\nu}_e$, $e^-e^+$, $\mu^-\mu^+$, $\tau^-\tau^+$, $\gamma\gamma$ and $Z^0Z^0$.
We also simulated 300 studies of Poisson background trials to produce $\pm 1\sigma$ and $\pm 2\sigma$ uncertainty bands. The results of these trials are shown in \cref{fig:LSmtd_bb_1of2} for the \LS{} catalog and in \cref{fig:GSmtd_bb_1of2} for the \GS{} catalog.
We see a general tendency of the observed limit to fall in line with the expected limits. This is especially true in the hard channels and even more so in the \LS{} catalog. The soft channels tend to see slight deviations within the 95\% band at the lowest masses.
At our middle and high range masses, the observed annihilation cross-section generally falls in line with the expectation in the \LS{} catalog. In the \GS{} catalog, we see the observed fall closest to the expected between $10$ and $100$ TeV.

HAWC reports no evidence for DM in dSphs observations, and sets competitive limits for heavy WIMP DM.
We set the most constraining limits on $\chi\chi \rightarrow b\bar{b}$ within the mass range of 30 TeV to 10 PeV.
Similarly, we define the constraints on $\chi\chi \rightarrow \tau^+\tau^-$ in the DM mass range between 4 TeV and 1 PeV.
Within the \LS{} catalog, the largest excess found in HAWC data was for DM annihilating to $b\overline{b}$ for $m_\chi=1$~TeV at a pre-trials significance of $2.17\sigma$.
In the \GS{} catalog, the largest excess found was also for DM annihilating to $b\overline{b}$ for $m_\chi=1$~TeV at a pre-trials significance of $2.53\sigma$.
Across both catalogs, Sextans and Hercules show excesses around $m_\chi=10$~TeV and $m_\chi=100$~TeV, respectively.
Additionally, Segue 1 maintains its position as one of the strongest contributors to the limits as a single source in both catalogs.
This is no surprise as it is a well studied ultra-faint dwarf and is located overhead in HAWC's FOV.
Notably, Willman 1 in the \LS{} catalog plays a large role in defining the upper limits in the high mass ranges.
This is due to its large and well-defined \J-factor.
As expected with lowered J-factors, the \GS{} catalog shows weakened combined limits in comparison to \LS{}. Despite this, the utilization of physically motivated priors makes it an important addition to this analysis.

\begin{figure}[htb]
\centering{
	\includegraphics[scale=0.3]{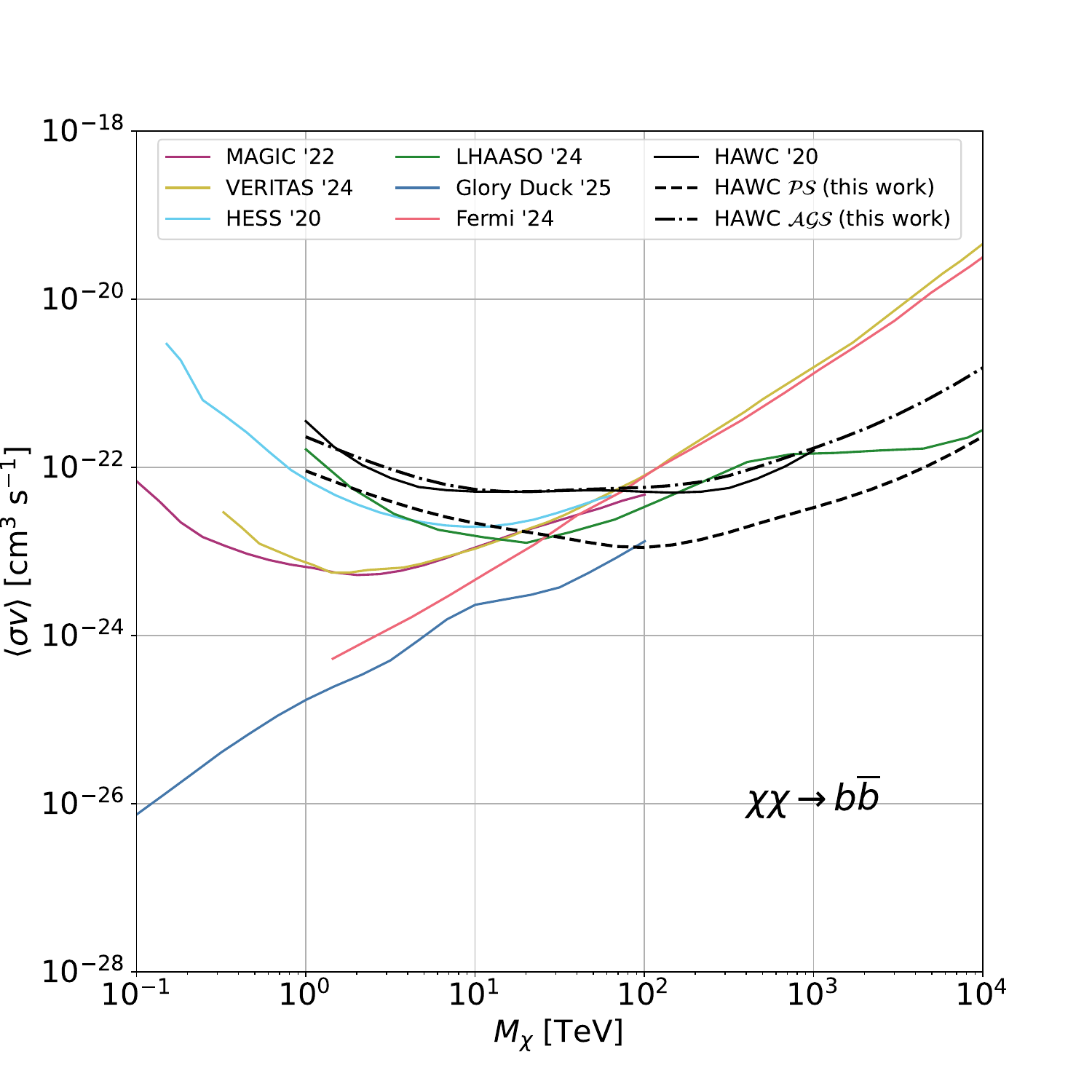}
	\includegraphics[scale=0.3]{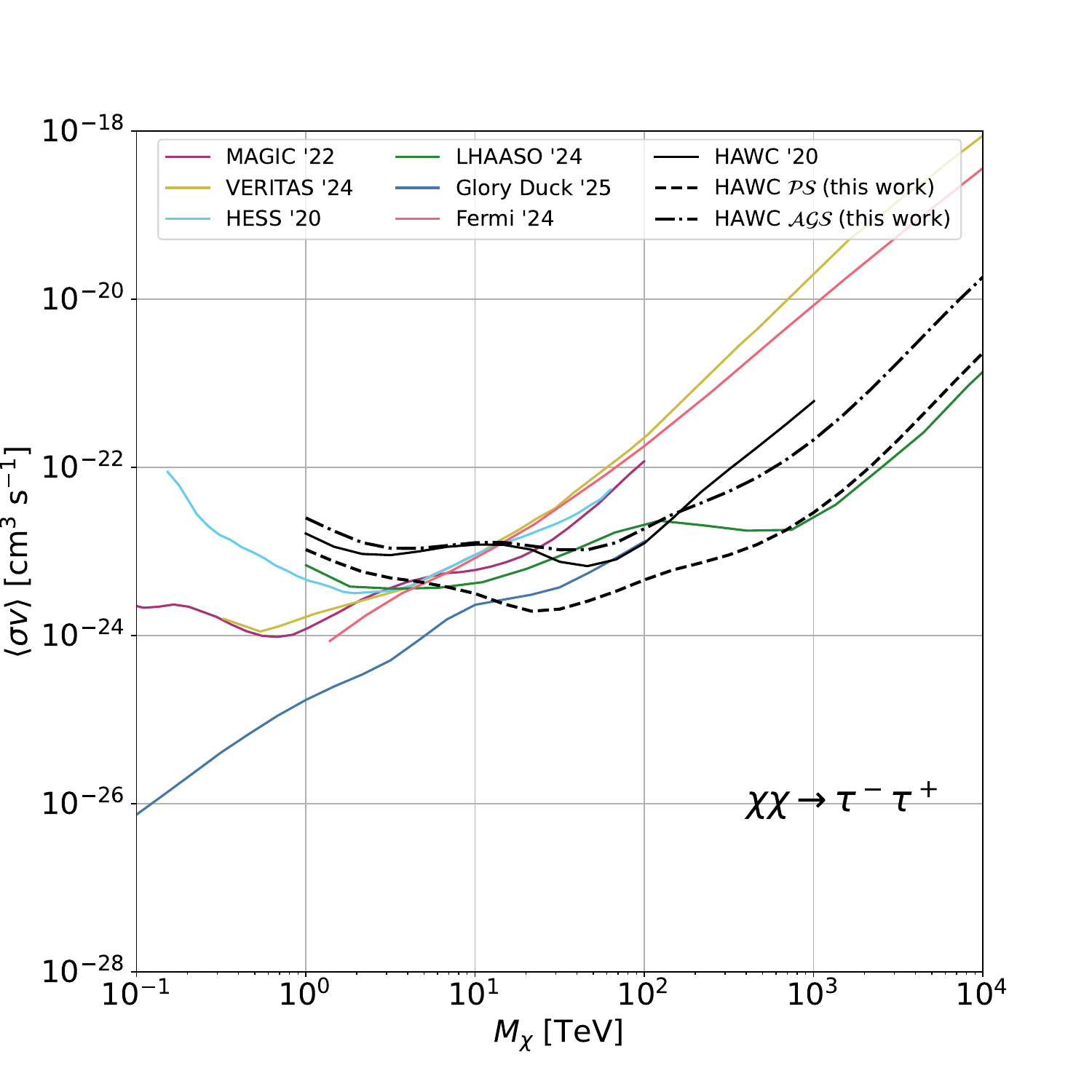}
    }
    \caption{Comparison of HAWC combined limits using \LS{} and \GS{} \J-factors to other DM searches for $\chi\chi \rightarrow b\overline{b}$ (left) and $\chi\chi \rightarrow \tau^-\tau^+$ (right). Each plot shows the 95\% confidence limits using \LS{} and \GS{} \J-factors in dashed and dash-dotted lines, respectively. Limit is compared to previous DM searches from HAWC \cite{HAWC_dm_gammalines} (black), Fermi-LAT \cite{Fermi_cascades} (red), Glory Duck multi-experiment analysis \cite{GloryDuck} (blue), H.E.S.S. \cite{HESSDM_2020} (sky-blue), MAGIC \cite{MAGIC_DM2022} (purple), LHAASO \cite{LHAASO_DM2024} (green), and VERITAS \cite{VERITAS_DM2024} (yellow) in colored solid lines.}
\label{fig:mtd_compare}
\end{figure}

In \cref{fig:mtd_compare}, this work sets strong limits in high DM regions when compared to other, similar DM searches towards dSphs.
Both catalogs have extended HAWC's mass range into the PeV range.
The \LS{} catalog is included as a point of comparison for the \GS{} catalog given the changes and improvements made with respec to HAWC's previous DM searches \cite{HAWC_dm_gammalines}. It displays strong limits in the high mass ranges when compared to other experiments. The limits from both catalogs provide information on DM annihilation from dSphs in the more general and frequently-used case of log-uniform priors (\LS{}) and the more specific kinematic priors (\GS{}).

Finally, we include data on the gamma ray flux as well as the flux upper bounds per energy bin for each source location in \cref{tab:gammaflux,tab:gammaupperbounds} respectively.
The gamma flux upper bounds are an important factor in reproducing our $\langle \sigma v\rangle$ limits.
HAWC detects mainly cosmic rays, making gamma-hadron separation an integral feature of data analysis.
The HAWC dataset used here makes gamma-hadron separation cuts with the use of ML, which has strengthened our precision significantly \cite{P5MLcuts}.
Both tables are organized following HAWC's binning scheme, defined in \cite{Abeysekara_2019}.
Each bin spans a quarter decade in log space across HAWC's full energy sensitivity range.
The 95\% confidence limits on the flux assuming a power law with spectral index $\Gamma=-2$ are reported in \cref{tab:gammaupperbounds}.

\section{Conclusion and Discussion}\label{sec:mtd_conclusion} %%%%%%%%%%%%%%%%%%%%%%%%%%%%

In this analysis, we have used observations of 17 dSphs from HAWC to perform a collective DM annihilation search towards dSphs.
The data were combined across sources to increase the sensitivity of the search.
Moreover, the current HAWC dataset yields significant improvement in sensitivity by means of more precise discrimination between cosmic and gamma rays.
We have observed no significant deviation from the null hypothesis, and so present our results in terms of upper limits on the velocity-weighted cross-section, \sv, for 11 potential DM annihilation channels across four decades of DM mass.

Finally, statistical studies with Poisson variation of HAWC's background are essential to a comprehensive understanding of our observed excesses. We have seen that across both catalogs, the observed limits generally fall in line with the expectations. While in some channels at low masses we see the observed annihilation cross-section falling slightly above the 95\% containment bands, these deviations correspond to significance well below 3$\sigma$. Thus, we observe no significant detection of WIMP dark matter with HAWC within our mass range and instead set limits on the annihilation cross-section. The \LS{} catalog sets some of the strongest current limits for WIMP DM annihilation from dSphs. The \GS{} catalog maintains similarly strong limits, especially for DM masses near 1 PeV.

\acknowledgments
We acknowledge the support from: the US National Science Foundation (NSF);
the US Department of Energy Office of High-Energy Physics;
the Laboratory Directed Research and Development (LDRD) program of Los Alamos National Laboratory;
Consejo Nacional de Ciencia y Tecnolog\'ia (CONACyT), M\'exico, grants LNC-2023-117, 271051, 232656, 260378, 179588, 254964, 258865, 243290, 132197, A1-S-46288, A1-S-22784, CF-2023-I-645, CBF2023-2024-1630, c\'atedras 873, 1563, 341, 323,
Red HAWC, M\'exico; DGAPA-UNAM grants IG101323, IG100726, IN111716-3, IN111419, IA102019, IN106521, IN114924, IN110521 , IN102223; VIEP-BUAP; PIFI 2012, 2013, PROFOCIE 2014, 2015;
the University of Wisconsin Alumni Research Foundation;
the Institute of Geophysics, Planetary Physics, and Signatures at Los Alamos National Laboratory;
Polish Science Centre grant, 2024/53/B/ST9/02671;
Coordinaci\'on de la Investigaci\'on Cient\'ifica de la Universidad Michoacana;
Royal Society - Newton Advanced Fellowship 180385;
Gobierno de Espa\'f1a and European Union-NextGenerationEU, grant CNS2023- 144099;
The Program Management Unit for Human Resources \& Institutional Development, Research and Innovation, NXPO (grant number B16F630069);
Coordinaci\'on General Acad\'emica e Innovaci\'on (CGAI-UdeG), PRODEP-SEP UDG-CA-499;
Institute of Cosmic Ray Research (ICRR), University of Tokyo. H.M. acknowledges support under grant number CBF2023-2024-1630. H.F. acknowledges support by NASA under award number 80GSFC21M0002. C.R. acknowledges support from National Research Foundation of Korea (RS-2023-00280210).
We also acknowledge the significant contributions over many years of Stefan Westerhoff, Gaurang Yodh and Arnulfo Zepeda Dom\'inguez, all deceased members of the HAWC collaboration.
Thanks to Scott Delay, Luciano D\'iaz and Eduardo Murrieta for technical support.

\begin{landscape}
\begin{table}[p]
\centering
\begin{adjustbox}{width=1.4\textheight}% angle=90}
	\small{\begin{tabular}{c|cccccccccccc}
	\hline
	\hline
	\CellTopTwo{}
	Source & 316-562 GeV & 562-1000 GeV & 1.00-1.78 TeV & 1.78-3.16 TeV & 3.16-5.62 TeV & 5.62-10.0 TeV & 10.0-17.8 TeV & 17.8-31.6 TeV & 31.6-56.2 TeV & 56.2-100 TeV & 100-177 TeV & 177-316 TeV \\
	&  \multicolumn{12}{c}{\scriptsize{$\times 10^{-13} [$(TeV$^{-1}$cm$^{-2}$s$^{-1}$)]}} \\
	\hline
\CellTopTwo{}
Boötes I & $0~^{+13}_{}$ & $7.52~^{+9.3}_{-7.5}$ & $0.567~^{+2.9}_{-0.57}$ & $0.643~^{+1.5}_{-0.64}$ & $0.515~^{+0.83}_{-0.52}$ & $0~^{+0.45}_{}$ & $0.595~^{+0.41}_{-0.41}$ & $0.04~^{+0.27}_{-0.04}$ & $0.105~^{+0.19}_{-0.1}$ & $0~^{+0.11}_{}$ & $0~^{+0.27}_{}$ & $0~^{+0.55}_{}$ \\ 
\CellTopTwo{}
Canes Venatici I & $26~^{+42}_{-26}$ & $4.02~^{+19}_{-4}$ & $0~^{+1.1}_{}$ & $0~^{+1.5}_{}$ & $0~^{+0.53}_{}$ & $0~^{+0.16}_{}$ & $0~^{+0.21}_{}$ & $0~^{+0.11}_{}$ & $0.0558~^{+0.25}_{-0.056}$ & $0.248~^{+0.35}_{-0.25}$ & $0~^{+0.27}_{}$ & $0.618~^{+0.96}_{-0.62}$ \\ 
\CellTopTwo{}
Canes Venatici II & $45.5~^{+30}_{-30}$ & $12.6~^{+14}_{-13}$ & $1.73~^{+3.9}_{-1.7}$ & $0~^{+1.3}_{}$ & $0.549~^{+1}_{-0.55}$ & $0.523~^{+0.57}_{-0.52}$ & $0~^{+0.31}_{}$ & $0~^{+0.088}_{}$ & $0.085~^{+0.23}_{-0.085}$ & $0.326~^{+0.28}_{-0.28}$ & $0~^{+0.33}_{}$ & $0~^{+0.48}_{}$ \\ 
\CellTopTwo{}
Coma Berencies & $42.4~^{+21}_{-21}$ & $0~^{+5.5}_{}$ & $0.251~^{+2.8}_{-0.25}$ & $0~^{+1.4}_{}$ & $0.509~^{+0.83}_{-0.51}$ & $0~^{+0.42}_{}$ & $0~^{+0.31}_{}$ & $0~^{+0.14}_{}$ & $0~^{+0.13}_{}$ & $0~^{+0.19}_{}$ & $0.722~^{+0.82}_{-0.72}$ & $0~^{+0.52}_{}$ \\ 
\CellTopTwo{}
Draco & $0~^{+55}_{}$ & $0~^{+35}_{}$ & $0~^{+8.6}_{}$ & $0~^{+7.9}_{}$ & $0~^{+3.1}_{}$ & $0~^{+0.98}_{}$ & $0.422~^{+1.1}_{-0.42}$ & $0.253~^{+0.81}_{-0.25}$ & $0~^{+0.59}_{}$ & $0~^{+0.41}_{}$ & $1.13~^{+1.3}_{-1.1}$ & $0~^{+0.63}_{}$ \\ 
\CellTopTwo{}
Draco II & $0~^{+7.7e+03}_{}$ & $0~^{+9.3e+02}_{}$ & $0.174~^{+7.5}_{-0.17}$ & $0~^{+2.2}_{}$ & $0~^{+2.6}_{}$ & $0~^{+1.1}_{}$ & $0~^{+0.5}_{}$ & $0~^{+0.95}_{}$ & $0.291~^{+1}_{-0.29}$ & $0~^{+0.55}_{}$ & $2.21~^{+2.6}_{-2.2}$ & $0~^{+1.7}_{}$ \\ 
\CellTopTwo{}
Hercules & $30.1~^{+22}_{-22}$ & $0~^{+2}_{}$ & $0~^{+1.9}_{}$ & $0~^{+0.82}_{}$ & $0~^{+0.39}_{}$ & $0.0404~^{+0.48}_{-0.04}$ & $0~^{+0.16}_{}$ & $0~^{+0.17}_{}$ & $0.474~^{+0.32}_{-0.32}$ & $0.0456~^{+0.21}_{-0.046}$ & $0~^{+0.29}_{}$ & $0~^{+0.48}_{}$ \\ 
\CellTopTwo{}
Leo I & $44.8~^{+22}_{-22}$ & $15.8~^{+9.7}_{-9.7}$ & $0~^{+1.9}_{}$ & $0~^{+1.3}_{}$ & $0.252~^{+0.86}_{-0.25}$ & $0~^{+0.26}_{}$ & $0~^{+0.26}_{}$ & $0~^{+0.11}_{}$ & $0~^{+0.11}_{}$ & $0~^{+0.11}_{}$ & $0~^{+0.29}_{}$ & $0~^{+0.52}_{}$ \\ 
\CellTopTwo{}
Leo II & $25.3~^{+20}_{-20}$ & $0~^{+6.2}_{}$ & $2.97~^{+2.8}_{-2.8}$ & $1.88~^{+1.4}_{-1.4}$ & $0.175~^{+0.81}_{-0.18}$ & $0~^{+0.22}_{}$ & $0~^{+0.18}_{}$ & $0~^{+0.2}_{}$ & $0~^{+0.098}_{}$ & $0~^{+0.088}_{}$ & $0~^{+0.32}_{}$ & $0.778~^{+0.99}_{-0.78}$ \\ 
\CellTopTwo{}
Leo IV & $0~^{+10}_{}$ & $0~^{+16}_{}$ & $0~^{+1.8}_{}$ & $2.63~^{+2.5}_{-2.5}$ & $2.7~^{+1.4}_{-1.4}$ & $0~^{+0.6}_{}$ & $0.292~^{+0.5}_{-0.29}$ & $0~^{+0.18}_{}$ & $0~^{+0.26}_{}$ & $0~^{+0.16}_{}$ & $0~^{+0.73}_{}$ & $0~^{+0.57}_{}$ \\ 
\CellTopTwo{}
Leo V & $0~^{+28}_{}$ & $0~^{+9.2}_{}$ & $3.78~^{+4.6}_{-3.8}$ & $1.29~^{+2.2}_{-1.3}$ & $1.55~^{+1.2}_{-1.2}$ & $0~^{+0.18}_{}$ & $0.117~^{+0.41}_{-0.12}$ & $0~^{+0.094}_{}$ & $0~^{+0.18}_{}$ & $0~^{+0.095}_{}$ & $0~^{+0.33}_{}$ & $0~^{+0.55}_{}$ \\ 
\CellTopTwo{}
Pisces II & $47.8~^{+28}_{-28}$ & $7.38~^{+13}_{-7.4}$ & $0.336~^{+3.7}_{-0.34}$ & $0~^{+0.4}_{}$ & $0~^{+0.4}_{}$ & $0~^{+0.54}_{}$ & $0.186~^{+0.4}_{-0.19}$ & $0~^{+0.23}_{}$ & $0~^{+0.14}_{}$ & $0~^{+0.22}_{}$ & $0~^{+0.3}_{}$ & $0.1~^{+1}_{-0.1}$ \\ 
\CellTopTwo{}
Segue 1 & $40.6~^{+20}_{-20}$ & $7.39~^{+9}_{-7.4}$ & $0~^{+2.7}_{}$ & $0.443~^{+1.4}_{-0.44}$ & $0.41~^{+0.8}_{-0.41}$ & $0.437~^{+0.49}_{-0.44}$ & $0~^{+0.17}_{}$ & $0~^{+0.088}_{}$ & $0~^{+0.085}_{}$ & $0~^{+0.12}_{}$ & $0~^{+0.3}_{}$ & $0~^{+0.53}_{}$ \\ 
\CellTopTwo{}
Sextans & $76.3~^{+51}_{-51}$ & $0~^{+4.3}_{}$ & $0~^{+2.3}_{}$ & $0~^{+1.9}_{}$ & $2.74~^{+1.5}_{-1.5}$ & $0.88~^{+0.78}_{-0.78}$ & $0.512~^{+0.51}_{-0.51}$ & $0~^{+0.31}_{}$ & $0~^{+0.21}_{}$ & $0~^{+0.22}_{}$ & $0~^{+0.33}_{}$ & $0~^{+0.56}_{}$ \\ 
\CellTopTwo{}
Ursa Major I & $0~^{+14}_{}$ & $51.3~^{+1.2e+02}_{-51}$ & $12.6~^{+20}_{-13}$ & $2.77~^{+6.7}_{-2.8}$ & $0~^{+1.1}_{}$ & $0~^{+0.26}_{}$ & $0~^{+0.26}_{}$ & $0.758~^{+0.69}_{-0.69}$ & $0.181~^{+0.44}_{-0.18}$ & $0.0426~^{+0.34}_{-0.043}$ & $0~^{+0.46}_{}$ & $0.291~^{+1.1}_{-0.29}$ \\ 
\CellTopTwo{}
Ursa Major II & $0~^{+1.9e+03}_{}$ & $0~^{+1.5e+02}_{}$ & $0~^{+20}_{}$ & $0~^{+4.6}_{}$ & $0~^{+8.1}_{}$ & $0~^{+1.6}_{}$ & $0~^{+0.71}_{}$ & $0~^{+0.31}_{}$ & $0.365~^{+0.95}_{-0.37}$ & $3.37~^{+1.9}_{-1.9}$ & $1.07~^{+1.8}_{-1.1}$ & $0~^{+1.2}_{}$ \\ 
\CellTopTwo{}
Willman 1 & $0~^{+16}_{}$ & $168~^{+1.1e+02}_{-1.1e+02}$ & $0~^{+6.8}_{}$ & $6.41~^{+6}_{-6}$ & $1.04~^{+2.9}_{-1}$ & $0~^{+0.9}_{}$ & $0~^{+0.24}_{}$ & $0.476~^{+0.67}_{-0.48}$ & $0~^{+0.18}_{}$ & $0~^{+0.22}_{}$ & $0~^{+0.33}_{}$ & $0~^{+0.73}_{}$ \\ 

	\hline
	\hline	
	\CellTopTwo{}
	\end{tabular}}
	\end{adjustbox}
\caption{Gamma-ray flux best fit for each source assuming a power law with $\Gamma=-2$. Lower error was removed on flux points equal to zero.}\label{tab:gammaflux}
\end{table}
\end{landscape}

\begin{landscape}
\begin{table}[p]
\centering
\begin{adjustbox}{width=1.4\textheight}% angle=90}
	\small{\begin{tabular}{c|cccccccccccc}
	\hline
	\hline
	\CellTopTwo{}
	Source & 316-562 GeV & 562-1000 GeV & 1.00-1.78 TeV & 1.78-3.16 TeV & 3.16-5.62 TeV & 5.62-10.0 TeV & 10.0-17.8 TeV & 17.8-31.6 TeV & 31.6-56.2 TeV & 56.2-100 TeV & 100-177 TeV & 177-316 TeV \\
	&  \multicolumn{12}{c}{\scriptsize{[$\log_{10}$(TeV$^{-1}$cm$^{-2}$s$^{-1}$)]}} \\
	\hline
\CellTopTwo{}
Bootes I & $-11.89$ & $-11.77$ & $-12.46$ & $-12.67$ & $-12.87$ & $-13.35$ & $-13.00$ & $-13.51$ & $-13.53$ & $-13.97$ & $-13.56$ & $-13.26$ \\ 
\CellTopTwo{}
Canes Venatici I & $-11.17$ & $-11.64$ & $-12.98$ & $-12.82$ & $-13.28$ & $-13.80$ & $-13.68$ & $-13.96$ & $-13.52$ & $-13.22$ & $-13.57$ & $-12.80$ \\ 
\CellTopTwo{}
Canes Venatici II & $-11.12$ & $-11.58$ & $-12.25$ & $-12.89$ & $-12.81$ & $-12.96$ & $-13.51$ & $-14.06$ & $-13.50$ & $-13.22$ & $-13.48$ & $-13.32$ \\ 
\CellTopTwo{}
Coma Berencies & $-11.20$ & $-12.26$ & $-12.51$ & $-12.87$ & $-12.87$ & $-13.38$ & $-13.51$ & $-13.85$ & $-13.88$ & $-13.72$ & $-12.81$ & $-13.29$ \\ 
\CellTopTwo{}
Draco & $-11.26$ & $-11.46$ & $-12.07$ & $-12.10$ & $-12.51$ & $-13.01$ & $-12.82$ & $-12.97$ & $-13.23$ & $-13.39$ & $-12.61$ & $-13.20$ \\ 
\CellTopTwo{}
Draco II & $-9.11$ & $-10.03$ & $-12.12$ & $-12.66$ & $-12.59$ & $-12.97$ & $-13.30$ & $-13.02$ & $-12.89$ & $-13.26$ & $-12.32$ & $-12.76$ \\ 
\CellTopTwo{}
Hercules & $-11.29$ & $-12.69$ & $-12.71$ & $-13.09$ & $-13.41$ & $-13.28$ & $-13.80$ & $-13.76$ & $-13.10$ & $-13.58$ & $-13.53$ & $-13.32$ \\ 
\CellTopTwo{}
Leo I & $-11.18$ & $-11.59$ & $-12.72$ & $-12.90$ & $-12.95$ & $-13.59$ & $-13.59$ & $-13.95$ & $-13.97$ & $-13.95$ & $-13.54$ & $-13.29$ \\ 
\CellTopTwo{}
Leo II & $-11.34$ & $-12.21$ & $-12.24$ & $-12.48$ & $-13.01$ & $-13.67$ & $-13.74$ & $-13.69$ & $-14.01$ & $-14.06$ & $-13.49$ & $-12.75$ \\ 
\CellTopTwo{}
Leo IV & $-11.99$ & $-11.79$ & $-12.75$ & $-12.29$ & $-12.39$ & $-13.22$ & $-13.10$ & $-13.75$ & $-13.59$ & $-13.81$ & $-13.13$ & $-13.25$ \\ 
\CellTopTwo{}
Leo V & $-11.56$ & $-12.03$ & $-12.08$ & $-12.46$ & $-12.56$ & $-13.73$ & $-13.28$ & $-14.03$ & $-13.75$ & $-14.02$ & $-13.48$ & $-13.26$ \\ 
\CellTopTwo{}
Pisces II & $-11.12$ & $-11.70$ & $-12.39$ & $-13.40$ & $-13.39$ & $-13.27$ & $-13.23$ & $-13.64$ & $-13.86$ & $-13.67$ & $-13.52$ & $-12.95$ \\ 
\CellTopTwo{}
Segue 1 & $-11.21$ & $-11.79$ & $-12.58$ & $-12.72$ & $-12.92$ & $-13.04$ & $-13.76$ & $-14.06$ & $-14.07$ & $-13.91$ & $-13.53$ & $-13.27$ \\ 
\CellTopTwo{}
Sextans & $-10.90$ & $-12.37$ & $-12.63$ & $-12.72$ & $-12.38$ & $-12.78$ & $-12.99$ & $-13.51$ & $-13.68$ & $-13.66$ & $-13.48$ & $-13.25$ \\ 
\CellTopTwo{}
Ursa Major I & $-11.86$ & $-10.76$ & $-11.49$ & $-12.02$ & $-12.97$ & $-13.58$ & $-13.59$ & $-12.84$ & $-13.21$ & $-13.42$ & $-13.34$ & $-12.85$ \\ 
\CellTopTwo{}
Ursa Major II & $-9.71$ & $-10.83$ & $-11.70$ & $-12.34$ & $-12.09$ & $-12.81$ & $-13.15$ & $-13.51$ & $-12.88$ & $-12.27$ & $-12.54$ & $-12.91$ \\ 
\CellTopTwo{}
Willman 1 & $-11.79$ & $-10.56$ & $-12.17$ & $-11.90$ & $-12.41$ & $-13.05$ & $-13.61$ & $-12.94$ & $-13.73$ & $-13.67$ & $-13.48$ & $-13.14$ \\ 
	\hline
	\hline	
	\CellTopTwo{}
	\end{tabular}}
	\end{adjustbox}
\caption{95\% upper bounds on gamma flux for each bin and source assuming a power law with $\Gamma=-2$.}\label{tab:gammaupperbounds}
\end{table}
\end{landscape}

\bibliographystyle{JHEP}
\bibliography{refs}

\appendix
\section{Likelihood Methods}\label{sec:likelihood}%%%%%%%%%%%%%%%%%%%%%%%%%%%%%%%

The likelihood $\mathcal{L}$ of each source for events that are Poisson distributed can be calculated as follows:

\begin{equation}
	\mathcal{L} = \prod_i \frac{(B_i+S_i)^{N_i}e^{-(B_i+S_i)}}{N_i!}.
\end{equation}
Where $B_i$ is the number of background counts observed, $S_i$ is the sum of the expected number of signal counts, and $N_i$ is the total number of counts. Both the background events $B$ and the total number of counts $N$ can be determined from HAWC's data. The excess of signal events $S$ from DM-DM annihilation is determined by convolving \cref{eq:id_dm_flux} with HAWC's energy response and pixel point spread functions.

From here, the TS can be computed using the log-likelihood ratio test:
\begin{equation}
	\text{TS}_{\text{max}}=-2\ln \left( \frac{\mathcal{L}_0}{\mathcal{L}^\text{max}} \right).
\end{equation}
The null hypothesis likelihood $\mathcal{L}_0$ represents no DM detection and the best fit signal hypothesis $\mathcal{L}^\text{max}$ is where $\langle \sigma v \rangle$ maximizes the likelihood.

To determine the upper limit on $\langle \sigma v \rangle$ we define the parameter
\begin{equation}
	\text{TS}_{95} \equiv \sum_\text{bins} \left[ 2N\ln\left(1+\frac{\epsilon S_\text{ref}}{B}\right)-2\epsilon S_\text{ref} \right].
\end{equation}
Where $S_\text{ref}$ is the expected number of excess counts in a bin for DM annihilation from a dwarf at the annihilation cross-section $\langle\sigma v \rangle$ and $\epsilon$ is a scaling factor. The 95\% confidence level is then determined when
\begin{equation}
	2.71 = \text{TS}_\text{max}-\text{TS}_{95}.
\end{equation}

\newpage

\section{Remaining Spectra Models}\label{sec:spectralplots}
\begin{figure}[htb!]
\centering{
	\begin{tabular}{ccc}
	\includegraphics[width=0.3\textwidth]{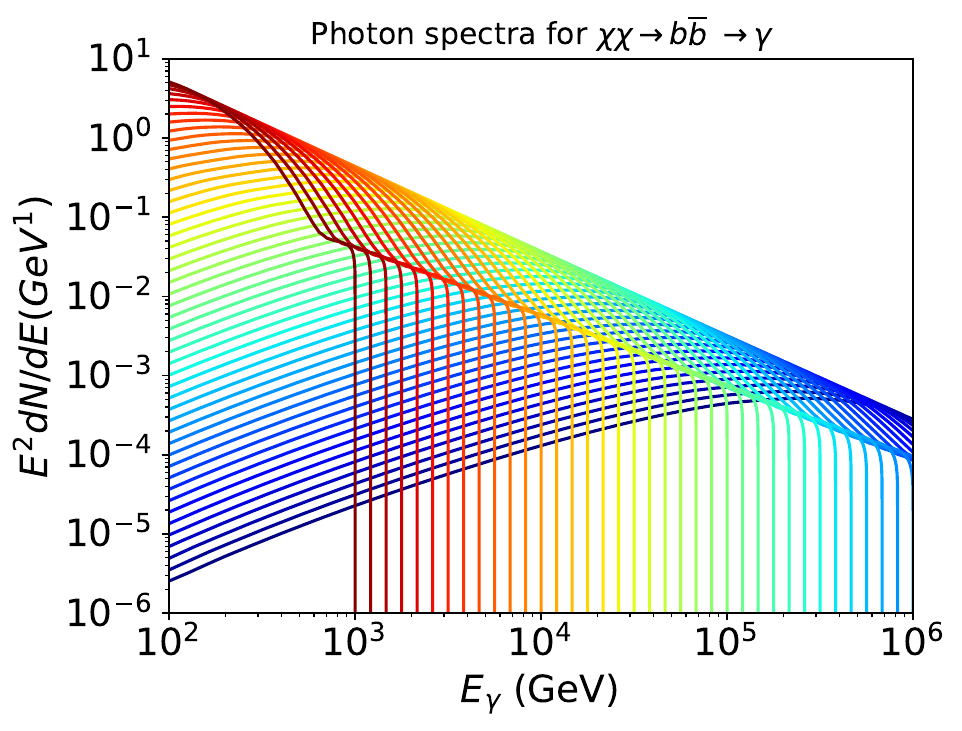} &
	\includegraphics[width=0.3\textwidth]{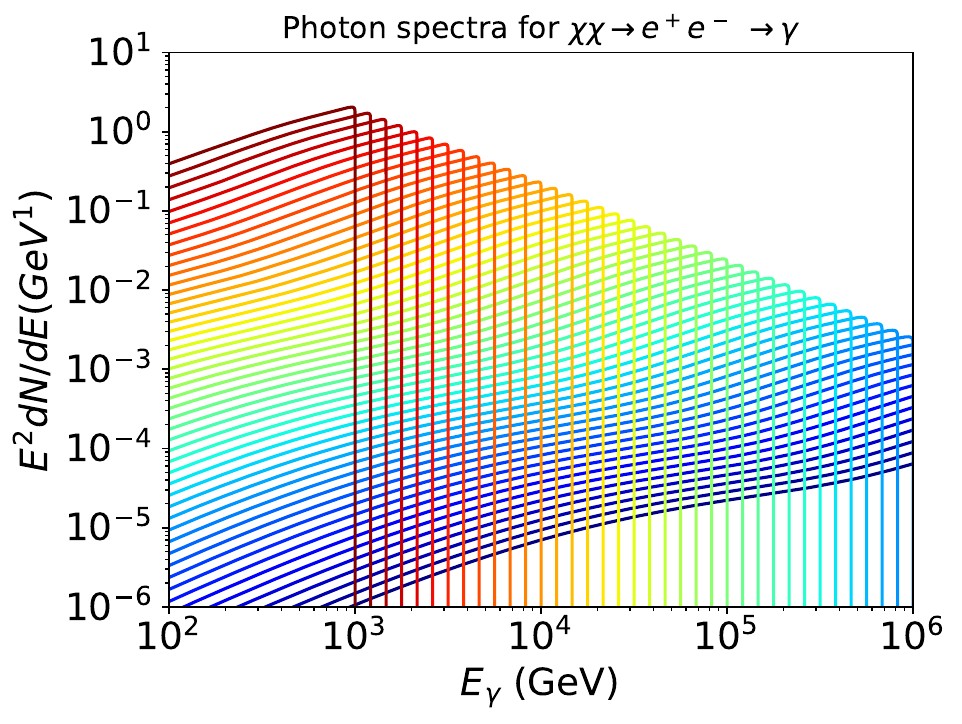} &
        	\includegraphics[width=0.3\textwidth]{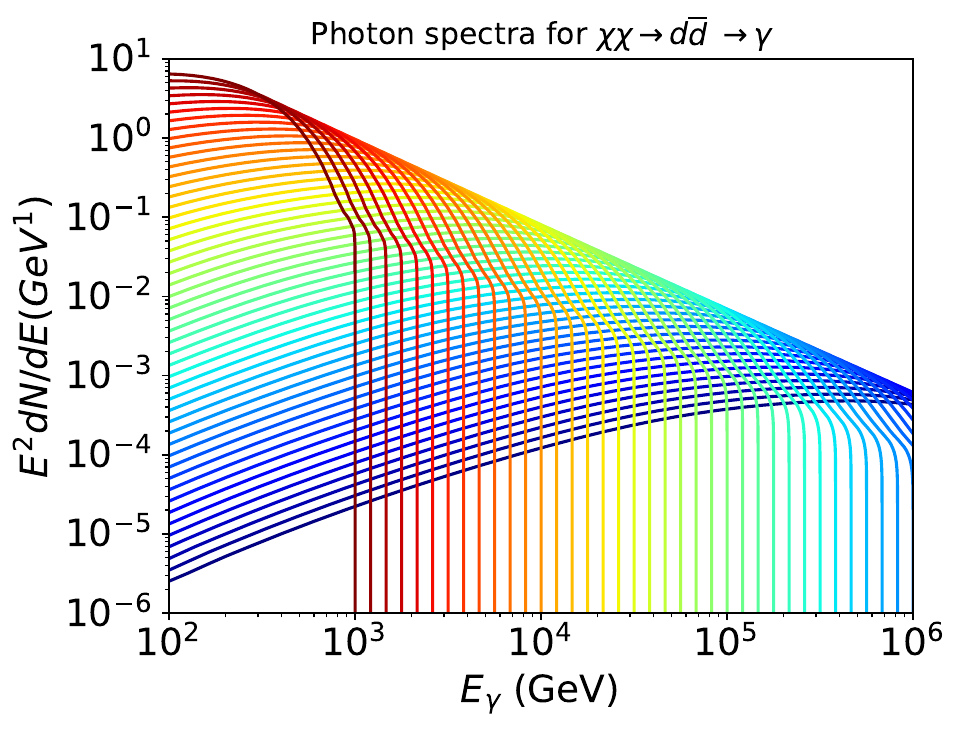} \\
     	\includegraphics[width=0.3\textwidth]{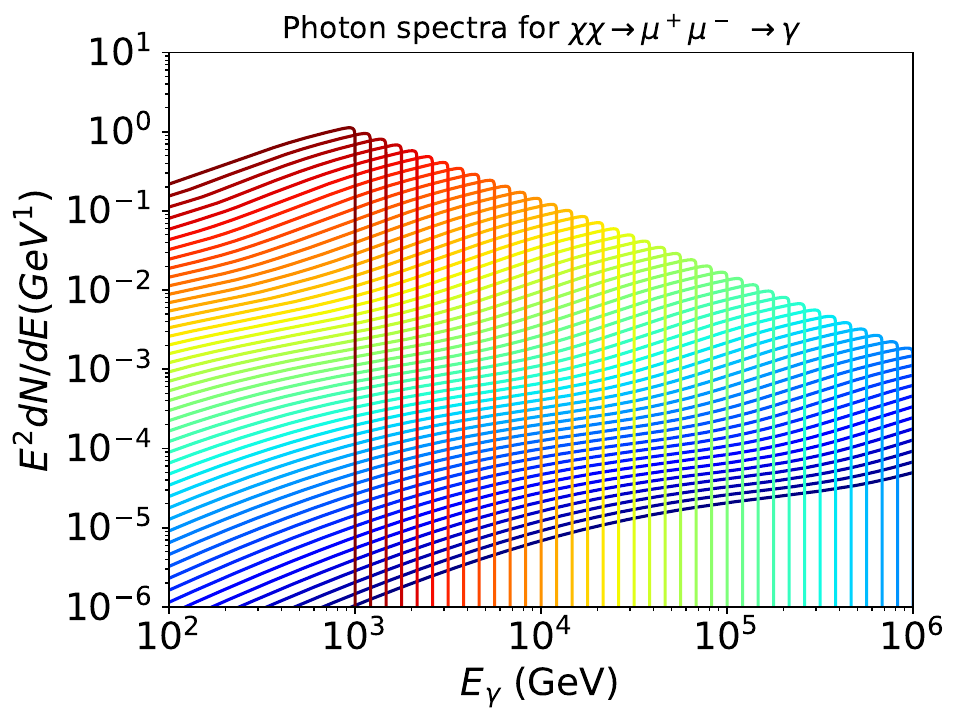} &
     	\includegraphics[width=0.3\textwidth]{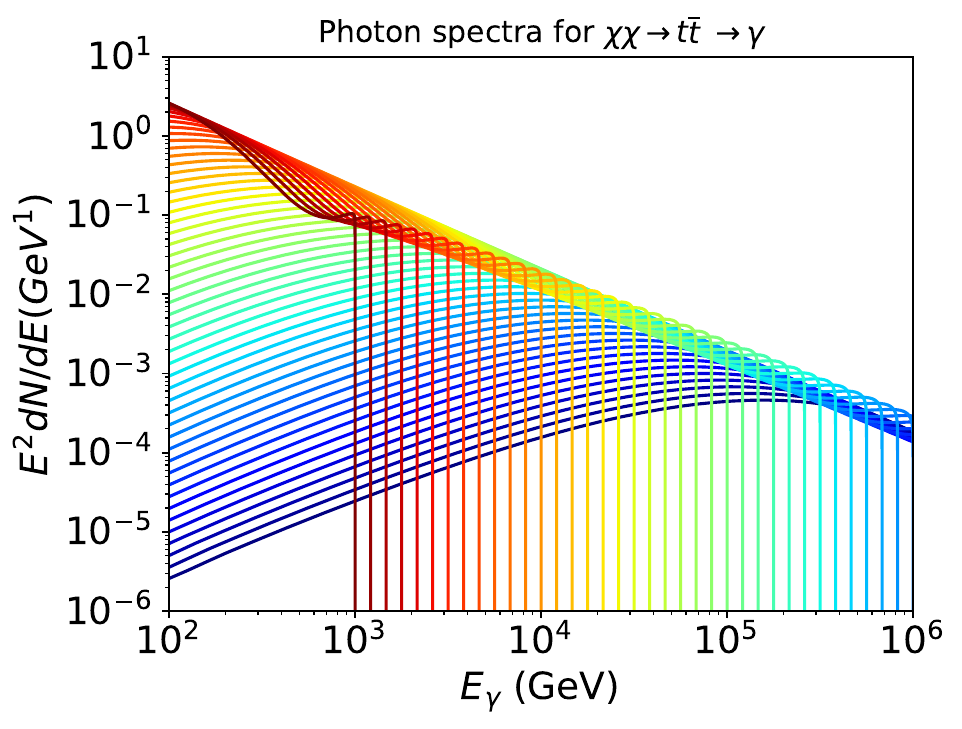} &
    	\includegraphics[width=0.3\textwidth]{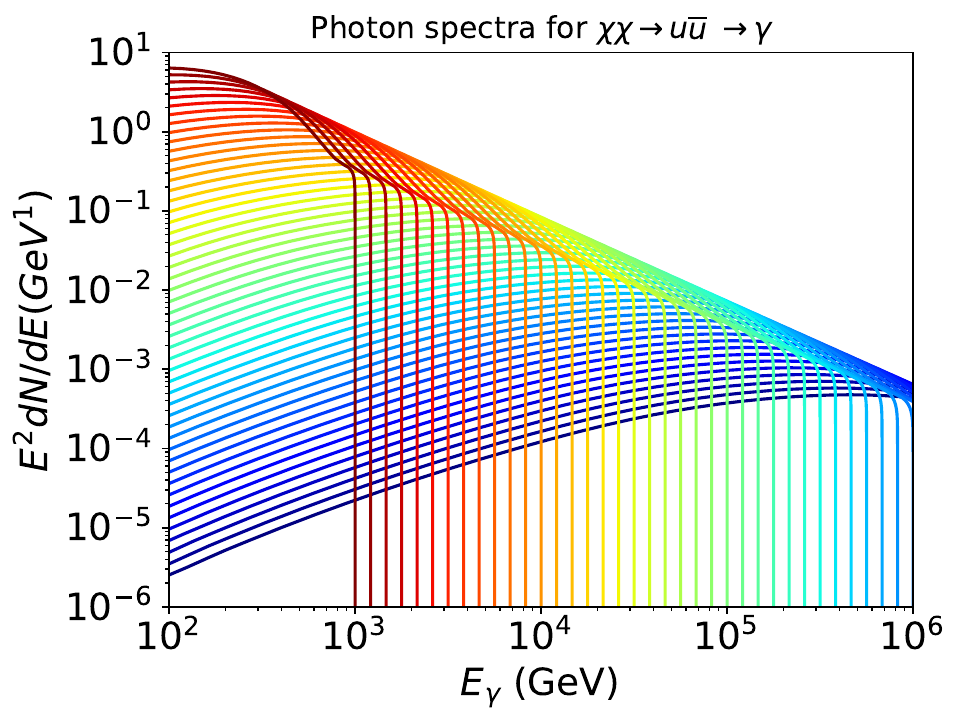} \\
   	\includegraphics[width=0.3\textwidth]{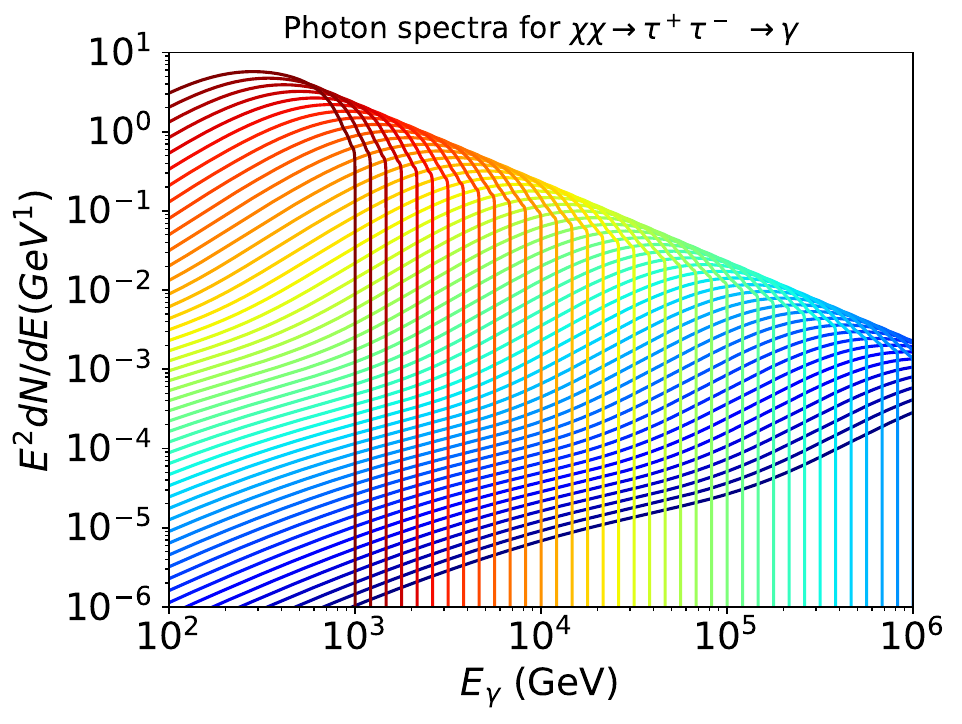} &
    	\includegraphics[width=0.3\textwidth]{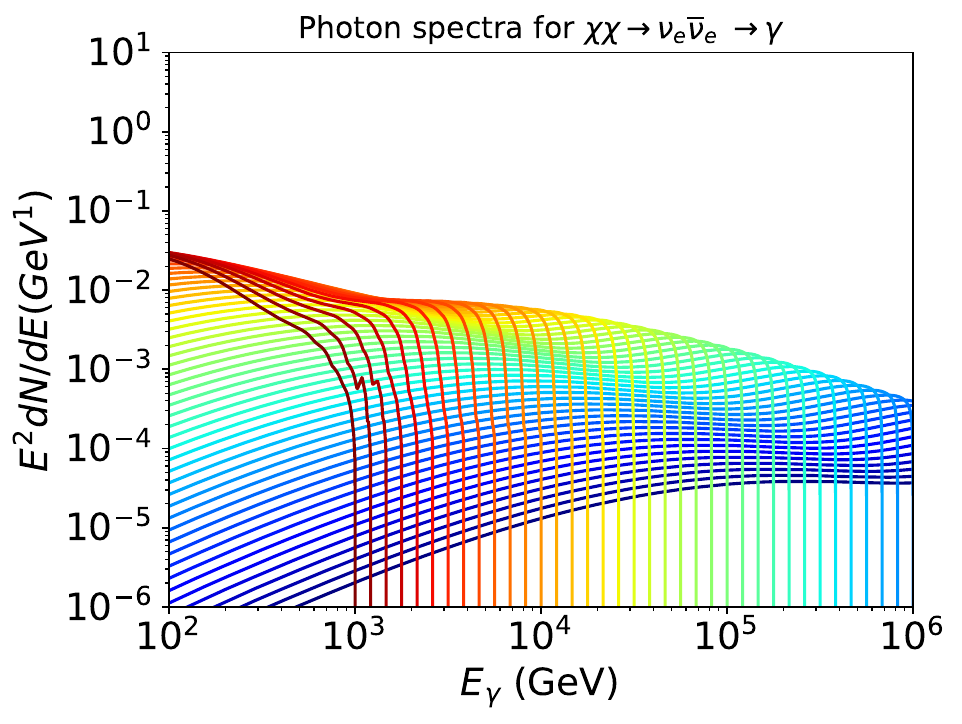} &
	\raisebox{1.2\height}{\includegraphics[width=0.2\textwidth]{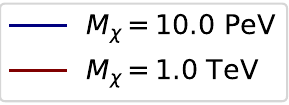}}
	\end{tabular}
    }
    \caption{Spectral models for remaining SM channels. These did not require any additional smoothing and are directly pulled from \cite{Rodd:HDM_spec}. Models that were manipulated further are in \cref{sec:mtd_particlephysics}.}
\label{fig:spectralmodel3of3}
\end{figure}

\newpage

\section{Differential \J-factor as a Function of $\theta$}\label{sec:djdomdists}
\begin{figure}[htb!]
	\centering{
	\begin{tabular}{ccc}
	\includegraphics[width=0.3\textwidth]{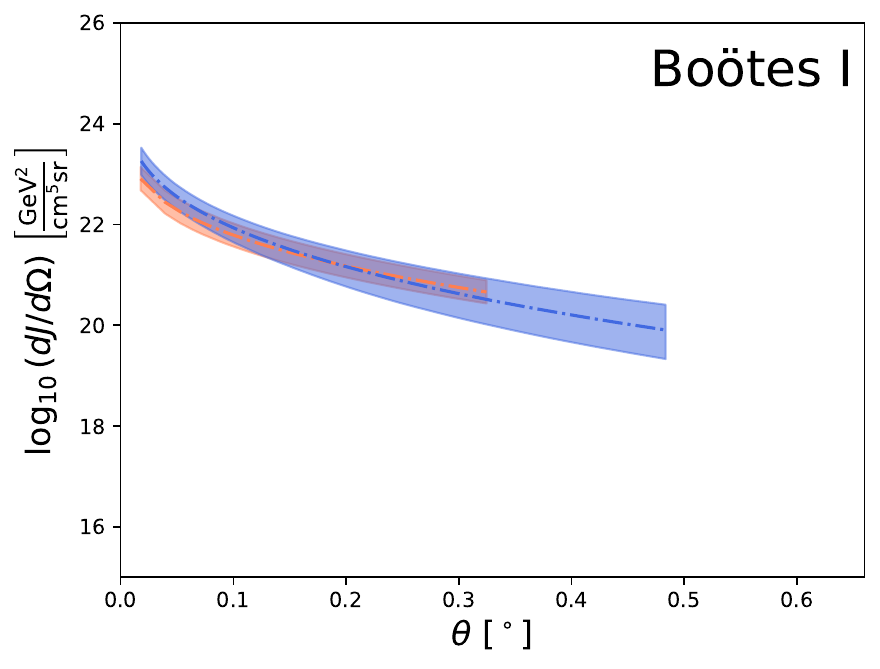} &
	\includegraphics[width=0.3\textwidth]{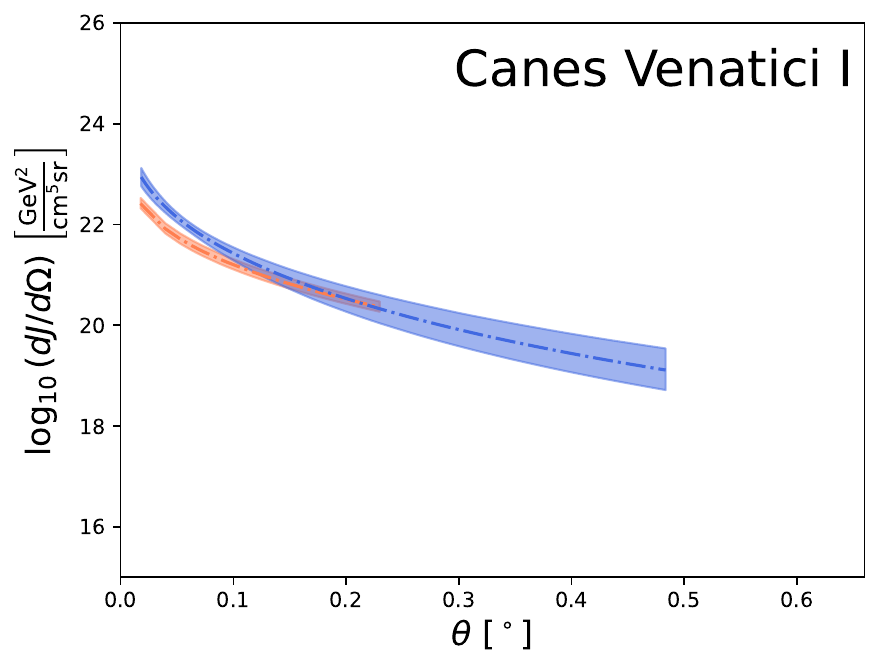} &
	\includegraphics[width=0.3\textwidth]{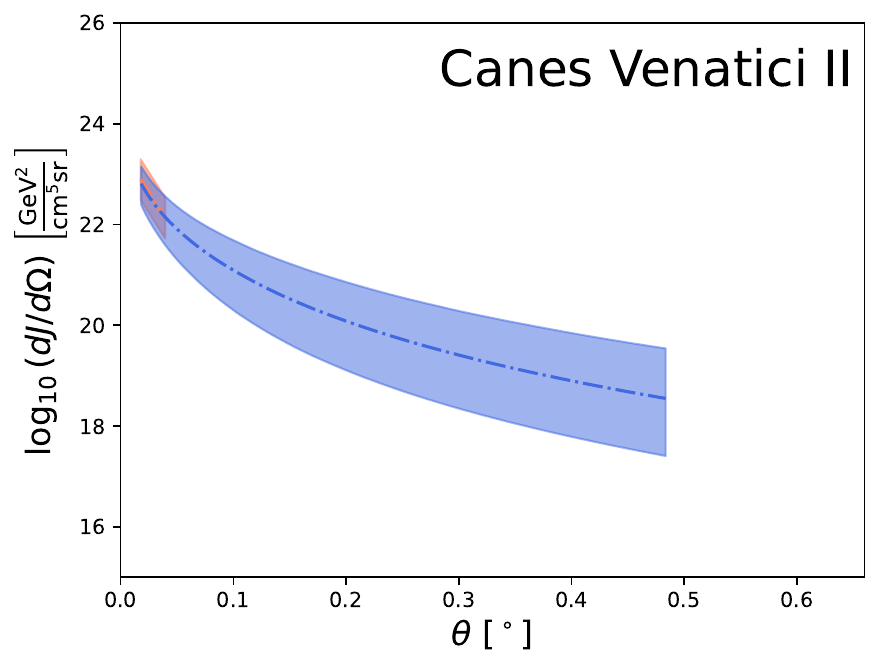} \\
	\includegraphics[width=0.3\textwidth]{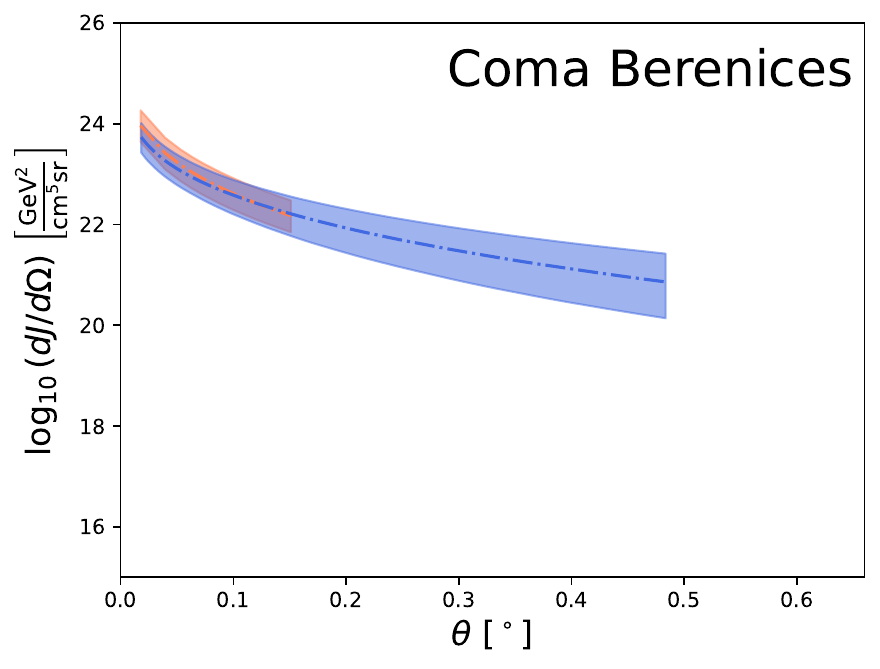} &
	\includegraphics[width=0.3\textwidth]{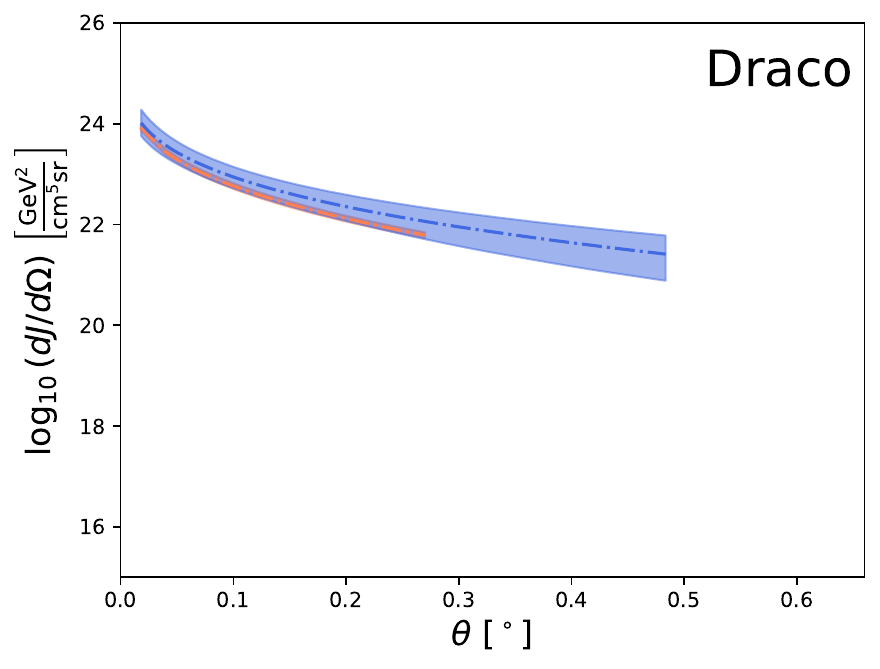} &
	\includegraphics[width=0.3\textwidth]{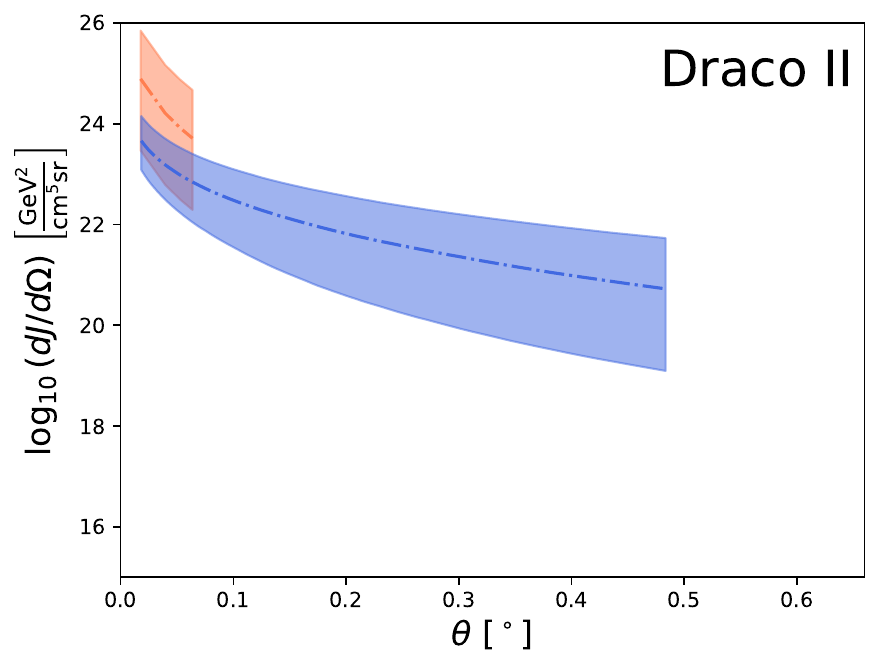} \\
	\includegraphics[width=0.3\textwidth]{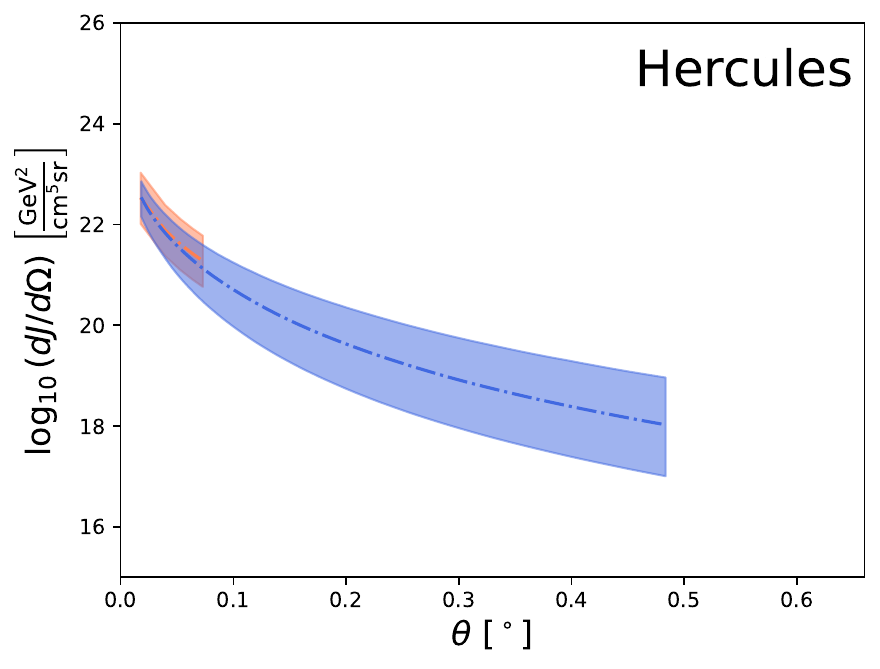} &
	\includegraphics[width=0.3\textwidth]{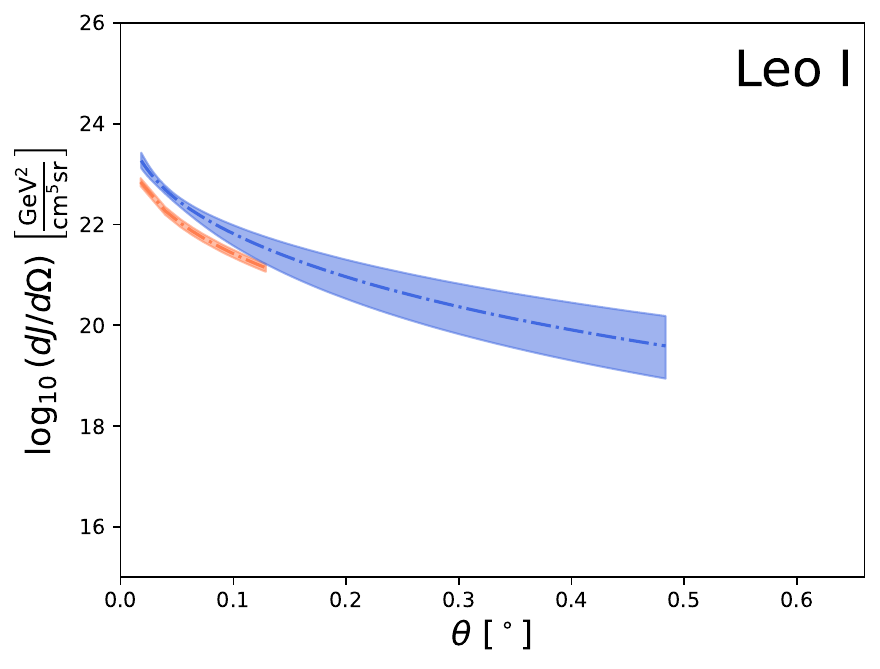} &
	\includegraphics[width=0.3\textwidth]{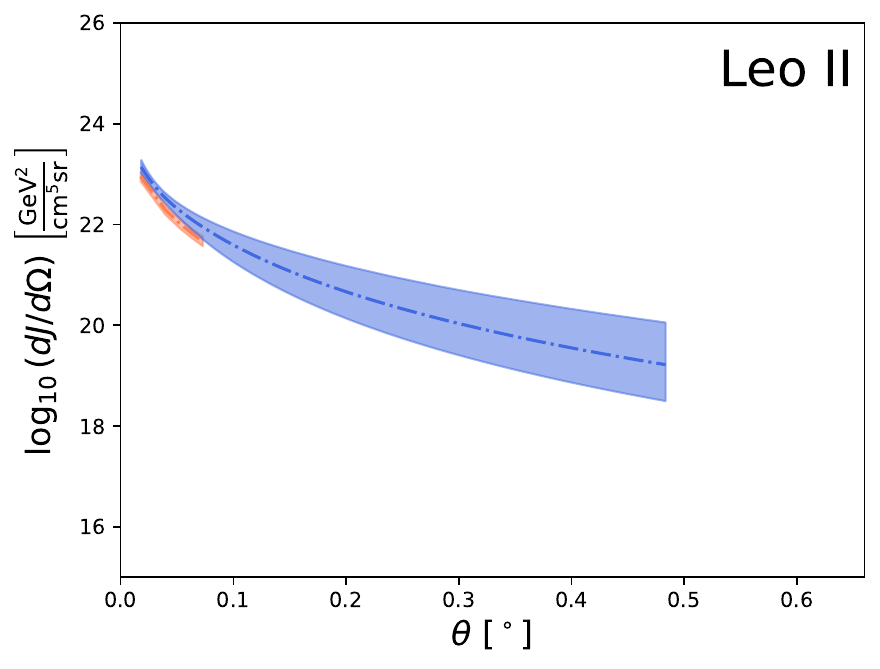} \\
	\includegraphics[width=0.3\textwidth]{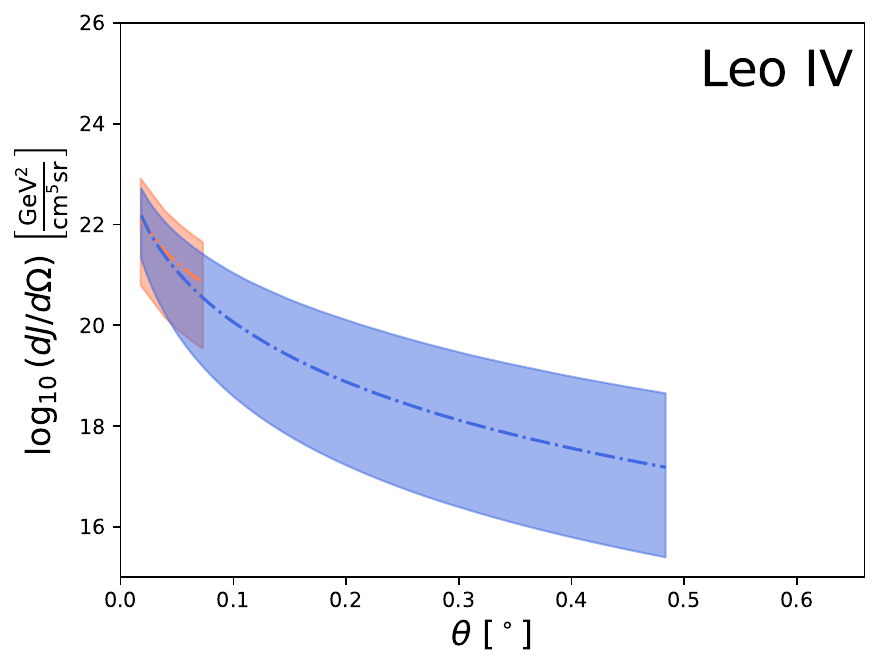} &
	\includegraphics[width=0.3\textwidth]{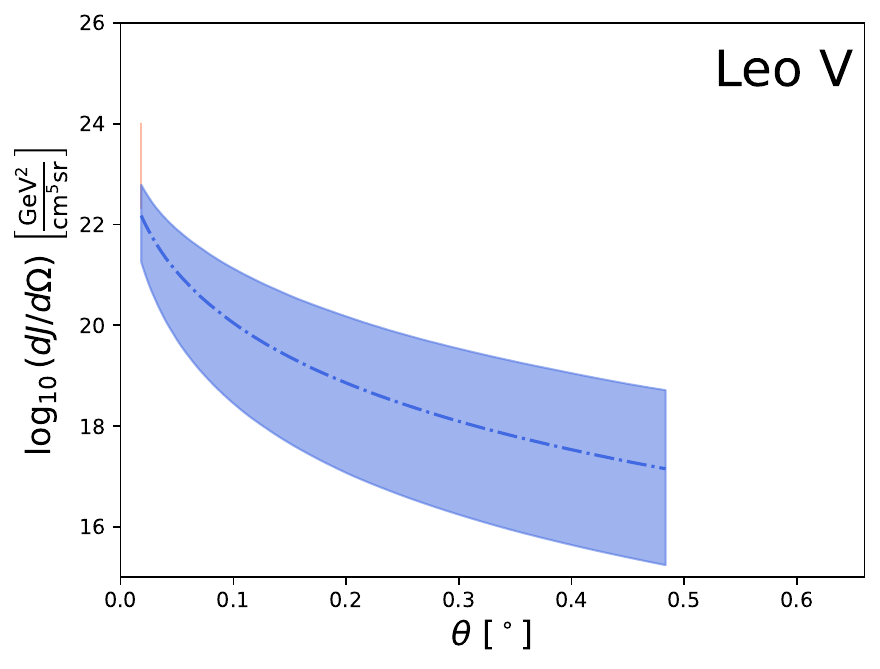} &
	\raisebox{.5\height}{\includegraphics[width=0.15\textwidth]{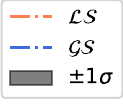}}
	\end{tabular}
	}
	 \caption{Differential \J-factor as a function of $\theta$ for all dwarfs in both catalogs. The dashed orange and blue lines represents the median $\frac{d\J}{d\Omega}$ for the \LS{} and \GS{} catalogs. The shaded bands around each are the $\pm1\sigma$ values. These are calculated from Gammapy's Jfactory class for \LS{} where each distribution extends to a truncation radius, $\alpha_c$, defined in \cref{tab:mtd_J_factor}, as determined in \cite{Ando_2020}. The Pisces II and Leo V distributions are not included here because they are treated as point sources due to their small truncation radius. For \GS{}, the uncertainty bands are calculated from 10,001 MC simulated parameters for \GS{} to a uniform radius of 0.5$^\circ$, as defined in \cite{DM_Strigari20}.}
 	\label{fig:jfact_dists}
\end{figure}

\begin{figure}[htbp!]
\centering{
	%\begin{tabular}{ccc}
	\includegraphics[width=0.3\textwidth]{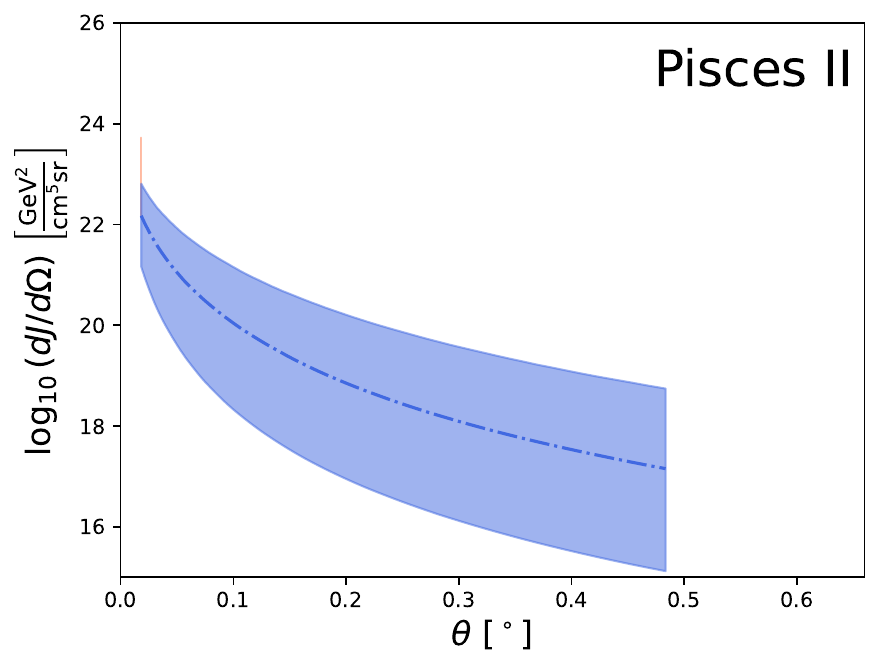}
	\includegraphics[width=0.3\textwidth]{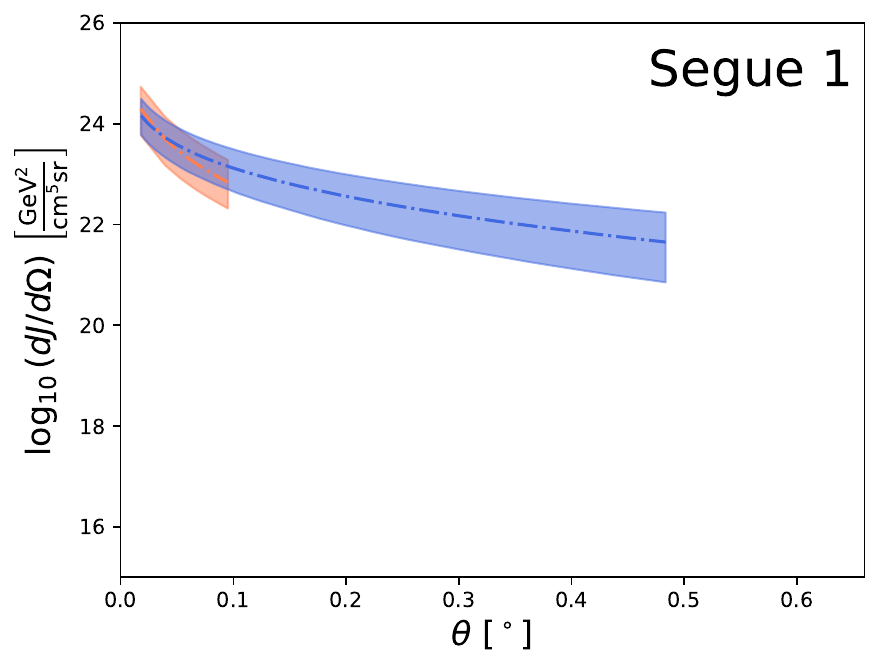}
	\includegraphics[width=0.3\textwidth]{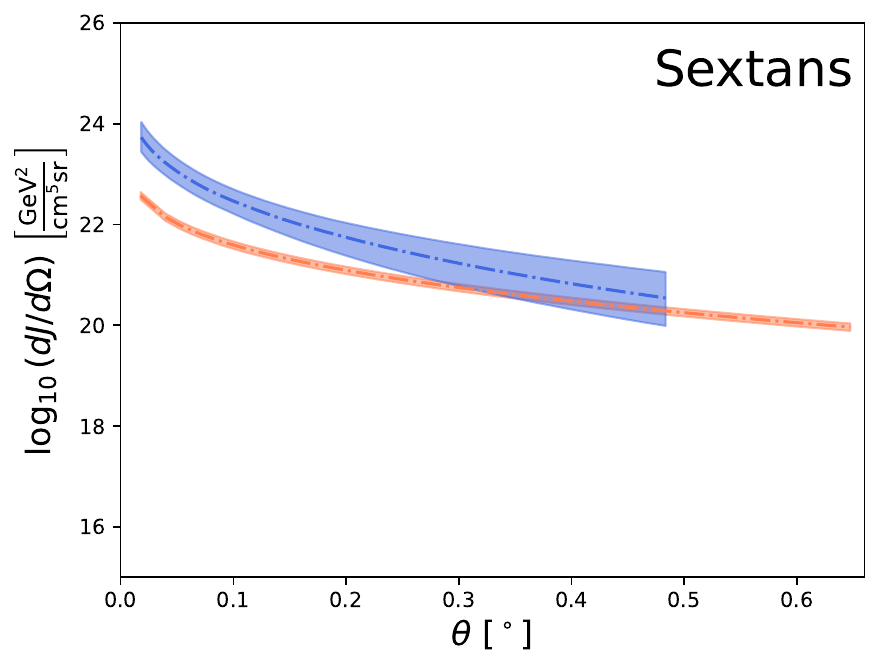}
	\includegraphics[width=0.3\textwidth]{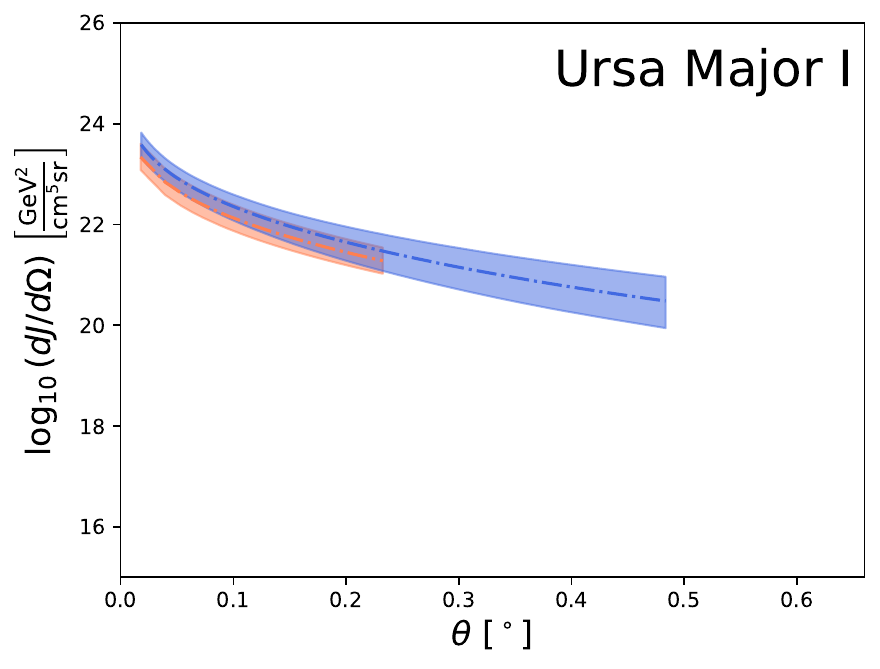}
	\includegraphics[width=0.3\textwidth]{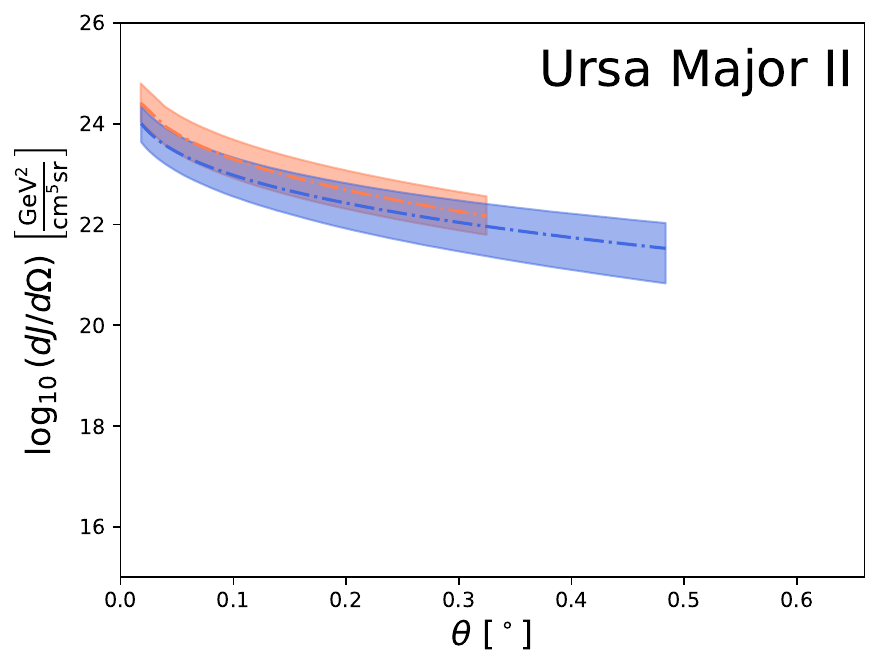}
	\includegraphics[width=0.3\textwidth]{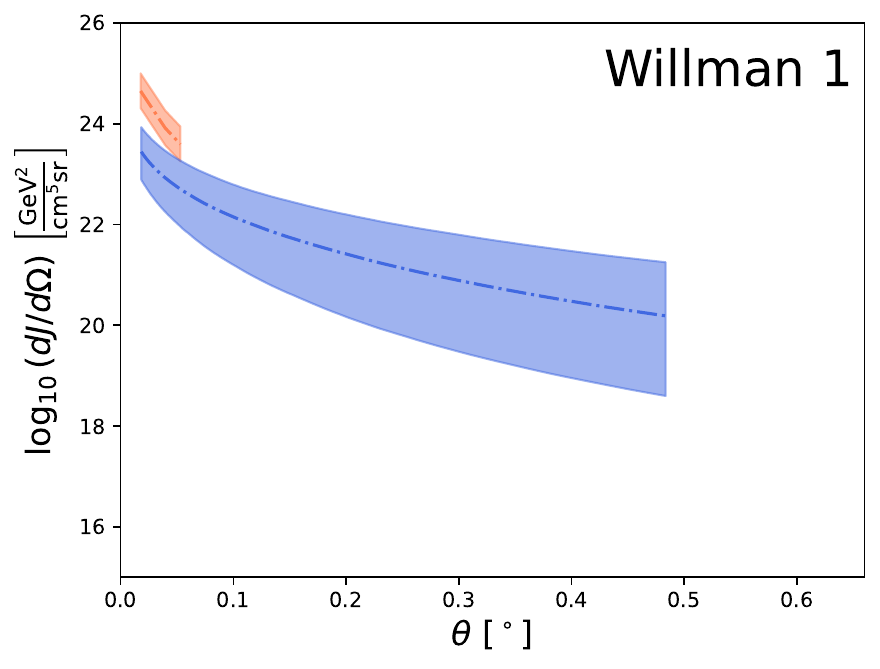}
	\includegraphics[width=0.15\textwidth]{figures/Jdists/legend.pdf}
	%\end{tabular}
	}
	\caption{Continued from \cref{fig:jfact_dists}}
 	\label{fig:jfac_dists2}
\end{figure}

%\paragraph{Note added.} This is also a good position for notes added after the paper has been written.

\end{document}